\documentclass[12pt]{article}
\usepackage{amssymb,amsmath,dsfont,verbatim}
\usepackage{colonequals}
\usepackage[normalem]{ulem}
\usepackage{slashed}
\usepackage{epsfig}
\usepackage{epstopdf}
\usepackage{latexsym}
\usepackage{graphicx}
\usepackage{booktabs}
\usepackage{bbm}
\usepackage{xspace}
\usepackage[utf8]{inputenc}
\usepackage[numbers,compress]{natbib}
\usepackage[T1]{fontenc}
\usepackage[margin=20pt,small]{caption}
\usepackage{subfig}
\usepackage{color}
\usepackage{float}
\usepackage{eso-pic}
\usepackage[
colorlinks=false,
linkcolor=darkblue,  
urlcolor=blue,    
filecolor=blue,
citecolor=red,
linktocpage=true,
pdfstartview=FitV,
bookmarksopen=true,
hidelinks
]{hyperref}
\numberwithin{equation}{section}

\newcommand*{\boxedcolor}{red}
\makeatletter
\renewcommand{\boxed}[1]{\textcolor{\boxedcolor}{%
		\fbox{\normalcolor\m@th$\displaystyle#1$}}}
\makeatother

\newcommand{\vev}[1]{\ensuremath{\langle #1 \rangle}\xspace}

\newcommand{\MM}{\mathcal M}

\newcommand{\op}{\ensuremath{\mathcal{O}}\xspace}

\newcommand{\D}{\Delta}
\newcommand{\hD}{\hat\D}

\newcommand{\Disp}{\text{D}}
\newcommand{\uhp}{{\mathbb{H}^+}}

\newcommand{\bdc}{\lambda}

\newcommand{\hC}{\hat{C}}

\newcommand{\tieta}{\tilde{\eta}}
\newcommand{\tcG}{\tilde{\cG}}
\newcommand{\cG}{{\mathcal{G}}}
\newcommand{\TTb}{T\bar{T}}
\newcommand{\AdStwo}{\text{AdS}_2}

 \def\im{{\rm Im}}
\newcommand{\id}{{\mathds 1}}

\newcommand{\hid}{\hat{\id}}
\newcommand{\dhD}{\delta\hD}

\definecolor{cardinal}{rgb}{0.6,0,0}
\definecolor{darkgreen}{rgb}{0,0.5,0}
\definecolor{golden}{rgb}{0.92, 0.7, 0}
\definecolor{midnight}{rgb}{0, 0, 0.5}
\definecolor{darkblue}{rgb}{0.2, 0, 0.8}

\newcommand{\reportnum}[2]{
	\AddToShipoutPictureBG*{%
		\AtPageUpperLeft{%
			\hspace{0.8\paperwidth}%
			\raisebox{#1\baselineskip}{%
				\makebox[0pt][l]{\textnormal{#2}}
	}}}%
}

\def\sl{\frak{sl}}

\begin{document}

\begin{titlepage}
\reportnum{-2}{DESY-23-141}
	\medskip
	\begin{center} 
		{\Large \bf A bootstrap study of minimal model deformations}

		\bigskip
		\bigskip
		\bigskip
		
		{\bf  António Antunes$^{1,2}$, Edoardo Lauria$^{3,4}$, and Balt C. van Rees$^4$\\ }
		\bigskip
		\bigskip
		${}^{1}$
		Deutsches Elektronen-Synchrotron DESY, \\
		Notkestr. 85, 22607 Hamburg, Germany
		\vskip 5mm
		${}^{2}$
	    Centro de F\'isica do Porto, Departamento de F\'isica e Astronomia,\\
		Faculdade de Ci\^encias da Universidade do Porto,\\
		Rua do Campo Alegre 687, 4169-007 Porto, Portugal
		\vskip 5mm
		${}^{3}$
		LPENS, Département de physique, École Normale Supérieure - PSL\\
		Centre Automatique et Systèmes (CAS), Mines Paris - PSL\\
		Université PSL, Sorbonne Université, CNRS, Inria, 75005 Paris\\
		\vskip 5mm
		${}^{4}$
		CPHT, CNRS, \'Ecole polytechnique, Institut Polytechnique de Paris,\\
		91120 Palaiseau, France
		\vskip 5mm
		\texttt{antonio.antunes@desy.de,~edoardo.lauria@minesparis.psl.eu} \\
		\texttt{balt.van-rees@polytechnique.edu} \\
	\end{center}
	
	\bigskip
	\bigskip
	
	\begin{abstract}
	\noindent 
	For QFTs in AdS the boundary correlation functions remain conformal even if the bulk theory has a scale. This allows one to constrain RG flows with numerical conformal bootstrap methods. We apply this idea to flows between two-dimensional CFTs, focusing on deformations of the tricritical and ordinary Ising model. We provide non-perturbative constraints for the boundary correlation functions of these flows and compare them with conformal perturbation theory in the vicinity of the fixed points. We also reproduce a completely general constraint on the sign of the $T\bar T$ deformation in two dimensions.
	\end{abstract}

	\noindent

\end{titlepage}

\setcounter{tocdepth}{2}	
\tableofcontents
\newpage

\section{Introduction}

The aim of this work is to constrain the physics of quantum field theories that undergo an RG flow between two non-trivial fixed points.

We largely focus on flows around the two lowest-lying diagonal minimal models in two spacetime dimensions: the tricritical Ising model with $m = 4$ and $c = 7/10$ and the Ising model with $m = 3$ and $c = 1/2$. Of particular interest is the flow \emph{between} these theories, which is triggered by the relevant $\phi_{(1,3)}$ deformation of the tricritical Ising model \cite{Zamolodchikov:1987ti,Zamolodchikov:1991vx}. More generally, the $\phi_{(1,3)}$ deformation of the $m$'th minimal model triggers a flow to the $m-1$'th minimal model. This is (a limit of) the integrable `staircase' flow of \cite{Zamolodchikov:1992ulx} which we will also briefly investigate.

Our method is to apply numerical conformal bootstrap techniques to the boundary correlation functions of the QFT on a hyperbolic background. For an RG flow parametrized by a scale $\mu$ in AdS with curvature radius $R$ this setup produces a one-parameter family of solutions of the boundary conformal crossing equations where  the OPE data depends on the dimensionless combination $\mu R$. We will aim to numerically constrain these families of consistent OPE data. A similar analysis was done earlier for deformations of the free massless scalar and the sine-Gordon RG flow in AdS$_2$ \cite{Antunes:2021abs}.

We get our most interesting results when the boundary correlation functions of the fixed point saturate (extrapolated) numerical bounds. This is because first-order corrections to the OPE data, which for specific deformations can be computed in conformal pertubation theory, can sometimes point \emph{into} the disallowed region. In such a case there is an inconsistency: the first-order correction to the OPE data may look totally innocuous, but in actuality the deformation cannot be exponentiated in that direction.

This situation occurs in particular for the $T\bar T$ deformation \cite{Zamolodchikov:2004ce,Smirnov:2016lqw,Cavaglia:2016oda} of a general two-dimensional CFT in AdS. From a first-order analysis we find that an irrelevant deformation of the form
\begin{equation}
	\label{ttbardeformationgeneral}
	\delta S = -\lambda \int_{\text{AdS}_2} T\bar T	 + (\text{more irrelevant deformations})\,,
\end{equation}
can only be consistent if 
\begin{equation}
	\label{ttbarconstraint}
	\lambda \geq 0\,.
\end{equation}
We note that in flat space a similar condition was found in \cite{Cardy:2015xaa,Delacretaz:2021ufg}, but the derivation from conformal bootstrap methods is new. Furthermore, our bound applies in an AdS space of arbitrary radius and sheds some light on the $\TTb$ deformation in curved space which can be of interest by itself \cite{Jiang:2019tcq,Brennan:2020dkw}.

We finally note that sign constraints for irrelevant couplings are reminiscent of the older causality constraints of \cite{Adams:2006sv} for effective field theories. The simplest of these, a bound on the $(\partial \phi)^4$ coupling for a massless shift-invariant scalar, was reproduced in two dimensions from QFT in AdS in \cite{Antunes:2021abs}. These bounds have also been vastly generalized with numerical methods \cite{Caron-Huot:2020cmc,Caron-Huot:2021rmr} and they were `uplifted' to AdS in \cite{Caron-Huot:2021enk} using the techniques of analytic functionals and conformal dispersion relations \cite{Carmi:2019cub,Mazac:2019shk,Caron-Huot:2020adz}. In those cases the IR theory however always consisted of a free massless field. Our QFT in AdS approach allows one to also constrain the irrelevant couplings around a general IR CFT.

The universality of our bootstrap bounds also comes with a potential downside. Consider a CFT deformed by two operators $\op_1$ and $\op_2$ with dimensionful couplings $\lambda_1$ and $\lambda_2$, so the dimensionless boundary OPE data becomes a function of
\begin{equation}
	g_1 \colonequals \lambda_1 R^{d-\Delta_1}\qquad \text{and} \qquad g_2 \colonequals \lambda_2 R^{d-\Delta_2}\,.
\end{equation}
With our numerical methods we can hope to carve out the embedding of the (physically allowed region in the) $(g_1,g_2)$ plane in the space of all boundary OPE data. On this plane there are however distinguished curves which correspond to the actual RG flows. For example, if both couplings are relevant then the fixed point is approached along curves that correspond to straight lines in the plane spanned by $(g_1^{1/(d- \Delta_1)},g_2^{1/(d -\Delta_2)})$. Without further assumptions these RG flow lines will however remain invisible in the bootstrap analysis. As an example we will find below a bound that is saturated to first order by a straight line in the $(g_1,g_2)$ plane instead of an actual RG flow. Another possibility, namely a plane where $g_1$ is relevant and $g_2$ is irrelevant, will also feature several times in our analysis.

In the next section we review some background material on two-dimensional BCFT, with particular emphasis on the different boundary conditions for the critical and tri-critical Ising models. This serves as the starting point for the RG flow in AdS, since for a conformally invariant system, physics in AdS and in the BCFT are related by a Weyl transformation.

In section \ref{sec:pertres}, we make use of the results of \cite{Lauria:2023uca} to study conformal perturbation theory in AdS at leading order. We derive the modification to boundary CFT data when the bulk is perturbed by a general Virasoro primary or by a special Virasoro descendant of the identity: The $\TTb$ operator. 

In section \ref{sec:numerical} we use the numerical conformal bootstrap of the boundary four-point functions to bound bulk RG flows. We compare to the results of section \ref{sec:pertres} in the perturbative regime finding saturation of the bounds for a general Virasoro primary deformation and a sign constraint on the $\TTb$ coupling. We then focus on the flow between the tricritical and critical Ising models, where the bootstrap carves out an allowed region with several interesting features, some of which can be identified with the physical RG flow with simple boundary conditions. We also perform a detailed numerical analysis of deformations in the vicinity of the UV and IR BCFTs. Finally, we consider bounds on the values of the correlator and its derivatives which turn out to be saturated by a different choice of boundary conditions for the tricritical Ising model and suggest a generalization to the full `staircase' RG flow.

We conclude in section \ref{sec:outlook} where we list some possible future directions. Some additional technical details on boundary correlation functions and bulk conformal perturbation theory, as well as a review of the `staircase' model are left to the various appendices.

\subsubsection*{Ancillary material}
With the submission of this work we have included a \texttt{Mathematica} notebook \texttt{ancillary.nb} which contains all the substantial computations done in this work.

\section{Some BCFT background}
\label{sec:setup}

Recall that correlation functions of a \emph{conformal} field theory in AdS are just boundary conformal field theory (BCFT) correlation functions up to a simple Weyl rescaling.\footnote{Note that in going from the upper half-plane to $\AdStwo$ the stress-energy tensor is unchanged (see e.g. \cite{Hogervorst:2021spa}).} In this section we therefore review some BCFT background material that will be important in the sequel. A more detailed discussion and references to the original literature can be found for example in the books~\cite{Ginsparg:1988ui,DiFrancesco:1997nk,Mussardo:2010mgq,Recknagel:2013uja}. Our conventions are collected in appendix \ref{app:conventions}.

Below we will parametrize the upper half-plane $\uhp$ with a complex coordinate $z = x + i y$ where $y \geq 0$. We use conventions where the one-point function coefficients of global bulk primaries are \cite{McAvity:1993ue,McAvity:1995zd,Liendo:2012hy}:
\begin{align}\label{oneptdef}
	\langle\phi(z,\bar z)\rangle_{\text{BCFT}}=\frac{B_{\phi}}{(2y)^{\D}}\,.
\end{align}

\subsection{The displacement and its square}\label{universalco}
In two-dimensional BCFTs the stress-energy tensor obeys the boundary condition \cite{Cardy:1984bb,Cardy:1989ir,Cardy:2004hm}
\begin{align}
	T(z)=\bar{T}({\bar z})\,,\quad \im~z=0\,.
\end{align}
By the Ward identities the boundary spectrum therefore necessarily features a displacement operator, defined as:
\begin{align}\label{bOPET}
	\Disp (x) = T(x+i y)\rvert_{y=0}\,.
\end{align}
The displacement is a parity-even boundary global primary with scaling dimension $\hD_\Disp = 2$. In terms of the one remaining Virasoro representation it is a level-two Virasoro descendant of the boundary identity $\hid$.

Correlation functions of $\Disp(x)$ can be obtained from correlation functions with the stress-energy tensor $T(z)$ on the upper half-plane by restricting all $T$-insertions to the real axis. We have, for example 
\begin{align}\label{D4ptfunction}
	&\qquad\qquad\langle \Disp(x_1) \Disp(x_2)\rangle_{\uhp} = \frac{c/2}{x_{12}^4}\,, \quad \langle \Disp(x_1) \Disp(x_2) \Disp(x_3)\rangle_{\uhp} = \frac{c}{(x_{12})^{2}(x_{23})^{2}(x_{31})^{2}}\,,\nonumber\\
	&\langle \Disp(x_1) \Disp(x_2) \Disp(x_3) \Disp(x_4)\rangle_{\uhp}= \frac{c^2/4}{(x_{12})^4 (x_{34})^4}\left[1+\eta ^4+\left(\frac{\eta }{\eta -1}\right)^4+\frac{8\eta ^2 ((\eta -1) \eta +1)}{c(\eta -1)^2}\right]\,,
\end{align}
with the four-point cross-ratio given by 
\begin{align}\label{eta_def_line}
	\eta= \frac{x_{12}x_{34}}{x_{13}x_{24}}\,,\quad 0<\eta<1\,.
\end{align}
More examples are discussed in appendix A.3 of \cite{Lauria:2023uca}.

The self-OPE of $\Disp(x)$ is just obtained from restricting the self-OPE of $T(z)$ to the boundary:
\begin{align}\label{DDOPmain}
	\Disp(x)\Disp(0)=\frac{c/2}{x^4}\hid+\frac{2}{x^2}\Disp(0)+\frac{1}{x} \Disp'(0)+\frac{3}{10}\Disp''(0)+\Disp^2(0)+O(x^2)\,,
\end{align}
where $'$ indicates derivatives along the boundary and we omitted higher-order contributions. Here we see a new operator $\Disp^2(x)$, which is the unique boundary global primary with $\hD_{\Disp^2} = 4$ in the identity module.

Both $\Disp(x)$ and $\Disp^2(x)$ will play a central role in our analysis. 

\subsection{Minimal models}
The $m$'th unitary diagonal minimal model, or more precisely $\mathcal{M}_{m+1,m}$, have central charge
\begin{equation}\label{centralmain}
	c=1-\frac{6}{m(m+1)}\,,\qquad {m=3,\,4,\,5,\ldots\,}~.
\end{equation}
We will mostly be interested in the Ising model with $m = 3$ and the tricritical Ising model with $m = 4$.

For a given $m$ the (bulk) Virasoro primaries are $\phi_{(r,s)}(z,\bar z)$ with integer $r$ and $s$ obeying the constraints
\begin{align}\label{VirasoroPrimariesrsmain}
	1\leq r\leq m-1\,,\quad 1\leq s\leq m\,,\quad (r,s) \cong (m-r,m+1-s)\,.
\end{align}
They have quantum numbers
\begin{align}
	\D_{r,s}=h_{r,s}+\bar{h}_{r,s}\,,\quad \ell_{r,s}=h_{r,s}-\bar{h}_{r,s}=0\,,
\end{align}
where
\begin{equation}\label{VirasoroPrimariesmain}
	h_{r,s}= \frac{\bigl( (m+1)r-ms \bigr)^2-1}{4m (m+1)}\,.
\end{equation}
The $m$'th diagonal minimal model enjoys a $\mathbb{Z}_2$ symmetry under which the charge of a Virasoro primary with labels $(r,s)$ is~\cite{Cardy:1986gw,Cappelli:1986hf,Ruelle:1998zu}
\begin{align}\label{rsparity}
	\epsilon^{(m)}_{(r,s)}=(-1)^{(m+1) r+m s+1}\,.
\end{align}
The fusion rules between the bulk operators read
\begin{align}\label{fusionrulesholo}
	\phi_{(r,s)} \times \phi_{(r',s')} = \sum_{(r'',s'') \,\in\, \mathcal S(r,s;r',s')} \phi_{(r'',s'')}\,,
\end{align}
where, for given positive integers $(r,s)$ and $(r',s')$, we define the set
\begin{multline}
	\mathcal S(r,s;r',s') = \\ \{(r'',s'')| r_\text{min} \leq r'' \leq r_\text{max} \land s_\text{min} \leq s'' \leq s_\text{max} \land r + r' + r'' \text{ odd} \land s + s' + s'' \text{ odd}\}\,,
\end{multline}
with
\begin{align}
	r_\text{min} &= |r - r'| + 1\,, & r_\text{max} &= \min(r+r'-1,\;2m-r-r'-1)\,,\nonumber\\
	s_\text{min} &= |s - s'| + 1\,, & s_\text{max} &= \min(s+s'-1,\;2m-s-s'+1)\,.
\end{align}

\subsection{Minimal model boundary conditions}
The  `elementary' conformal boundary conditions (which have a unique identity operator) for the minimal models are the so-called Cardy states \cite{Cardy:1984bb,Cardy:1989ir,Cardy:2004hm}. Like the Virasoro primaries, they are also labeled with two integers $(a_1,a_2)$ that obey:
\begin{align}\label{bclabelmain}
	{\bf a}= (a_1,a_2)_m\,,\quad 1\leq a_1\leq m-1\,,\quad 1\leq a_2\leq m\,,\quad (a_1,a_2) \cong (m-a_1,m+1-a_2)\,.
\end{align}

The one-point function coefficients in eq.~\eqref{oneptdef} are completely determined by the Cardy state~\cite{Cardy:1991tv,Lewellen:1991tb}. The explicit formula for $\phi_{(r,s)}$ in boundary condition $\bf a$ is:
\begin{align}\label{bulkboundaryidentity}
	&B_{(r,s)}^{\bf a} = \frac{S_{(a_1,a_2)}^{(r,s)} \sqrt{S_{(1,1)}^{(1,1)}}}{S_{(a_1,a_2)}^{(1,1)} \sqrt{S_{(1,1)}^{(r,s)}}}\,,\nonumber\\
		S_{(a_1,a_2)}^{(r,s)}=\sqrt{\frac{8}{m (m+1)}}&(-1)^{1+a_1s+a_2r}\,
	\sin\left(\tfrac{m+1}{m}\pi a_1r\right)
	\sin\left(\tfrac{m}{m+1}\pi a_2s\right)\,.
\end{align}

For a given boundary condition $\bf a$ there are boundary Virasoro primaries $\psi_{(r,s)}(x)$ with scaling dimensions
\begin{align}
	\hD_{r,s}=h_{r,s}\,.
\end{align}
The labels $(r,s)$ here do not only obey the constraints \eqref{VirasoroPrimariesrsmain}, but are also restricted to be such that they appear in the $\phi_{\bf a}\times \phi_{\bf a}$ OPE. In other words, $\psi_{(r'',s'')}$ only exists in the $(a_1,a_2)$ boundary condition if $(r'',s'') \in \mathcal S(a_1,a_2;a_1,a_2)$.

Another selection rule is as follows. If we send the bulk operator $\phi_{(r,s)}(z,\bar z)$ to the boundary then the bulk-boundary operator expansion generally contains a subset of the full set of boundary operators: the operator $\psi_{(r'',s'')}$ can only appear if $(r'',s'') \in \mathcal S(r,s;r,s)$ as well.

\subsection{The Ising CFT}
We now review the consistent boundary conditions for the Ising CFT which is the $m=3$ diagonal minimal model. It has $c=1/2$ and is characterized by the following set of scalar Virasoro primaries:
\begin{center}
	\begin{tabular}{|c|c|c|}
		\hline
		\hline
		$\D$&$\text{Symbol}$& $(r,s)$\\
		\hline
		\hline
		0&$\id$& $(1,1)$ or $(2,3)$\\
		\hline
		1/8&$\sigma$& $(1,2)$ or $(2,2)$\\
		\hline
		1&$\epsilon$& $(1,3)$ or $(2,1)$\\
		\hline
		\hline
	\end{tabular}
\end{center}
The non-trivial fusion rules are
\begin{align}\label{ising_fusions}
	\epsilon \times \epsilon = \id\,,\quad	
	\sigma \times \epsilon = \sigma\,,\quad
	\sigma \times \sigma =\id+ \epsilon\,.
\end{align}
We recall that the bulk theory is invariant under a $\mathbb{Z}_2$ global symmetry under which $\sigma$ is odd and $\epsilon$ is even.

\subsubsection{The $(1,2)_3$ BCFT}
Out of the three elementary conformal boundary conditions, the ones labelled by $(1,1)_3$ and $(1,3)_3$ are $\mathbb{Z}_2$-breaking while the one labelled by $(1,2)_3$ is $\mathbb{Z}_2$-preserving.\footnote{A conformal boundary condition is $\mathbb{Z}_2$-invariant when all bulk one-point functions of $\mathbb{Z}_2$-odd operators vanish.} We will focus on the latter here.

First, in Table~\ref{tbl:smallconfbcis} we report the spectrum of allowed boundary Virasoro primaries, as well as non-vanishing bulk one-point functions.

\begin{table}[H]
	\begin{center}
		$\begin{array}{|c|c|c|c|}
			\hline
			\hline
			(a_1,a_2)_m &\text{Boundary spectrum}&  \mathbb{Z}_2& \hD \\
			\hline
			\hline
			(1,2)_3& \hid&+1& 0  \\
			&\psi_{(1,3)}\simeq \psi_{(2,1)}&-1& 1/2  \\
			\hline
			\hline
		\end{array}
\qquad \qquad
		\begin{array}{|c|c|c|}
			\hline
			\hline
			(a_1,a_2)_m &\text{Bulk primary}& B_\phi^{\bf a}  \\
			\hline
			\hline
			(1,2)_3 &\id&1 \\
			 &\epsilon&-1\\
			\hline
			\hline
		\end{array}$
	\end{center}
	\caption{The $\mathbb{Z}_2$-preserving conformal boundary condition for the Ising model. In the first Table: spectrum of boundary Virasoro primaries. In the second Table, non-vanishing one-point functions $B_\phi^{\bf a}$ (see eq.~\eqref{oneptdef}) of bulk Virasoro primaries.}
	\label{tbl:smallconfbcis}
\end{table}

Let us discuss the reason behind the $\mathbb{Z}_2$ charge assignments in Table~\ref{tbl:smallconfbcis}. In a given $\mathbb{Z}_2$-preserving conformal boundary condition, a boundary global primary that appears in the bulk-boundary OPE of a $\mathbb{Z}_2$-even (odd) operator, is $\mathbb{Z}_2$-even (odd). The $\mathbb{Z}_2$ charge for the $\psi_{(1,3)}$ boundary operator in the $(1,2)_3$ conformal boundary condition can for example be determined from the bulk-boundary OPE:
\begin{align}\label{sigmabdising}
	\sigma(x+i y,x-i y) = \frac{B_{\sigma}^{\bf a}}{(2y)^{\D_{\sigma}}}\hid +\frac{B_{\sigma}^{{\bf a}\,(1,3)}}{(2y)^{\D_{\sigma}-\hD_{1,3}}}\psi_{(1,3)}(x)+\text{desc.}
\end{align}
After deriving the bulk two-point function of $\sigma$ one finds that~\cite{Lewellen:1991tb,Runkel:1998he,Liendo:2012hy} 
\begin{align}
	(B_{\sigma}^{{\bf a}\,(1,3)})^2=\frac{1}{\sqrt{2}}\,,
\end{align}
(see also our appendix~\ref{app:phi122pt} for an independent derivation of this result). This is not zero and therefore $\psi_{(1,3)}$ is $\mathbb{Z}_2$-odd. Consequently the boundary fusion rule must be
\begin{align}
	\psi_{(1,3)} \times \psi_{(1,3)} = \hid\,.
\end{align}
This is also the holomorphic counterpart of the fusion rules in eq.~\eqref{ising_fusions}, which is not surprising: boundary Virasoro primaries behave as holomorphic Virasoro primaries as far as Ward identities are concerned.

We do not preserve Virasoro symmetry along the RG flow so it is important to have an understanding of the decomposition of four-point functions into global conformal blocks. We will study the $\mathbb{Z}_2$-invariant four-point correlation functions between $\psi_{(1,3)}$ and $\Disp$. These have the following (schematic) OPEs:
\begin{align}\label{leadingOPEising}
	\psi_{(1,3)}\times \psi_{(1,3)} &\sim \hid + \Disp +\dots\,,\nonumber\\
	\psi_{(1,3)}\times \Disp &\sim \psi_{(1,3)}+\psi^{(4)}_{(1,3)}+ \psi^{(7)}_{(1,3)}+\dots\,,\nonumber\\
	\Disp\times \Disp &\sim \hid + \Disp +\Disp^2+\dots\,.
\end{align}
The superscript $(n)$ denotes level-$n$ Virasoro descendants which are global primaries. The quantum numbers of the operators that appear in eq.~\eqref{leadingOPEising} are then reported in Table~\ref{tbl:chargesIsing} (the analysis of the parity-odd channel in correlation functions with the displacement is worked out in appendix~\ref{app:displcorrelatorsDecn}).

\begin{table}[H]
	\begin{center}
		$\begin{array}{|c||c|c|c|}
			\hline
			& \hD &\mathbb{Z}_2 & P \\
			\hline\hline
			\psi_{(1,3)}& \frac{1}{2}&-1& +1\\
			\hline
			\Disp& 2&+1& +1\\
			\hline
			\psi^{(4)}_{(1,3)}& 4+\frac{1}{2}&-1& +1\\
			\hline
			\Disp^2& 4&+1& +1\\
			\hline
			\psi^{(7)}_{(1,3)}& 7+\frac{1}{2}&-1& -1\\
			\hline
		\end{array}$
	\end{center}
	\caption{Ising model with $(1,2)_3$ conformal boundary condition. Quantum numbers of the leading global boundary primaries appearing in the OPEs \eqref{leadingOPEising}.}
	\label{tbl:chargesIsing}
\end{table}

\subsection{The tricritical Ising CFT}
The tricritical Ising model is the $m=4$ diagonal minimal model. It has $c=7/10$ and is characterized by the following set of scalar Virasoro primaries:
\begin{center}
	\begin{tabular}{|c|c|c|}
		\hline
		\hline
		$\D$&$\text{Symbol}$& $(r,s)$\\
		\hline
		\hline
		0&$\id$& $(1,1)$ or $(3,4)$\\
		\hline
		1/5&$\epsilon$& $(1,2)$ or $(3,3)$\\
		\hline
		6/5&$\epsilon'$& $(1,3)$ or $(3,2)$\\
		\hline
		3&$\epsilon''$& $(1,4)$ or $(3,1)$\\
		\hline
		3/40&$\sigma$& $(2,2)$ or $(2,3)$\\
		\hline
		7/8&$\sigma'$& $(2,4)$ or $(2,1)$\\
		\hline
		\hline
	\end{tabular}
	\label{tricr_ops}
\end{center}
The non-trivial fusion rules are (see eq.~\eqref{fusionrulesholo})
\begin{align}\label{tricr_fusions}
	&\epsilon \times \epsilon =\id+\epsilon'\,,\quad
	&&\epsilon \times \epsilon'=\epsilon+\epsilon''\,,\qquad
	&&\epsilon \times \epsilon''=\epsilon'\,,\nonumber\\
	&\epsilon' \times \epsilon'=\id+\epsilon'\,,\quad
	&&\epsilon' \times \epsilon''=\epsilon\,,\qquad
	&&\epsilon'' \times \epsilon''=\id\,,\nonumber\\
	&\epsilon \times \sigma=\sigma+\sigma'\,,\quad
	&&\epsilon \times \sigma'=\sigma\,,\qquad
	&&\epsilon '\times \sigma=\sigma+\sigma'\,,\nonumber\\
	&\epsilon' \times \sigma'=\sigma\,,\quad
	&&\epsilon'' \times \sigma=\sigma\,,\qquad
	&&\epsilon'' \times \sigma'=\sigma'\,,\nonumber\\
	&\sigma \times \sigma=\id+\epsilon+\epsilon'+\epsilon''\,,\quad
	&&\sigma \times \sigma'=\epsilon+\epsilon'\,,\qquad
	&&\sigma' \times \sigma'=\id+\epsilon''\,.
\end{align}
The bulk theory is invariant under a $\mathbb{Z}_2$ global symmetry under which only $\sigma$ and $\sigma'$ are odd, see e.g.~\cite{Cardy:1986gw,Cappelli:1986hf}.

We will again focus on the elementary $\mathbb{Z}_2$-preserving conformal boundary conditions. There are two of these, with labels $(2,1)_4$ and $(2,2)_4$.

\subsubsection{The $(2,1)_4$ BCFT}
We present the basic observables for the $(2,1)_4$ BCFT in Table \ref{tbl:tric214}.

\begin{table}[H]
	\begin{center}
		$\begin{array}{|c|c|c|c|}
			\hline
			\hline
			(a_1,a_2)_m &\text{Boundary spectrum}&  \mathbb{Z}_2& \hD \\
			\hline
			\hline
			(2,1)_4& \hid&+1& 0  \\
			&\psi_{(3,1)}&-1& 3/2  \\
			\hline
			\hline
		\end{array}
		\qquad \qquad
		\begin{array}{|c|c|c|}
			\hline
			\hline
			(a_1,a_2)_m &\text{Bulk primary}& B_\phi^{\bf a} \\
			\hline
			\hline
			(2,1)_4 &\id & 1\\
			&\epsilon &-\sqrt{\frac{1}{2} \left(1+\sqrt{5}\right)} \\
			&\epsilon' &\sqrt{\frac{1}{2} \left(1+\sqrt{5}\right)} \\
			&\epsilon''&-1 \\
			\hline
			\hline
		\end{array}$
	\end{center}
	\caption{The boundary spectrum and bulk one-point functions (see eq.~\eqref{oneptdef} for conventions) for the $(2,1)_4$ BCFT.}
	\label{tbl:tric214}
\end{table}

In this boundary condition the non-trivial boundary Virasoro primary is $\psi_{(3,1)}$. It is $\mathbb{Z}_2$-odd because of the bulk-boundary OPE
\begin{align}\label{sigmaprimetricbOPE}
	\sigma'(x+i y,x-i y) =\frac{B_{\sigma'}^{\bf a}}{(2y)^{\D_{\sigma'}}}\hid +\frac{B_{\sigma'}^{{\bf a}\,(3,1)}}{(2y)^{\D_{\sigma'}-\hD_{3,1}}}\psi_{(3,1)}(x)+\text{desc.}\,
\end{align}
with non-zero coefficient~\cite{Runkel:1998he}
\begin{align}
	(B_{\sigma'}^{{\bf a}\,(3,1)})^2=\frac{7}{4 \sqrt{2}}\,.
\end{align}
We have reproduced this result by studying the bulk two-point function of $\sigma'$ in appendix~\ref{app:phi212pt}. 

We will below be interested in the global conformal block decomposition of the $\mathbb{Z}_2$-invariant four-point correlation functions between $\psi_{(3,1)}$ and $\Disp$. At tree-level, the leading OPEs are (schematically)
\begin{align}\label{leadingOPEstric}
	\psi_{(3,1)}\times \psi_{(3,1)} &\sim \hid + \Disp +\dots\,,\nonumber\\
	\psi_{(3,1)}\times \Disp &\sim \psi_{(3,1)}+\psi^{(2)}_{(3,1)} + \psi^{(5)}_{(3,1)}+\dots\,,\nonumber\\
	\Disp\times \Disp &\sim \hid + \Disp +\Disp^2+\dots\,.
\end{align}
The superscript $(n)$ denotes level-$n$ Virasoro descendants which are global primaries. 
The quantum numbers of the operators that appear in \eqref{leadingOPEstric} are reported in Table~\ref{tbl:chargesTric}. The analysis of the parity-odd channel in correlation functions with the displacement is reviewed in appendix~\ref{app:displcorrelatorsDecn}.
\begin{table}[H]
	\begin{center}
		$\begin{array}{|c||c|c|c|}
			\hline
			& \hD &\mathbb{Z}_2 & P \\
			\hline\hline
			\psi_{(3,1)}& \frac{3}{2}&-1& +1\\
			\hline
			\Disp& 2&+1& +1\\
			\hline
			\psi^{(2)}_{(3,1)}& 2+\frac{3}{2}&-1& +1\\
			\hline
			\Disp^2& 4&+1& +1\\
			\hline
			\psi^{(5)}_{(3,1)}& 5+\frac{3}{2}&-1& -1\\
			\hline
		\end{array}$
	\end{center}
	\caption{Tricritical Ising model with $(2,1)_4$ conformal boundary condition. Quantum numbers of the leading global boundary primaries appearing in the OPEs \eqref{leadingOPEstric}.}
	\label{tbl:chargesTric}
\end{table}

\subsubsection{The $(2,2)_4$ BCFT}
We again present the main observables in Table \ref{tbl:tric224}.

\begin{table}[H]
	\begin{center}
		$\begin{array}{|c|c|c|c|}
			\hline
			\hline
			(a_1,a_2)_m &\text{Boundary spectrum}&  \mathbb{Z}_2& \hD \\
			\hline
			\hline
			(2,2)_4&\hid&+1& 0  \\
			&\psi_{(1,2)}\simeq \psi_{(3,3)}&-1& 1/10  \\
			&\psi_{(1,3)}&+1& 3/5  \\
			&\psi_{(3,1)}&-1& 3/2  \\
			\hline
			\hline
		\end{array}
		\qquad
		\begin{array}{|c|c|c|}
			\hline
			\hline
			(a_1,a_2)_m &\text{Bulk primary}& B_\phi^{\bf a} \\
			\hline
			\hline
			(2,2)_4 &\id & 1\\
			&\epsilon & \sqrt{-2+\sqrt{5}}\\
			&\epsilon' &-\sqrt{-2+\sqrt{5}}\\
			&\epsilon'' &-1 \\
			\hline
			\hline
		\end{array}$
	\end{center}
	\caption{The boundary spectrum and bulk one-point functions for the $(2,2)_4$ BCFT.\label{tbl:tric224}}
\end{table}

As for the $\mathbb{Z}_2$ charges in Table~\ref{tbl:tric224}: first, the $\psi_{(3,1)}$ boundary operator is again odd because it appears in the $\sigma'$ bulk-boundary operator expansion, just as in the $(2,1)_4$ BCFT. For $\psi_{(1,3)}$ in $(2,2)_4$ one can consider instead
\begin{align}
	\epsilon'(x+i y,x-i y) =\frac{B_{\epsilon'}^{\bf a}}{(2y)^{\D_{\epsilon'}}}\hid +\frac{B_{\epsilon'}^{{\bf a}\,(1,3)}}{(2y)^{\D_{\epsilon'}-\hD_{1,3}}}\psi_{(1,3)}(x)+\text{desc.}
\end{align}
From the bulk two-point function of $\epsilon'$ in appendix~\ref{app:tric} we find that $\psi_{(1,3)}$ is $\mathbb{Z}_2$-even, since~\cite{Runkel:1998he}
\begin{align}
	(B_{\epsilon'}^{{\bf a}\,(1,3)})^2\simeq 0.663053\,.
\end{align}
For $\psi_{(3,3)}$ in $(2,2)_4$, instead of computing the bulk-boundary OPE of $\sigma$ (which is complicated), we can investigate the boundary four-point correlation function with $\psi_{(1,3)}$ and $\psi_{(3,1)}$
\begin{align}\label{13314pt}
	\langle \psi_{(1,3)}(x_1) \psi_{(3,1)}(x_2) \psi_{(1,3)} (x_3)\psi_{(3,1)}(x_4) \rangle\,.
\end{align}
Since $\psi_{(1,3)}$ and $\psi_{(3,1)}$ are (respectively) parity-even and odd, the s-channel blocks expansion of the above expression can contain at most $\psi_{(3,1)}$ and $\psi_{(3,3)}$. On the other hand for the OPE coefficients we have
\begin{align}
	\bdc_{(1,3)(3,1)(3,1)}^{{\bf a}} = \bdc_{(3,1)(3,1)(1,3)}^{{\bf a}}= 0\,,
\end{align}
since the self-OPE of $\psi_{(3,1)}$ does not contain $\psi_{(1,3)}$ -- see appendix~\ref{app:bd4pt31}. Being \eqref{13314pt} non-vanishing, this correlator must contain $\psi_{(3,3)}$, which therefore must be $\mathbb{Z}_2$-odd.  Proceeding this way, we again end up reconstructing the holomorphic counterpart of the fusion rules in eq.~\eqref{tricr_fusions}:
\begin{align}\label{tricr_fusions_bd}
	&\psi_{(3,3)} \times \psi_{(3,3)} = \hid+\psi_{(1,3)}\,,\quad	
	&&\psi_{(3,3)} \times \psi_{(1,3)}= \psi_{(3,3)}+\psi_{(3,1)}\,,\nonumber\\
	&\psi_{(3,3)} \times \psi_{(3,1)} = \psi_{(1,3)}\,,\quad 
	&&\psi_{(1,3)}\times \psi_{(1,3)}= \hid+\psi_{(1,3)}\,,\nonumber\\
	&\psi_{(1,3)}\times \psi_{(3,1)}= \psi_{(3,3)}\,,\quad
	&&\psi_{(3,1)} \times \psi_{(3,1)} = \hid\,.
\end{align}

We are interested in the global primary operators appearing in the following OPEs:
\begin{align}\label{leadingOPEstric22}
	\psi_{(3,3)} \times \psi_{(3,3)} &\sim \hid + \psi_{(1,3)} + \Disp + \psi^{(2)}_{(1,3)} + \dots \, , \nonumber \\
	\psi_{(1,3)} \times \psi_{(1,3)} &\sim \hid+ \psi_{(1,3)} + \Disp +  \psi^{(2)}_{(1,3)}+ \dots \, , \nonumber \\
	\psi_{(3,3)} \times \psi_{(1,3)} &\sim \psi_{(3,3)} + \psi_{(3,1)} +\dots +  \psi^{(3)}_{(3,3)} + \dots \, ,\nonumber \\
	\psi_{(3,1)} \times \psi_{(3,1)} &\sim \hid + \Disp + \dots \,,\nonumber\\
	\psi_{(3,3)} \times \psi_{(3,1)} &\sim \psi_{(1,3)}+ \psi^{(2)}_{(1,3)}  \dots \,,\nonumber\\
	\psi_{(1,3)} \times \psi_{(3,1)} &\sim \psi_{(3,3)}+ \dots\,.
\end{align}
The superscript $(n)$ still denotes level-$n$ Virasoro descendants which are global primaries. In the third and fifth lines of eq.~\eqref{leadingOPEstric22} we have also omitted leading parity-even descendants of $\psi_{(3,3)}$ and $\psi_{(3,1)}$, which are subleading with respect to $\psi^{(2)}_{(1,3)}$. The quantum numbers of the operators in eq.~\eqref{leadingOPEstric22} are reported in Table~\ref{tbl:chargesTric22} (the analysis of the parity-odd channel in correlation functions with the displacement is worked out in appendix~\ref{app:displcorrelatorsDecn}).
\begin{table}[H]
	\begin{center}
		$\begin{array}{|c||c|c|c|}
			\hline
			& \hD &\mathbb{Z}_2 & P \\
			\hline\hline
			\psi_{(3,3)}& \frac{1}{10}&-1& +1\\
			\hline
			\psi_{(1,3)}& \frac{3}{5}&+1& +1\\
			\hline
			\psi_{(3,1)}& \frac{3}{2}&-1& +1\\
			\hline
			\Disp& 2&+1& +1\\
			\hline
			\psi^{(2)}_{(1,3)}& 2+\frac{3}{5}&+1& +1\\
			\hline
			\psi^{(3)}_{(3,3)}& 3+\frac{1}{10}&-1& -1\\
			\hline
		\end{array}$
	\end{center}
	\caption{Tricritical Ising model with $(2,2)_4$ conformal boundary condition. Quantum numbers of the leading global boundary primaries appearing in the OPEs \eqref{leadingOPEstric22}.}
	\label{tbl:chargesTric22}
\end{table}

\section{The AdS background at zero and one loop}\label{sec:pertres}

In this section we consider perturbed two-dimensional CFTs in AdS and present the result of several one-loop computations. We will compare these results with the numerical bootstrap analysis afterwards.

We will work in Poincar\'e coordinates of $\AdStwo$ with curvature radius $R$ so the metric reads
\begin{equation}
	ds^2 = g_{\mu \nu} dx^\mu dx^\nu = \frac{R^2}{y^2} (dy^2 + d x^2)~\,.
\end{equation}
We will sometimes use complex coordinates $z = x + i y$, $\bar z = x - i y$. For bulk (global) primary operators $\phi(z,\bar z)$ with scaling dimension $\D$ the Weyl rescaling rule is
\begin{equation}
	\vev{\phi(x+i y, x - i y)\ldots}_\text{AdS} = (y/R)^\D \vev{\phi(x+i y, x - i y)\ldots}_\text{BCFT}\,.
\end{equation}
For example, since one-point functions in BCFT must take the form given in equation \eqref{oneptdef}, it follows that the one-point functions in AdS are simply constant, $(2R)^{\D}\vev{\phi} = B_\phi$, in agreement with general covariance. Boundary operators $\psi(x)$, on the other hand, remain untouched under the Weyl rescaling:
\begin{equation}
	\vev{\psi(x)\ldots}_\text{AdS} = \vev{\psi(x)\ldots}_\text{BCFT}\,.
\end{equation}

Suppose we now switch on a deformation of a 2d BCFT in $\AdStwo$ by a local operator $\phi(x)$. The correlation functions in the deformed theory can be computed perturbatively by expanding 
\begin{equation}\label{genRG}
	\vev{\ldots \exp\left( - g_{\phi}  R^{\D_\phi - 2} \int d^2 x \sqrt{g}\, \phi(x)+{\text{counterterms}} \right)}\,,
\end{equation}
in the dimensionless coupling $g_{\phi}$. The bare deformation generically induces both UV and IR divergences. The UV divergences are essentially the same as in flat space, even though new counterterms involving the AdS curvature may be needed. The IR divergences can be cured by including bulk counterterms evaluated at a cut-off surface near the boundary. As discussed for example in~\cite{Lauria:2023uca}, these counterterms can generally be chosen to preserve boundary conformal invariance and then the `boundary follows the bulk'.\footnote{An exception occurs when the boundary has a marginal operator that can be switched on along the RG flow. In that case the bulk RG will induce a boundary RG flow and potentially destabilize the boundary condition, see the discussions in~\cite{Hogervorst:2021spa,Ankur:2023lum,Lauria:2023uca,Copetti:2023sya}.} This is in contrast to flat-space RG flows emanating from BCFTs where bulk and boundary can flow independently, see for instance \cite{Fredenhagen:2006dn,Gaberdiel:2008fn,Fredenhagen:2009tn}.

In this section we will compute boundary OPE data to one loop in conformal perturbation theory. In subsections~\ref{TTbpred} and~\ref{sec:reviewDD2} we take the undeformed theory to be a generic local 2d BCFTs with bulk central charge $c$, and consider first a $\TTb$ deformation and then a generic Virasoro primary deformation. In subsections \ref{reldefIs} and \ref{reldeftric} we will again focus on the first two diagonal minimal models.

\subsection{\texorpdfstring{$\TTb$}{TTb} deformed CFTs}\label{TTbpred}
Consider the (perturbative) $\TTb$ deformation of a CFT in $\AdStwo$. In Poincaré coordinates:
\begin{align}\label{TTbads}
	\delta S =g_{\TTb} R^{2} \int \frac{dx\,dy}{y^2}\,\,\TTb(x+ iy,x-iy)+\text{counterterms}\,.
\end{align}
The $\TTb$ insertion for a CFT on $\AdStwo$ is a Weyl rescaling away from the $\TTb$ insertion on the upper half-plane, which in turn is obtained from an insertion of $T(z)T(z')$ on the complex plane, with $z'=z^*$, i.e.
\begin{align}\label{prescr_TTb}
	\langle \dots T\bar{T} (z) \rangle & \equiv \lim_{z'\rightarrow z}  \langle \dots T(z')\bar{T} (z) \rangle~\nonumber\\
	&=  \lim_{z'\rightarrow z} \langle \dots T(z') T (z^*) \rangle =  \langle \dots T(z) T (z^*) \rangle\,,
\end{align}
on the flat upper half-plane.

As we turn on the interaction, operators will generically get anomalous dimensions. Starting from correlation functions with $\TTb$ insertions computed in ref.~\cite{Lauria:2023uca}, in appendix~\ref{app:pertTTb} we compute the anomalous dimensions of $\Disp$ and $\Disp^2$ under the deformation of eq.~\eqref{TTbads}, at the first order in the coupling. We show that\footnote{Everywhere in this paper we will use hats to denote scaling dimensions of undeformed BCFT$_2$ boundary operators, and remove them when the BCFT$_2$ is deformed.}
\begin{align}\label{anDD2TTb}
	\D_{\Disp}(g_{\TTb})&=2+g_{\TTb}\,\dhD_{\Disp}+O(g_{\TTb}^2)\,,& &\dhD_{\Disp} = \pi\,,\nonumber\\
	\D_{\Disp^2}(g_{\TTb})&=4+g_{\TTb}\,\dhD_{\Disp^2}+O(g_{\TTb}^2)\,,& &\dhD_{\Disp^2} = 6\pi\,.
\end{align}
The undeformed BCFT$_2$ might feature a boundary Virasoro primary $\psi$ with tree-level dimension $\hD_\psi$. The $\TTb$ deformation then results in the following anomalous dimension for $\psi$, as again shown in appendix~\ref{app:pertTTb}
\begin{align}\label{anphioTTb}
	\D_\psi(g_{\TTb})&=\hD_\psi+g_{\TTb}\,\dhD_\psi+O(g_{\TTb}^2)\,,& &\dhD_{\psi} = \frac{\pi}{2}\hD_\psi (\hD_\psi-1)\,.
\end{align}
We note that $\dhD_{\psi}\sim \hD_\psi^2$ at large $\hD_\psi$, generalizing the expectation from AdS effective field theory \cite{Heemskerk:2009pn,Fitzpatrick:2010zm}.
In appendix~\ref{app:pertTTb} we also compute the following boundary correlation functions
\begin{align}
	\langle \Disp(1)\Disp(x)\Disp(0) \rangle\,,\quad \langle \psi(1)\psi(x)\Disp(0) \rangle\,,\quad
	\langle \Disp(1)\Disp(x)\Disp^2(0) \rangle\,,\quad \langle \psi(1)\psi(x)\Disp^2(0) \rangle\,,
\end{align}
at one loop in the $\TTb$ deformation. For unit-normalized boundary operators we find the OPE coefficients:
\begin{align}\label{OPEcoeffsTTb}
	\bdc_{\Disp\Disp\Disp}(g_{\TTb})&=\frac{2 \sqrt{2}}{c} \left(1-\frac{\pi (c-24)}{8}g_{\TTb}+O(g_{\TTb}^2)\right)\,,\nonumber\\
	\bdc_{\psi\psi\Disp}(g_{\TTb})&=\frac{\sqrt{2} \hD_\psi}{\sqrt{c}}\left(1-\pi  \left(1+\frac{c}{24}-2 \hD_\psi\right)g_{\TTb}+O(g_{\TTb}^2)\right)\,,\nonumber\\
	\bdc_{\psi\psi\Disp^2}(g_{\TTb})&=\frac{\sqrt{\frac{2}{5}} \hD_\psi (5 \hD_\psi+1)}{\sqrt{c (5 c+22)}}\left(1-\frac{\pi}{60}  (5 c-240 \hD_\psi+262)g_{\TTb}+O(g_{\TTb}^2)\right)\,,\nonumber\\
	\bdc_{\Disp\Disp\Disp^2}(g_{\TTb})&=\frac{1}{c}{\sqrt{\frac{2}{5}} \sqrt{c (5 c+22)}}{}\left(1+\frac{109 \pi }{30}g_{\TTb}+O(g_{\TTb}^2)\right)\,.
\end{align}

\subsection{Deformations by a bulk Virasoro primary}\label{sec:reviewDD2}
If the bulk theory supports a scalar bulk Virasoro primary with scaling dimension $\D_{\phi}$, we can turn on the following deformation
\begin{align}\label{defgenconformalbc}
	\delta S = g_{\phi} R^{\D_\phi - 2}  \int \frac{dx dy}{y^2}\,\phi(x+ iy,x-iy)+\text{counterterms}\,.
\end{align}
We assume that $\phi$ does not contain any marginal boundary global primary in its bulk-boundary OPE. The undeformed theory features again both $\Disp$ and $\Disp^2$. The anomalous dimensions of these operators under the deformation of eq.~\eqref{defgenconformalbc} at the first order in the coupling read
\begin{align}
	\D_{\Disp}(g_\phi)&=2+g_{\phi}\,\dhD_{\Disp}+O(g_{\phi}^2)\,,\quad \nonumber\\
	\D_{\Disp^2}(g_\phi)&=4+g_{\phi}\,\dhD_{\Disp^2}+O(g_{\phi}^2)\,,
\end{align}
with~\cite{Lauria:2023uca} (see also appendix~\ref{app:pertVir} for a derivation)
\begin{align}\label{first_order_exp_D_and_D2}
	\dhD_{\Disp}&=\frac{B_\phi}{2^{\D_{\phi}}}\frac{4\pi}{c}   (\D_\phi -2) \D_\phi \,,\nonumber\\
	\dhD_{\Disp^2}&=\frac{B_\phi}{2^{\D_\phi}}  \frac{2\pi \D_\phi (\D_\phi -2) (20 c+25 (\D_\phi -2) \D_\phi +64)}{c (5 c+22)}\,.
\end{align}
Here $B_\phi$ is the tree-level one-point function coefficient for $\phi$, see eq.~\eqref{oneptdef}. Note that the one-loop anomalous dimensions vanish if $\D_\phi = 2$ which is due to the preservation of bulk conformal invariance at this order.

In appendix~\ref{app:pertVir} we also compute
\begin{align}
	\langle \Disp(1)\Disp(x)\Disp(0) \rangle\,,\quad 
	\langle \Disp(1)\Disp(x)\Disp^2(0) \rangle\,,
\end{align}
at one-loop in the Virasoro deformation above. For unit-normalized boundary operators we find
\begin{align}
	\bdc_{\Disp\Disp\Disp}(g_{\phi})&=\frac{2 \sqrt{2}}{c} \left(1+\frac{B_\phi}{2^{\D_\phi } }\frac{3\pi}{c} (\D_\phi -2) \D_\phi^2\,g_{\phi}+O(g_{\phi}^2)\right)\,,\\
	\bdc_{\Disp\Disp\Disp^2}(g_{\phi})&=\frac{1}{c}{\sqrt{\frac{2}{5}} \sqrt{c (5 c+22)}}{}\left(1+\frac{B_\phi}{2^{\D_\phi }}\frac{\pi(\D_\phi -2) \D_\phi  (5 \D_\phi +2) (25 \D_\phi +336)}{30 c (5 c+22)}g_{\phi}+O(g_{\phi}^2)\right)\,.\nonumber
\end{align}

\subsection{Deformations of the Ising model}\label{reldefIs}
Below we will need the first-order data of both the relevant $\phi_{(1,3)}$ deformation and the leading irrelevant $T\bar T$ deformation of the $(1,2)_3$ BCFT in AdS. Our deformation therefore reads:
\begin{align}\label{epsilonprimepertgenising}
	\delta S &= g_{(1,3)}R^{\D_{(1,3)}-2} \int d^2 x \sqrt{g}\,\phi_{(1,3)}(x+ i y, x - iy)+g_{\TTb} R^{2} \int d^2 x \sqrt{g}\,\TTb(x+ i y, x - iy)\,,\nonumber\\
	&\qquad+\text{counterterms}\,.
\end{align} 
Note that this deformation preserves the $\mathbb Z_2$ global symmetry.

The one-loop anomalous dimensions for the boundary operators $\psi_{(1,3)}$, $\Disp$ and $\Disp^2$ under the bulk deformations of eq.~\eqref{epsilonprimepertgenising} then read:
\begin{align}\label{an21ising}
	\dhD_{(1,3)} &= 2\pi g_{(1,3)}-\pi g_{\TTb}/8\,,\nonumber\\
	\dhD_{\Disp} &= 4\pi g_{(1,3)}+\pi g_{\TTb}\,,\nonumber\\
	\dhD_{\Disp^2} &= 4\pi g_{(1,3)}+6\pi g_{\TTb}\,.
\end{align}
For the contributions of the $\phi_{(1,3)}$ deformation to the anomalous dimensions of $\psi_{(1,3)}$ we have used the result of ref.~\cite{Lauria:2023uca} while for that of $\Disp, \Disp^2$ we have used eq.~\eqref{first_order_exp_D_and_D2}. For the one-loop anomalous dimensions under the $\TTb$ deformation we have just used eq.~\eqref{anphioTTb}.

\subsection{Deformations of the tricritical Ising model}\label{reldeftric}
In both the $(2,1)_4$ or $(2,2)_4$ BCFTs we will consider a simultaneous deformation with two relevant and two irrelevant operators, as follows:
\begin{align}\label{epsilonprimepertgen}
	\delta S_{\text{rel}} &=  \int d^2 x \sqrt{g}\,\left[g_{(3,3)}R^{\D_{(3,3)}-2}\phi_{(3,3)}(x+ i y, x - iy)+g_{(1,3)}R^{\D_{(1,3)}-2}\phi_{(1,3)}(x+ i y, x - iy)\right]\,,\nonumber\\
	\delta S_{\text{irrel}} &=  \int d^2 x \sqrt{g}\,\left[g_{(3,1)}R^{\D_{(3,1)}-2}\phi_{(3,1)}(x+ i y, x - iy)+g_{\TTb}R^2\TTb(x+ i y, x - iy)\right]\,.
\end{align}
For each allowed global boundary primary with tree-level dimension $\hD_\psi$ we will compute, at the leading order in the deformation
\begin{align}
	\D_{\psi}(g_{(i)})= \hD_\psi+\sum_ig_{(i)}\dhD^{(i)}_{\psi}+\dots,\quad g_{(i)} =\{g_{(3,3)},~g_{(1,3)},~g_{(3,1)},~g_{(3,3)},~g_{\TTb}\}\,.
\end{align}
For the contributions of the $\phi_{(1,3)}$, $\phi_{(3,1)}$ deformation to the anomalous dimensions of $\psi_{(r,s)}$ we can use the result of ref.~\cite{Lauria:2023uca}, while the $\phi_{(1,2)}$ deformation is studied in our appendix~\ref{app:phi12def}. For the one-loop anomalous dimensions under the $\TTb$ deformation we use eq.~\eqref{anphioTTb}, while for that of $\Disp, \Disp^2$ under a generic bulk Virasoro primary we use eq.~\eqref{first_order_exp_D_and_D2}.

In Table~\ref{tbl:an21} we report the one-loop anomalous dimensions for the boundary operators $\psi_{(3,1)}$, $\Disp$ and $\Disp^2$ in the $(2,1)_4$ boundary condition, and in Table \ref{tbl:an22} we list the one-loop anomalous dimensions for the boundary operators $\psi_{(r,s)}$, $\Disp$ and $\Disp^2$ in the $(2,2)_4$ boundary condition.

\begin{table}[H]
	\begin{center}
		$\begin{array}{|c||c|c|c|c|}
			\hline
			&g_{(3,3)}\phi_{(3,3)} & g_{(1,3)}\phi_{(1,3)}  & g_{(3,1)}\phi_{(3,1)}  & g_{\TTb} \TTb \\
			\hline
			\hline
			\dhD_{(3,1)}^{(i)}&\beta'/2&-\beta'&\frac{45\pi}{14}&\frac{3\pi}{8}\\
			\dhD_{\Disp}^{(i)} &-\frac{3 \beta'}{5}&-\frac{4\beta'}{5}&-\frac{15 \pi }{7}&\pi\\
			\dhD_{\Disp^2}^{(i)} &-\frac{69\beta'}{85} &-\frac{72 \beta'}{85} &-\frac{45 \pi }{7}&6\pi\\
			\hline
		\end{array}$
		\begin{align}\label{abcdef21}
			\beta'&= \frac{12\pi}{7} 2^{4/5}  B_{(1,3)}^{\bf a} \approx 11.9275~
		\end{align}
	\end{center}
	\caption{Anomalous dimensions of leading boundary primaries in $(2,1)_4$. Rows: contribution to $\dhD_{(r,s)}$ from each deformation. The number $\beta'$ is defined in equation \eqref{abcdef21}. The corresponding value of $B^{\bf a}_{(1,3)}$ is from Table~\ref{tbl:tric214}.}
	\label{tbl:an21}
\end{table}

\begin{table}[H]
	\begin{center}
		$\begin{array}{|c||c|c|c|c|}
			\hline
			&g_{(3,3)}\phi_{(3,3)} & g_{(1,3)}\phi_{(1,3)}  & g_{(3,1)}\phi_{(3,1)}  & g_{\TTb} \TTb \\
			\hline
			\hline
			\dhD_{(3,3)}^{(i)} &-\sigma&-\delta&\frac{3\pi}{14}&-\frac{9\pi}{200}\\
			\dhD_{(1,3)}^{(i)} &\zeta&\alpha&\frac{6\pi}{7}&-\frac{3 \pi }{25}\\
			\dhD_{(3,1)}^{(i)}&-\beta/2&\beta&\frac{45\pi}{14}&\frac{3\pi}{8}\\
			\dhD_{\Disp}^{(i)} &-\frac{3 \beta }{5}&\frac{4 \beta }{5}&-\frac{15 \pi }{7}&\pi\\
			\dhD_{\Disp^2}^{(i)} &-\frac{69 \beta }{85}&\frac{72 \beta }{85} &-\frac{45 \pi }{7}&6\pi\\
			\hline
		\end{array}$
	\end{center}
	\caption{Anomalous dimensions of leading boundary primaries in $(2,2)_4$. Rows: contribution to $\dhD_{(r,s)}$ from each deformation. The numbers $\alpha,\beta,\delta,\sigma,\zeta$ are defined in \eqref{abcdef}.}
	\label{tbl:an22}
\end{table}

The parameters $\alpha,\beta,\delta,\sigma,\zeta$ of Table~\ref{tbl:an22} are defined as follows
\begin{align}\label{abcdef}
	\alpha&=-B^{\bf a}_{(1,3)} \left(\frac{36\pi}{35} 2^{4/5} \kappa_1+\frac{25 \left(\sqrt{5}+1\right) \sqrt{\pi }  \Gamma \left(-\frac{3}{10}\right) \Gamma \left(\frac{7}{10}\right)}{224\ 2^{4/5} \Gamma \left(-\frac{12}{5}\right) \Gamma \left(\frac{13}{10}\right)}\kappa_2\right)\approx 6.3241\,,\nonumber\\
	\beta&=-\frac{12 \pi }{7}  2^{4/5} B^{\bf a}_{(1,3)} \approx 4.55592\,,\nonumber\\
	\delta &=-B^{\bf a}_{(1,3)} \left(-\frac{6}{35} 2^{4/5} \pi  \kappa_3+\frac{\left(\sqrt{5}+1\right) \sqrt{\pi } \Gamma \left(\frac{7}{10}\right)^2}{2\ 2^{4/5} \Gamma \left(\frac{3}{5}\right) \Gamma \left(\frac{13}{10}\right)}\kappa_4 \right)\approx 0.702678\,,\nonumber\\
	\sigma&=\frac{2\pi}{35}  2^{4/5}  B^{\bf a}_{(1,2)}  \kappa_5\approx 0.209884\,,\nonumber\\
	\zeta&= B^{\bf a}_{(1,2)}\left(-\frac{12\pi}{35} 2^{4/5} \kappa_3+\frac{3\pi 2^{2/5} \Gamma \left(\frac{7}{10}\right) \Gamma \left(\frac{11}{10}\right)}{\Gamma \left(\frac{1}{5}\right) \Gamma \left(\frac{13}{10}\right)^2}\kappa_4\right)\approx 1.40536\,,
\end{align}
with ${\bf a} = (2,2)_4$. Correspondingly, the values of $B^{\bf a}_{(1,2)}$, $B^{\bf a}_{(1,3)}$ are shown in Table~\ref{tbl:tric224}, while $\kappa_i$ are the following numbers:
\begin{align}
	\kappa_1 &\equiv \, _4F_3\left(-\frac{1}{5},\frac{2}{5},\frac{1}{2},\frac{13}{5};\frac{3}{5},\frac{17}{10},2;1\right)\,,\nonumber\\
	\kappa_2 &\equiv\,_4F_3\left(-\frac{9}{10},-\frac{3}{10},-\frac{1}{5},\frac{19}{10};-\frac{1}{10},\frac{3}{10},\frac{13}{10};1\right)\,,\nonumber\\
	\kappa_3 &\equiv \, _3F_2\left(\frac{2}{5},\frac{1}{2},\frac{9}{5};\frac{17}{10},2;1\right)\,,\nonumber\\
	\kappa_4 &\equiv \, _3F_2\left(-\frac{3}{10},-\frac{1}{5},\frac{11}{10};\frac{3}{10},\frac{13}{10};1\right)\,,\nonumber\\
	\kappa_5&\equiv \, _3F_2\left(\frac{1}{2},\frac{4}{5},\frac{7}{5};\frac{17}{10},2;1\right)\,.
\end{align}

\section{Numerical Results}
\label{sec:numerical}

In this section we show results from the numerical bootstrap and compare them to the perturbative predictions of the previous section. Our numerical setup is standard by now and makes use of semi-definite optimization solver \texttt{SDPB} \cite{SimmonsDuffin:2015,Landry:2019}.

In subsection \ref{sec:Displacement}, we analyze the four-point function $\vev{\Disp\Disp\Disp\Disp}$ of the displacement operator $\Disp$, showing that it universally saturates the bound on a ratio of OPE coefficients for any CFT in the bulk. We then consider perturbative bulk deformations, imputing data from section \ref{sec:pertres} and observe that the bounds remain saturated when the deforming operator is a bulk Virasoro primary, while a $\TTb$ deformation is not always allowed: only one sign of the coupling is consistent.

In subsection \ref{ss:trictocrit}, we focus on the specific RG flow between the tricritical Ising model with the $(2,1)_4$ BCFT in the UV and the Ising model with the $(1,2)_3$ BCFT in the IR. To do this, we first bootstrap the four-point function $\vev{\psi \psi \psi \psi}$ of the lightest $\mathbb{Z}_2$-odd operator $\psi$. Afterwards we upgrade the setup to a mixed correlator system where we include $\vev{\psi \psi \Disp \Disp}$ and $\vev{\Disp\Disp\Disp\Disp}$. The allowed region in a subspace of scaling dimensions turns out to have several sharp features, some of which correspond to the RG flow of interest.

In subsection \ref{ss:Bootspert}, the same three-correlator setup is used to explore the vicinity of the UV and IR endpoints of the previous flow, establishing saturation of the bounds in more detail. We also explore other relevant and irrelevant deformations of these CFTs, notably again finding inconsistency of the $\TTb$ deformation for the wrong sign of the coupling from a different setup.

Finally, in subsection \ref{ss:corrmax}, we explore the more complicated $(2,2)_4$ BCFT setup which turns out to saturate a bound on the space of values of the correlator and its derivatives around the crossing symmetric point. The same bound is saturated by the Ising model with the $(1,2)_3$ BCFT, suggesting a second RG flow with the same IR fixed point. This is the simplest example in the infinite family of `staircase' RG flows between the diagonal minimal models.
 
\subsection{Universal bounds from displacement four-point function}
\label{sec:Displacement}

The displacement $\Disp$ is a universal boundary operator of any BCFT in $d+1$ bulk dimensions with a stress-energy tensor. At the BCFT point it has (protected) scaling dimension equal to $d + 1$ in our case. For $d = 1$ its four-point correlation function is completely fixed in terms of the bulk central charge $c$, see equation \eqref{D4ptfunction}.

As soon as we turn on a (covariant) deformation in the bulk of AdS, the displacement is no longer protected and we expect its (conformal) correlation functions to depend non-trivially on the RG trajectory. This correlator should remain crossing symmetric along the AdS deformation, and so we can use the numerical bootstrap to constrain it.

Now consider the following situation. Suppose that we have found a bound that happens to be \emph{saturated} by the unperturbed BCFT correlator (we will soon show an example of this). Then what happens if we turn on a deformation in $\AdStwo$? One potential constraint arises if, for a particular sign of the perturbation, the first-order prediction points \emph{into} the disallowed region, since then only deformations with the opposite sign can be consistent.\footnote{This idea was used earlier in \cite{Antunes:2021abs} to constrain the sign of an irrelevant $(\partial \phi)^4$ deformation in two bulk dimensions.} In this subsection we will show that the sign of the $T\bar T$ deformation is constrained in exactly this way.

As a warm-up exploration, let us focus on the unperturbed theory and look for a bound saturated by the four-point correlation function of the displacement operator. We recall that $\hD_{\Disp}= 2$, $\hD_{\Disp^2}=4$ and that, from eq.~\eqref{DDOPmain},
\begin{equation}
	\Disp \times \Disp \sim \hid + \Disp + \Disp^2 + \dots\,.
\end{equation}
The first bound that comes to mind to a bootstrapper is gap maximization, and so we can try to maximize the gap after $\Disp$. As it turns out, the bound in this case approaches $2\hD_{\Disp}+1=5$, which is the gap of the generalized free fermion solution. In fact this can be proved rigorously: the same functional that proves that this is the maximal allowed gap in a general correlator \cite{Mazac:2018ycv} applies here because it also happens to be positive at $\hD_{\Disp} = 2$ where we have an additional conformal block. We must conclude that the exchange of $\hD_{\Disp^2}=4$ in our correlator means that it is far from extremizing the gap.

The next bound that comes to mind is OPE coefficient maximization. One could for example attempt to find upper bounds on $\bdc_{\Disp\Disp\Disp}$ or $\bdc_{\Disp\Disp\Disp^2}$ (independently), but this cannot work because of the following. The displacement four-point function in eq.~\eqref{D4ptfunction} admits a positive conformal block decomposition for arbitrary $c$, with the leading OPE coefficients reading:
\begin{equation}
	\label{eq:CDDDnorm}
	(\bdc_{\Disp\Disp\Disp})^2=\frac{8}{c}\,,\quad (\bdc_{\Disp\Disp\Disp^2})^2=2 + \frac{11}{10} (\bdc_{\Disp\Disp\Disp})^2 \,,
\end{equation} 
where we took $\Disp$ and $\Disp^2$ to have unit normalized two-point functions. One can formally take $c\to0$ in the equation above, and this provides a legitimate solution to crossing with arbitrarily large OPE coefficients.\footnote{Alternatively, note that the unit-normalized four-point function of $\Disp$ contains a GFB piece and a $c$-dependent piece that makes the results above unbounded. This piece is exactly equal to the fully connected Wick contractions of $\langle \phi^2 \phi^2\phi^2\phi^2 \rangle$ in a GFB theory. This correlator has a positive conformal block decomposition with no identity operator, and so it can be multiplied by an arbitrarily large number, leading to the unboundedness property.}

\begin{figure}
	\centering
	\includegraphics[width=0.55\textwidth]{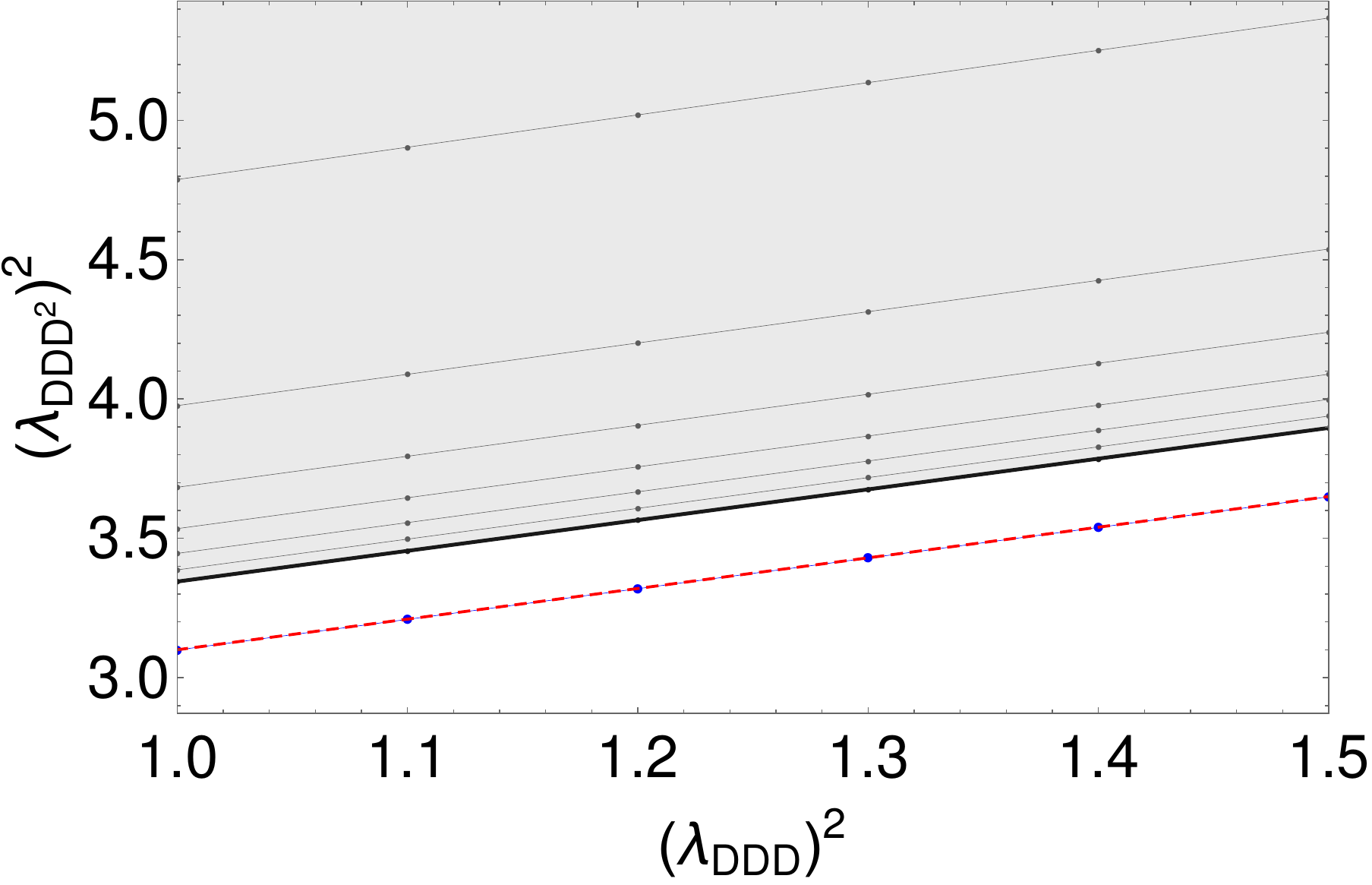} 
	\caption{Upper bound on the squared OPE coefficient $(\bdc_{\Disp\Disp\Disp^2})^2$ as a function of $(\bdc_{\Disp\Disp\Disp})^2$. The grey dots denote bounds at increasing derivative order, which are then extrapolated to the blue dots. The red line corresponds to the analytic results of eq.~\eqref{eq:CDDDnorm}. The extrapolated bounds agree with the analytic result to the third decimal place.}
	\label{fig:CDDD2vsCDDD}
\end{figure}

On the other hand, what happens if we fix $(\bdc_{\Disp\Disp\Disp})^2$ and maximize $(\bdc_{\Disp\Disp\Disp^2})^2$? In that case we do find a non-trivial upper bound: it is displayed in fig.~\ref{fig:CDDD2vsCDDD} and converges nicely to the relation in eq.~\eqref{eq:CDDDnorm}, when extrapolated to an infinite number of derivatives. It would be interesting to prove this property using for instance extremal functionals~\cite{Mazac:2018ycv,Mazac:2016qev,Mazac:2018mdx,Ghosh:2023lwe}.

\begin{figure}
	\centering
	\hspace{50pt}\includegraphics[width=0.8\textwidth]{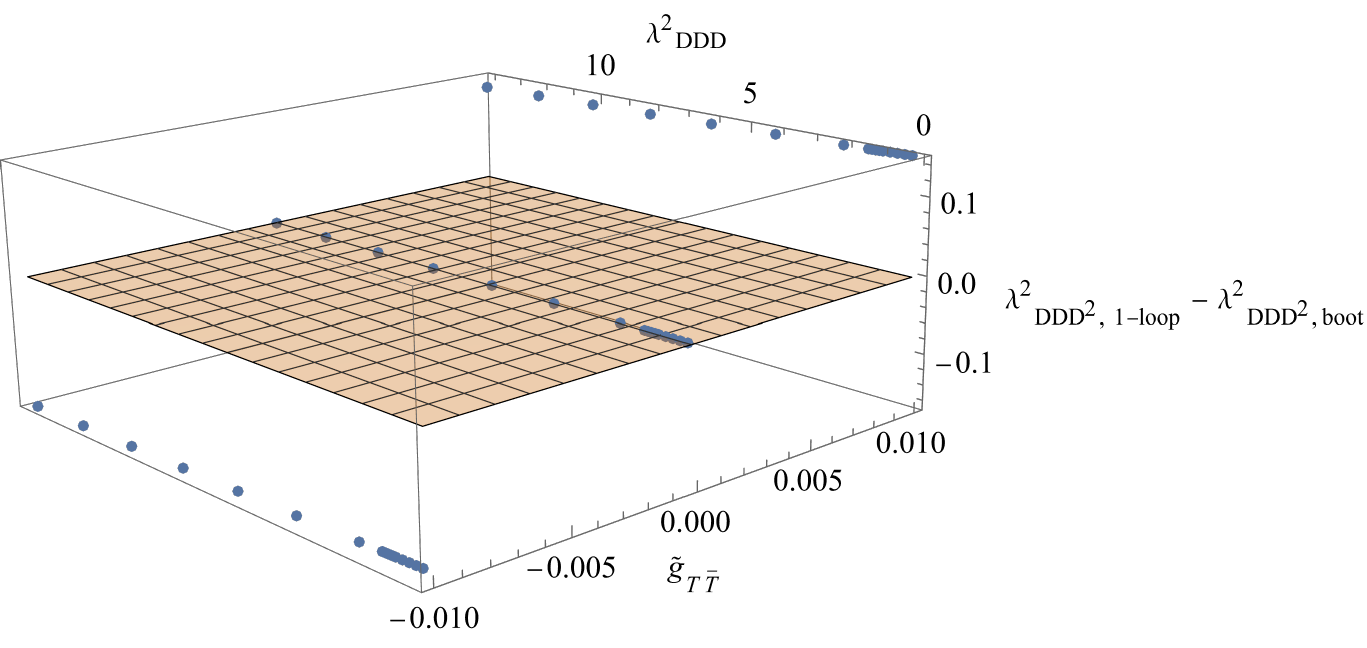} 
	\caption{ Difference between the perturbative prediction and the extrapolated bound on the $(\bdc_{\Disp\Disp\Disp^2})^2$ for varying $c$ and $\D_{\Disp}, \D_{\Disp^2}$, following the perturbative prediction parametrized by $\tilde{g}_{\TTb}$. The yellow surface serves as a guide to the eye, excluding points above it.}
	\label{fig:CDDDTTb}
\end{figure}

Having obtained a bound saturated by the `unperturbed' correlator, we can start exploring how it changes as we turn on a bulk deformation. Let us focus on the $\TTb$ deformation of section~\ref{sec:reviewDD2}. Guided by one-loop perturbation theory, we can explore the following direction in the space of CFT data
\begin{equation}
	\D_\Disp-2=\pi g_{\TTb}\equiv \tilde{g}_{\TTb} \,, \quad \D_{\Disp^2}=4 +6 \tilde{g}_{\TTb} \,, 
\end{equation}
and maximize $(\bdc_{\Disp\Disp\Disp^2})^2$, as a function of $\tilde{g}_{\TTb}$ and $(\bdc_{\Disp\Disp\Disp})^2$. Figure~\ref{fig:CDDDTTb} shows the difference between the perturbative prediction 
\begin{align}
	(\bdc_{\Disp\Disp\Disp^2})^2=2 + \frac{11}{10} (\bdc_{\Disp\Disp\Disp})^2 + \left(\frac{2510+209(\bdc_{\Disp\Disp\Disp})^2}{150}\right)\tilde{g}_{\TTb}\,,
\end{align}
and the extrapolated (upper) bound. Consistent theories must lie below the plotted surface, and so for any given central charge $c$ we observe that only the negative sign of the $\TTb$ deformation is allowed.\footnote{Notice that our sign convention here for the $\TTb$ deformation is opposite with respect to the one that we used in the Introduction. It agrees with our general convention for AdS deformations, as defined in eq.~\eqref{TTbads}.}

We emphasize that the scale of the variation with $\tilde{g}_{\TTb}$ is much larger than the accuracy of the extrapolation, so we believe that our numerical results are robust. This sign constraint is a known property of the $\TTb$ deformation, see e.g.~\cite{Cardy:2015xaa,Delacretaz:2021ufg}, which we have re-discovered using 1d numerical bootstrap.\footnote{For a positive $\TTb$ coupling, the anomalous dimensions grow very quickly, destroying the good Regge behavior of the four-point function. This is related to bulk causality, which is similar to the analysis of \cite{Delacretaz:2021ufg} which showed that the same sign of the $\TTb$ coupling leads to a superluminal sound speed.}

\begin{figure}
	\centering
	\hspace{50pt}\includegraphics[width=0.8\textwidth]{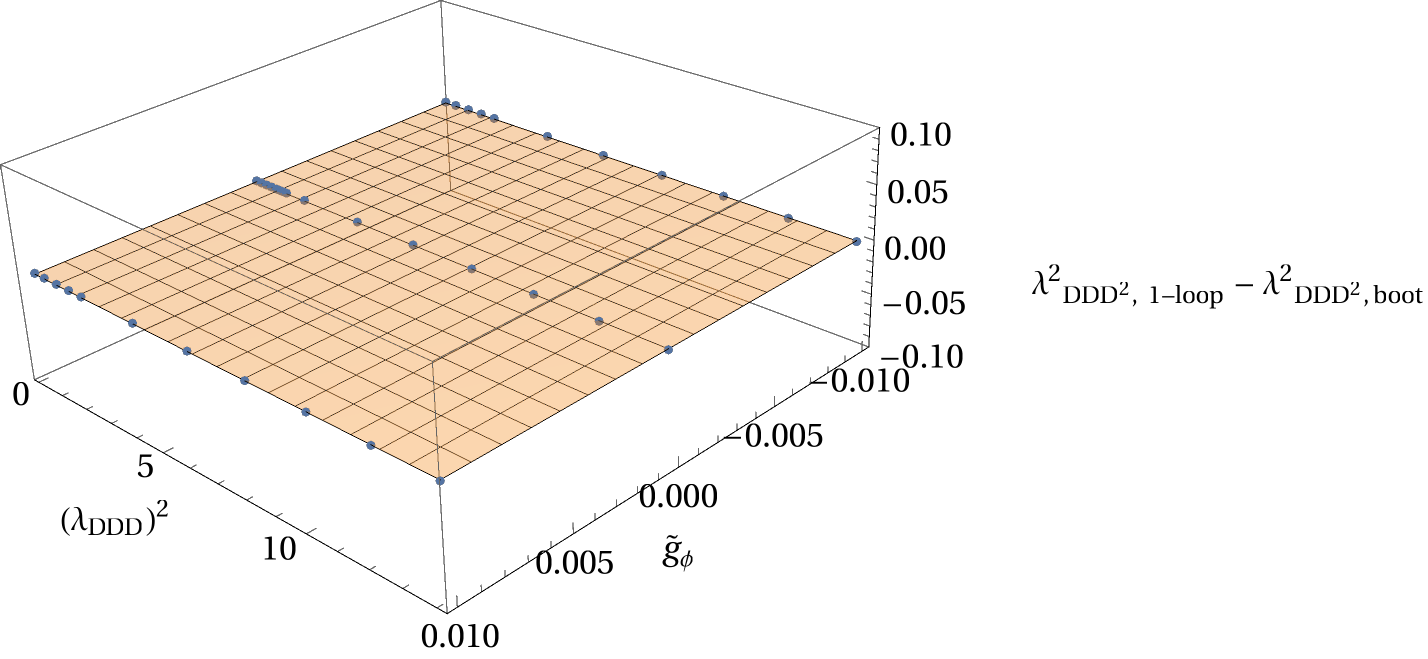} 
	\caption{Difference between the perturbative prediction and the extrapolated bound on the $(\bdc_{\Disp\Disp\Disp^2})^2$ OPE coefficient for varying $c$ and $\D_{\Disp}, \D_{\Disp^2}$, following the perturbative prediction parametrized by $\tilde{g}_{\phi}$ with $\D_\phi=1.5$. 
		The yellow surface serves as a guide to the eye, excluding points above it.}
	\label{fig:CDDDphi1p5}
\end{figure}

We can play a similar game for deformations by a generic scalar Virasoro primary, for which the first-order analysis was done in section~\ref{sec:reviewDD2}. For a deformation $\phi$ with scaling dimension $\D_\phi$, the perturbative results can be rewritten as
\begin{align}
	\D_{\Disp}-2&\equiv \tilde{g}_\phi\,,\quad \D_{\Disp^2}= 4 + \tilde{g}_\phi \left(\frac{64+ 160(\bdc_{\Disp
			\Disp\Disp})^{-2} +25\D_\phi(\D_\phi-2)}{44+80 (\bdc_{\Disp
			\Disp\Disp})^{-2}}\right)\,,\nonumber\\
	(\bdc_{\Disp\Disp\Disp^2})^2&=2 + \frac{11}{10} (\bdc_{\Disp\Disp\Disp})^2 + \tilde{g}_{\phi}\frac{(672-250\D_\phi+125\D_\phi^2)(\bdc_{\Disp\Disp\Disp})^2}{1200}\,.
\end{align}
We have explored a few values of $\D_\phi$, including both relevant and irrelevant deformations. This time, the same game leads to a very different result. To within our numerical precision, all these deformations saturate the numerical upper bounds, and therefore no universal constraint on the sign of the coupling seems to be possible.\footnote{In a CFT one can always replace a local primary operator with minus itself, at the expense of flipping also the relevant OPE coefficients. The invariant statement we are making is that the sign of the product $g_\phi B_\phi$ is not constrained by our analysis.}
Instead, we can now track the perturbative RG flows to leading order by following the numerical bounds, see fig.~\ref{fig:CDDDphi1p5} for an illustrative example with $\D_\phi=1.5$.\footnote{The fact that, unlike for $\TTb$, no sign constraints arise for general deformations at the leading order is consistent with causality constraints. More precisely, any causality violation by irrelevant couplings, while linear for $\TTb$, is at least quadratic for a generic interaction~\cite{Delacretaz:2021ufg}.}

\subsection{Bootstrapping the Tricritical to Critical Ising RG flow}\label{ss:trictocrit}
In this section we consider the RG flow induced by the $\phi_{(1,3)}$ deformation of the tricritical Ising model in $\AdStwo$. In the complex plane, this is the famous tricritical-to-critical Ising RG flow \cite{Zamolodchikov:1987ti,Zamolodchikov:1991vx}. These flows have many interesting properties: in flat space they preserve integrability, and as such can be studied through the TBA equations \cite{Zamolodchikov:1991vx}. They can also be studied perturbatively by means of a finite $m$ extrapolation of large $m$ perturbation theory \cite{Zamolodchikov:1987ti}.\footnote{See \cite{Antunes:2022vtb} for a recent discussion of large $m$ perturbation theory.} In the presence of a boundary some highly non-trivial boundary dynamics emerges \cite{Fredenhagen:2009tn}.

In the following we will use the numerical conformal bootstrap to obtain new non-perturbative constraints on the boundary OPE data for these flows in AdS$_2$. We will always assume the boundary conditions to be $\mathbb{Z}_2$-preserving.

\subsubsection{Single correlator bound}\label{ss:singlecorr}
We consider the four-point correlation function of a $\mathbb{Z}_2$-odd (global) boundary primary $\psi$ with scaling dimension $\D_\psi$
\begin{align}
	\langle \psi(x_1)\psi(x_2)\psi(x_3)\psi(x_4) \rangle = \frac{\mathcal{G}(\eta)}{x_{12}^{2\D_\psi} x_{34}^{2\D_\psi}}\,.
\end{align}
As we vary $\D_\psi$ along the RG, $\psi$ can interpolate between $\psi_{(3,1)}$ in the $(2,1)_4$ boundary condition for tricritical Ising (UV) and $\psi_{(1,3)}$ in the $(1,2)_3$ boundary condition for Ising (IR). Its scaling dimension correspondingly is expected to decrease from $3/2$ to $1/2$ along the flow.

We take the $\psi\times\psi$ OPE to be, schematically and up to subleading $\mathbb{Z}_2$-even exchanges
\begin{align}
	\psi \times \psi \sim \hid+ \Disp  + \Disp^2+\dots\,.
\end{align}
Note that the scaling dimensions $\D_{\Disp}$ and $\D_{\Disp^2}$ must equal 2 and 4 both at the UV and at the IR fixed point, but along the flow they are not protected.

\begin{figure}
	\centering
	\includegraphics[width=0.7\textwidth]{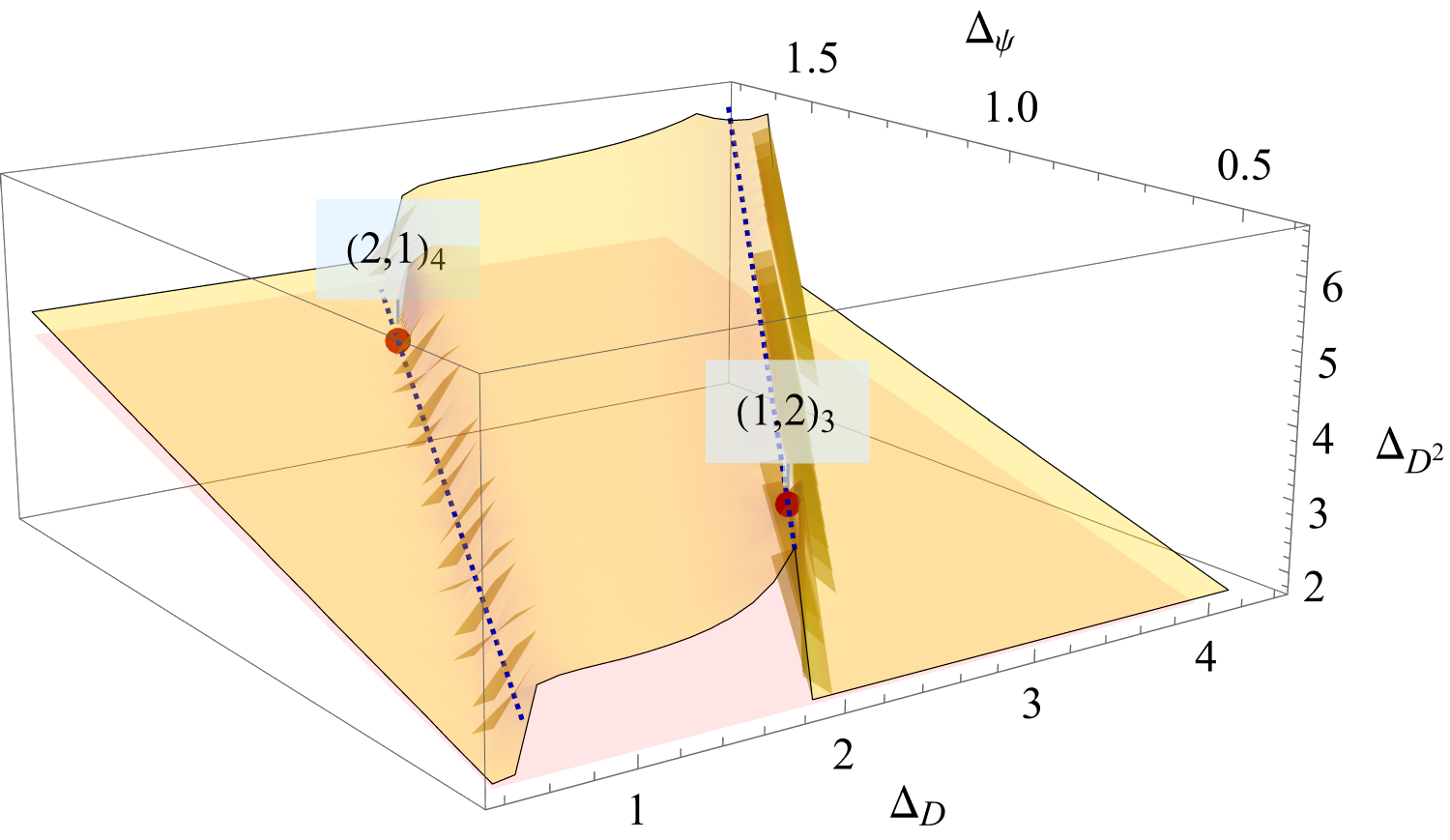} 
	\caption{The light yellow surface is the upper bound on the scalar gap $\Delta_{\Disp^2}$ as a function of $\Delta_\psi,\Delta_\Disp$ obtained from the single correlator bootstrap with $\Lambda=25$ derivatives.}
	\label{3d_gap_bounds}
\end{figure}

\begin{figure}
	\centering
	\includegraphics[width=0.55\textwidth]{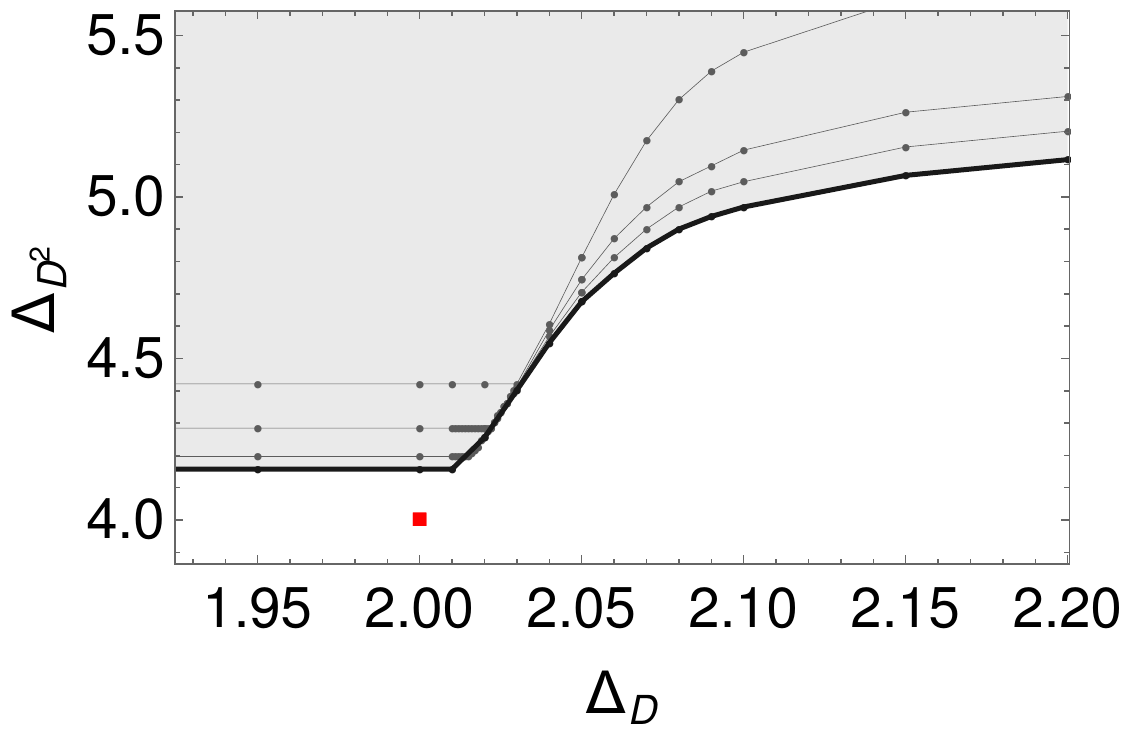}
	\caption{Bounds on the $\mathbb{Z}_2$-even gap from the numerical bootstrap for $\D_\psi = 3/2$. We  show the results obtained with different number of derivatives, from $\Lambda=15$ to $\Lambda=45$ in steps of 10. The red dot corresponds to the $(2,1)_4$ b.c. of tricritical Ising.}
	\label{gap_TriIsing_deltaD_deltaGap}
\end{figure}

We first searched for the maximal gap $\D_{\Disp^2}>\D_{\Disp}$ as a function of $\D_\psi$ and $\D_\Disp$. The results are shown in fig.~\ref{3d_gap_bounds}, which provides an overview of the landscape in which the RG flow is embedded. A section of this 3d plot at $\D_\psi = 3/2$ is shown in fig.~\ref{gap_TriIsing_deltaD_deltaGap}.
The interesting `bump' in the plot of fig.~\ref{3d_gap_bounds} is delimited by:
\begin{itemize}
	
	\item[(i)] A `floor' spanned by the generalized free fermion solution with $\D_{\Disp^2} = 2\D_\psi+1$ for any $\D_\psi$ and $\D_{\Disp}$, shown as a pink surface in the figure.
	
	\item[(ii)] The dashed blue line at the top of the cliff, which is another generalized free fermion solution with $(\D_\psi, \D_{\Disp} = 2\D_\psi+1, \D_{\Disp^2} = 2\D_\psi+3)$. The $(1,2)_3$ boundary condition for Ising corresponds to $\D_\psi = 1/2 $ along this line and is signaled by a red dot.
	
	\item[(iii)] The dashed blue line at the bottom of the cliff, which corresponds to the following four-point correlation function
	\begin{equation}
		\label{solution_without_identity}
		\mathcal{G}(\eta)= \frac{\eta^{4 \D_\psi/3}}{(1-\eta)^{2 \D_\psi/3}}\,.
	\end{equation}
	This solution has $\D_\Disp = \frac{4}{3}\D_\psi$ and $\D_{\Disp^2}=\D_\Disp +2$, for arbitrary $\D_\psi$, and has previously appeared in \cite{Antunes:2021abs,Esterlis:2016psv,Cordova:2022pbl}.\footnote{This correlator maximizes the gap without an identity exchange. See also \cite{Antunes:2023dlk} for a discussion of gap maximization without identity in the context of six-point correlators.} It coincides with the $(2,1)_4$ boundary condition at $\D_\psi=3/2$, which is the other red dot in fig.~\ref{gap_TriIsing_deltaD_deltaGap}.
\end{itemize}

The upshot of this analysis is that the UV and IR endpoints of the RG flow are close to saturating the bound. In between these points, the full RG flow is guaranteed to lie at or below the yellow surface.

\subsubsection{Mixed correlator system with $\psi$  and $\Disp$: the anteater}\label{ss:mixedcorr}
To further constrain the RG flows around the tricritical and ordinary Ising model, let us assume that $\psi$ and $\Disp$ are the leading boundary primaries in the $\mathbb{Z}_2$-odd and $\mathbb{Z}_2$-even sectors, respectively. Having analyzed their individual four-point functions in sections \ref{sec:Displacement} and \ref{ss:singlecorr},
we now consider the following mixed system of correlators
\begin{equation}\label{sysmix}
	\langle\psi\psi\psi\psi\rangle\,, \quad 
	\langle\Disp\Disp\Disp\Disp\rangle\,, \quad
\langle\psi\Disp\psi\Disp\rangle	\,,\quad
\langle\psi\psi\Disp\Disp\rangle\,.
\end{equation}
Below, we shall refer to this setup simply as ``the mixed correlator system''. Our assumptions on the various OPEs are then as follows:
\begin{align}
	\psi \times \psi &\sim \hid + \Disp + \Disp^2 + \tt{gap}_{(+,+)}+\dots \, , \nonumber \\
	\Disp\times \Disp &\sim \hid + \Disp + \Disp^2 + \tt{gap}_{(+,+)}+ \dots \, , \nonumber \\
	\psi\times \Disp &\sim \psi + \tt{gap}_{(-,+)}+ \dots +\tt{gap}_{(-,-)} \dots \,.
\end{align}
Here  ${\tt gap}_{({\mathbb Z}_2, P)}$ is the gap in the sector with a given ${\mathbb Z}_2$-charge and parity $P$. We recall that $P$ is the analog of spin for the one-dimensional conformal group and takes the value $+1$ for even operators and $-1$ for odd operators. Both $\Disp$ and $\psi$ are P even, but the $\psi \times \Disp$ OPE can contain operators of either parity so we can impose two different gaps.

We refer to appendix K of \cite{Homrich:2019cbt} for a detailed discussion of the conformal block decomposition of the four four-point functions and the otherwise standard numerical setup. In particular it is discussed there how $\langle\psi\Disp\psi\Disp\rangle$ and $\langle\psi\psi\Disp\Disp\rangle$ are not related by analytic continuation and allow to distinguish between operators with different parity.

\begin{figure}[h]
	\centering
	\subfloat[]{
		\begin{minipage}{0.3\textwidth}
			\includegraphics[width=1\textwidth]{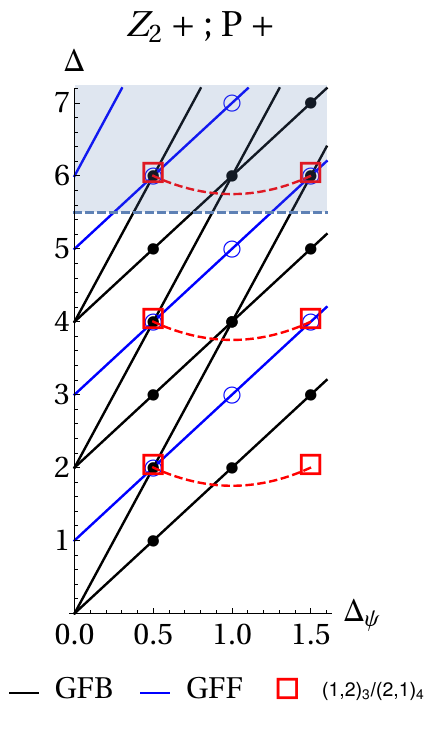} 
	\end{minipage}}
	\hspace{0.2cm}
	\subfloat[]{
		\begin{minipage}{0.3\textwidth}
			\includegraphics[width=1\textwidth]{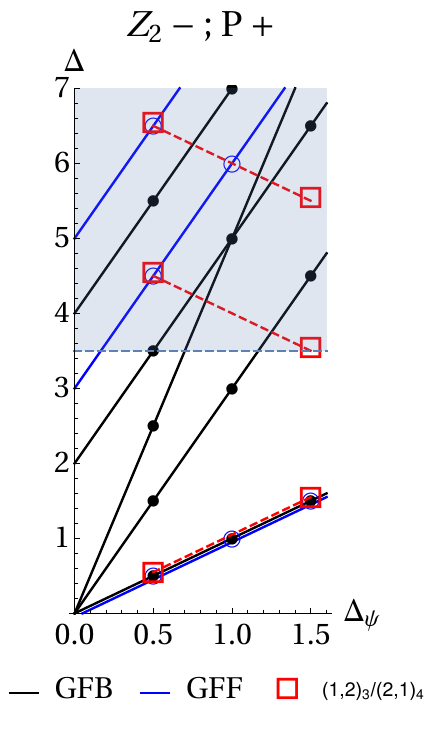} 
	\end{minipage}}
	\hspace{0.2cm}
	\subfloat[]{
		\begin{minipage}{0.3\textwidth}
			\includegraphics[width=1\textwidth]{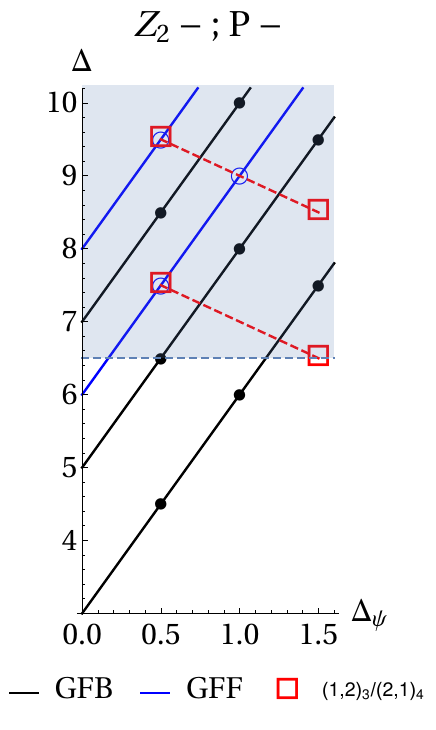} 
	\end{minipage}}~
	\caption{Spectra of GFF (blue lines and circles), GFB (black lines and circles) and minimal model boundary conditions (red squares) for the different $\mathbb{Z}_2$ and $P$ sectors: (a) $\mathbb{Z}_2:+,P:+$; (b) $\mathbb{Z}_2:-,P:+$; (c) $\mathbb{Z}_2:-,P:-$. The red dashed lines are suggestive of the qualitative behavior that the different dimension might take along the RG flow. The blue dashed lines and the above shaded region correspond to the gap assumptions specified in Table \ref{tbl:gapstric}.} 
	\label{fig:spectra}
\end{figure}

In one dimension our bounds are at risk of being saturated by a generalized free solution. To gain some perspective we therefore plot the spectra of the generalized free fermion (GFF), the generalized free boson (GFB) and our minimal model boundary conditions in figure \ref{fig:spectra}. This then leads us to consider the gaps shown in Table \ref{tbl:gapstric}. Note that operators with $(\mathbb{Z}_2, P) = (+,-)$ are not exchanged here.
\begin{table}[H]
	\begin{center}
		\begin{tabular}{c||c|c}
			${\tt gap}_{(\mathbb{Z}_2,P)}$& $P \, +$ & $P \, -$ \\
			\hline
			\hline
			$\mathbb{Z}_2 \, +$ & $5.5$ & -\\
			\hline
			$\mathbb{Z}_2 \, -$ & $3.5$ & $6.5$ \\
		\end{tabular}
	\end{center}
	\caption{Our assumed values for ${\tt gap}_{({\mathbb Z}_2, P)}$.}
	\label{tbl:gapstric}
\end{table}

\begin{figure}
	\centering
	\includegraphics[width=0.57\textwidth]{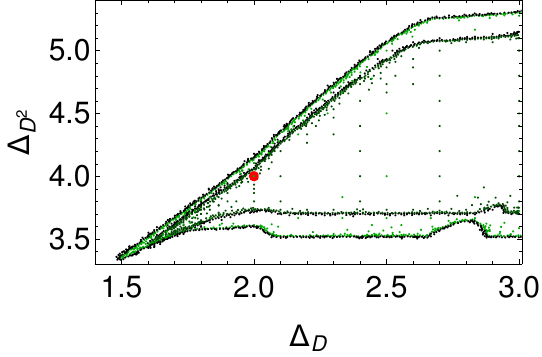} 
	\caption{Bounds on $(\D_{\Disp},\D_{\Disp^2})$ for $\D_{\psi}=3/2$ at $\Lambda=45$.  The dark green and black points respectively denote allowed and excluded points. We also show allowed points at the lower derivative order $\Lambda=30$ in lighter green and black.
		The red circle pinpoints the $(2,1)_4$ b.c., i.e. the eye of the `anteater', which is attached to a very sharp `nose'. }
	\label{fig:anteater1p5}	
\end{figure}
\begin{figure}
	\centering
	\subfloat[]{
		\begin{minipage}{0.5\textwidth}
				\hspace{-0.8cm}\includegraphics[width=1\textwidth]{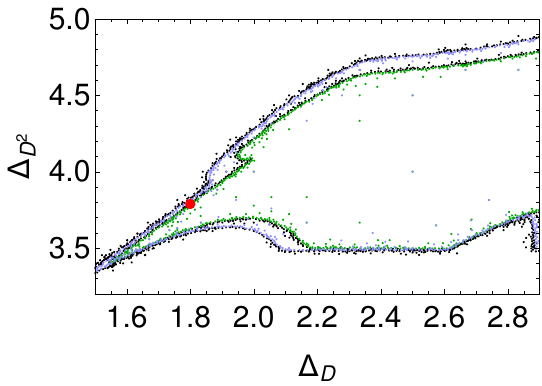} 
	\end{minipage}}
	\subfloat[]{
		\begin{minipage}{0.5\textwidth}
				\hspace{-0.8cm}\includegraphics[width=1\textwidth]{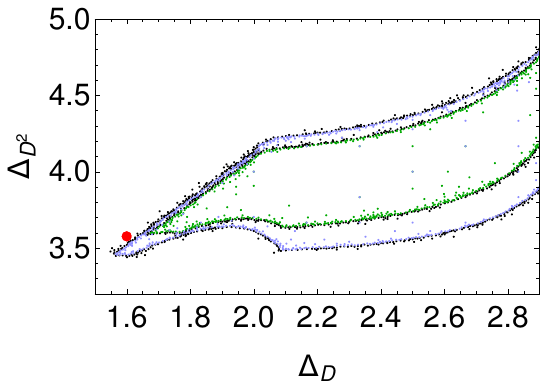} 
	\end{minipage}}
	\hspace{0.5cm}
	\subfloat[]{
		\begin{minipage}{0.5\textwidth}
			\hspace{-0.9cm}	\includegraphics[width=1\textwidth]{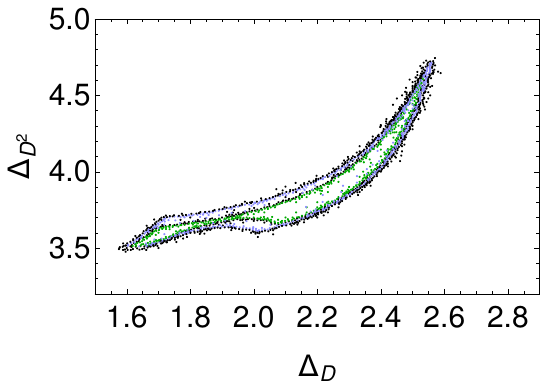} 
	\end{minipage}}
	\subfloat[]{
		\begin{minipage}{0.5\textwidth}
			\hspace{-0.9cm}	\includegraphics[width=1\textwidth]{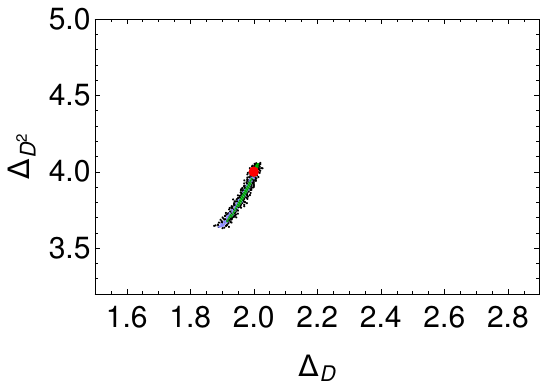} 
	\end{minipage}}~
	\caption{The anteater for different values of $\D_\psi$: (a) 1.25; (b) 1.0; (c) 0.75; (d) 0.5, interpolating between the UV and IR fixed points. The green allowed points were obtained at $\Lambda=30$ and the blue ones at $\Lambda=20$. The red circles in (a-b) denote the leading order $\phi_{1,3}$ deformation of the $(2,1)_4$ b.c, eqs.(\ref{eq:tricritphi13}, \ref{RGpred13tricD2}) below, evaluated at the corresponding value of $\D_\psi$. The `anteater' shrinks as we approach the IR, and forms a small island around the $(1,2)_3$ b.c. of Ising, which is denoted by the red circle in (d).}
	\label{fig:anteater}
\end{figure}

Below these gaps we assume two operators in the $(\mathbb{Z}_2,P) = (+,+)$ sector ($\Disp$ and $\Disp^2$) and one operator in the $(\mathbb{Z}_2,P) = (-,+)$ sector ($\psi$). Altogether this means that these assumptions rule out the GFB except when $\D_\psi \geq 1.17$, but not the GFF which has a very sparse spectrum. We can nevertheless hope that the RG flow saturates our bounds at least somewhere in the $(\D_{\Disp},\D_{\Disp^2})$ plane as we vary $\D_\psi$.

We compute allowed points in the space $(\D_{\Disp},\D_{\Disp^2})$ as we vary $\D_\psi$ within the interval $ [1/2, 3/2] $. We then use these points to delineate the allowed region using Delaunay triangulation. The numerically allowed region, the `anteater', is shown in figures \ref{fig:anteater1p5} and \ref{fig:anteater} for different values of $\D_\psi$.

A few comments are in order:
\begin{itemize}
	\item[(i)] In the UV, the $(2,1)_4$ boundary condition for the tricritical Ising model appears to saturate the bounds, see the red circle in fig.~\ref{fig:anteater1p5}.
	\item[(ii)] As we go with the flow towards lower values of $\D_\psi$, the allowed region shrinks, becoming substantially thinner, see figures~\ref{fig:anteater}(a-c).
	\item[(iii)] In the IR, the $(1,2)_3$ boundary condition for the Ising model is almost isolated by our gap assumptions, up to a small lobe shown in fig.~\ref{fig:anteater}(d). As we discuss below, this can be understood by combining a $\phi_{(1,3)}$ with a $\TTb$ deformation.
	\item[(iv)] Overall, we see the bounds have not fully converged as we increase $\Lambda$. In particular, the sharper features such as the `nose' of the anteater might still shrink significantly. The size of the nose is also sensitive to the gap assumptions, as we will discuss further below.
\end{itemize}

{\tiny }We are going to explore these features more closely in the next sections, by focusing on specific perturbative RG flows.

\subsection{Bootstrapping perturbative RG flows}\label{ss:Bootspert}
In this section we refine the `agnostic' bootstrap employed in section \ref{ss:trictocrit} by focusing on some of the $\mathbb{Z}_2$-preserving perturbative RG flows of section~\ref{sec:pertres}. Specifically, we again adopt the mixed-correlator system, but this time we bound CFT data along a specific RG trajectory by inputting the one-loop predictions for $\D_\psi$ and $\D_\Disp$ and comparing the slope in the bound on $\D_{\Disp^2}$ at the fixed point to predictions from perturbation theory.

\subsubsection{Deformations of the Ising Model with $(1,2)_3$ boundary condition}

We begin by studying the vicinity of the Ising model with $(1,2)_3$ boundary condition using the single-correlator bootstrap setup of section~\ref{ss:singlecorr}.
\paragraph{(I) The relevant deformation\\}

Turning on $\phi_{(1,3)}$ in the bulk of $\AdStwo$ corresponds to giving a mass to the free massless Majorana fermion in the dual description of the Ising model. In terms of boundary correlators, $\psi$ becomes a GFF whose scaling dimension $\D_\psi$ smoothly moves away from $1/2$, the Ising value. Since both signs of the fermion mass are allowed, there is no constraint on the sign of $\phi_{(1,3)}$.
This expectation is corroborated by our bootstrap study. One can search for the maximal gap (after $\Disp$) in the four-point correlation function of $\psi$, along the direction suggested by the one-loop results of eq.~\eqref{an21ising}:
\begin{align}
	\D_\psi  = 1/2 + \tilde{g}_{(1,3)}\,,\quad \D_\Disp = 2 +2\tilde{g}_{(1,3)}\,.
\end{align}
The upper bound is saturated for both signs of the coupling by
\begin{align}
	\D_{\Disp^2} = 4 +2\tilde{g}_{(1,3)}=2\D_\psi +3\,,
\end{align}
which is just the GFF value. This solution remains valid until $\tilde{g}_{(1,3)}$ is such that $\D_\psi$ decreases down to zero, below which unitary is necessarily violated.

\paragraph{(II) The leading irrelevant deformation\\}

\begin{figure}
	\centering
	\subfloat[]{
		\begin{minipage}{0.5\textwidth}
			\includegraphics[width=1\textwidth]{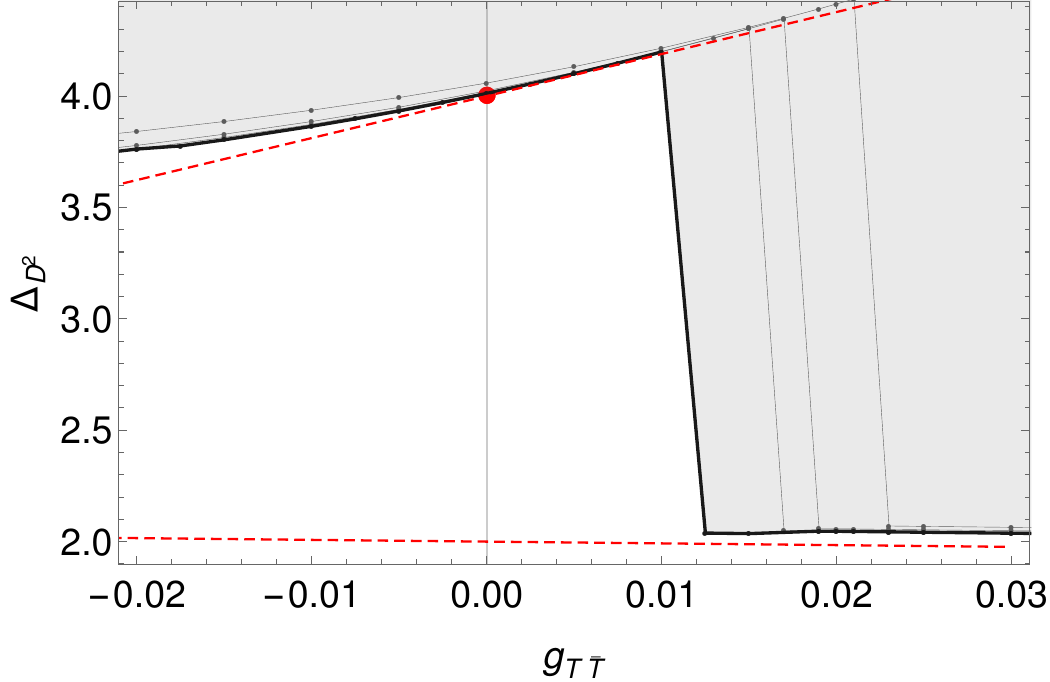} 
	\end{minipage}}
	\hspace{0.5cm}
	\subfloat[]{
		\begin{minipage}{0.4\textwidth}
			\includegraphics[width=1\textwidth]{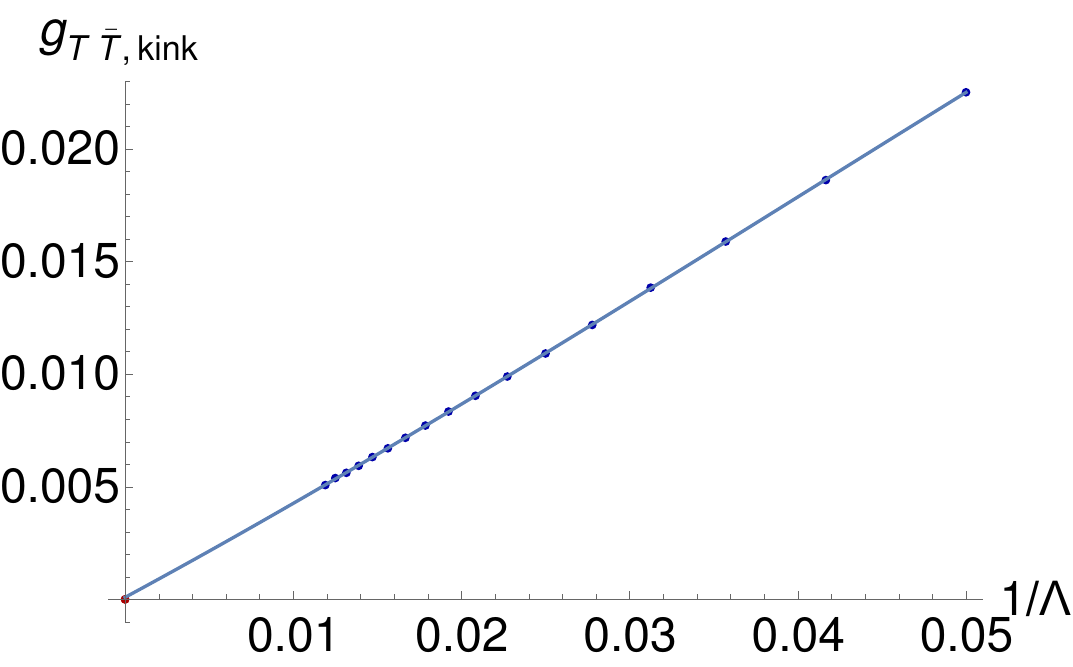} 
	\end{minipage}}~
	\caption{Comparison between the numerical upper bounds (solid lines) and the predictions from the first-order $\TTb$ deformation (top dashed line) near $(1,2)_3$, which is denoted by the red dot. In (a), to the right of the kink, the bound converges to a GFF with $\D_\psi=\frac{1}{2}-\frac{g_{\TTb}\pi}{8}$ (bottom dashed line). Figure (b) shows the convergence of the kink along the perturbation theory line towards $g_{\TTb}=0$. The data are fitted with a cubic, and the intercept is $\simeq 10^{-4}$.}
	\label{section_gap_bounds_vs_TTb_Ising}
\end{figure}

We can repeat the same gap maximisation along the $\TTb$ deformation of the Ising model $(1,2)_3$ conformal boundary condition, which from eq.~\eqref{an21ising} means that we take
\begin{align}
	\label{TTbarleadingorderinput}
	\D_\psi  = 1/2 -{g}_{\TTb}\pi/8\,,\quad \D_\Disp = 2 +{g}_{\TTb}\pi\,.
\end{align}
The comparison with the one-loop prediction for $\D_{\Disp^2}$ is shown in fig.~\ref{section_gap_bounds_vs_TTb_Ising}.
After extrapolating in the number of derivatives we see that the $g_{\TTb}>0$ region is completely excluded (on the basis of crossing and unitarity). This is consistent with the results of section~\ref{sec:Displacement}, but here the same conclusion is obtained from a different correlation function.

It is clear that this has to be the case.  For a single-correlator with external dimension $\D_\psi$, the maximal gap above the identity is $2\D_\psi+1$, see e.g.~\cite{Mazac:2016qev}. The Ising model $(1,2)_3$ boundary condition saturates this bound when $g_{\TTb}=0$, while the first-order $\TTb$ perturbation of eq.~\eqref{TTbarleadingorderinput} clearly violates it when $g_{\TTb}>0$. 
Note that, in this particular case, the $\TTb$ deformation can be written as a special higher derivative interaction around the free fermion, and this sign constraint can be understood using AdS$_2$ dispersion relations~\cite{Knop:2022viy}. 
At the first order this fermionic derivative interaction leads to a $u$-channel Regge behavior ($\eta \to i \infty$) of $\eta^{-2\D_\psi}\mathcal{G}(\eta)\sim a\eta^1$ \cite{Mazac:2018ycv}. This is precisely the maximal Regge behavior allowed in a planar CFT, corresponding to a Regge spin of 2 \cite{Mazac:2018ycv,Penedones:2019tng} (note that higher dimensional $u$-channel Regge limit bounds can consistently be studied in 1d by setting $z=\bar{z}$). When this Regge behavior is saturated, bounds from causality/chaos also constrain the sign of $a$ \cite{Maldacena:2015waa,Hartman:2015lfa}, reproducing the constraint we derived from the bootstrap. The same argument also applies for the bosonic $(\partial \phi)^4$ coupling discussed in \cite{Antunes:2021abs}. 

\paragraph{(III) The combined relevant-irrelevant deformation\\}
We can also consider the combined deformation where we switch on both $g_{\TTb}$ and $g_{(1,3)}$ to first order. Indeed, using the dimensionless couplings
\begin{equation}
		g_i \colonequals \lambda_i R^{d-\D_i}\,,
\end{equation}
it is natural to consider deformations of the form
\begin{equation}
	S_{CFT}+ \sum_i g_i \int_{AdS_2} d^2x \sqrt{g} R^{\D_i-2}\mathcal{O}_i(x) \,,
\end{equation}	
where one then studies perturbation theory in $g_i$, as we did in section \ref{sec:pertres}. This then leads to corrections to the boundary conformal data, for example
\begin{equation}
\D_{\psi}(g_{i})= \hD_\psi+\sum_i g_{i}\dhD^{i}_{\psi}+\dots\,.
\end{equation}
Therefore, we can consider fixing the ratio $g_i/g_j$ while keeping both couplings small to stay in the region of validity of perturbation theory. This dimensionless quantity parametrizes a certain family of deformations. However, as remarked in the Introduction, it is important to note that these ratios $g_i/g_j$ are not what usually parametrizes families of RG flows. This is typically done by fixing instead the quantities $g_i^{1/d-\D_i}/g_j^{1/d-\D_j}$, which are obtained from building a dimensionless quantity only in terms of the dimensionful couplings $\lambda_i$ without making use of the AdS radius. 

\begin{figure}[h]
	\centering
	\includegraphics[width=0.55\textwidth]{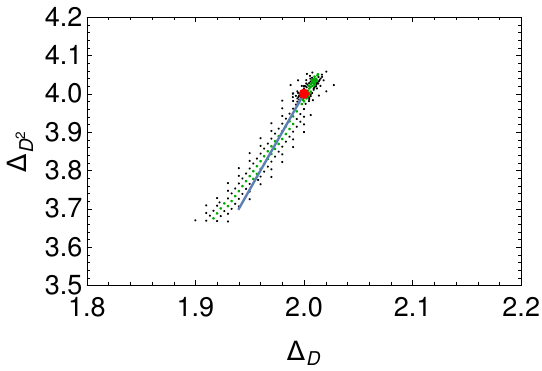} 
	\caption{A zoom-in of fig.~\ref{fig:anteater}(d). The Ising boundary condition is identified by the red dot and the blue line corresponds to the combined $g_{\TTb}+g_{(1,3)}$ deformation, tuned to keep $\D_\psi=1/2$.}
	\label{fig:isingislanddeform}
\end{figure}

We will hence study perturbations with $g_{\TTb}/g_{(1,3)}$ fixed and use the freedom in picking this ratio to keep the dimension $\D_\psi=1/2$ to leading order. Using equation \eqref{an21ising}, we can solve for one coupling in terms of the other: $g_{(1,3)}= g_{\TTb}/16$. To first order in the couplings this leads to the relation
\begin{equation}
\D_{\Disp^2}= 4 + 5(\D_\Disp -2) \,.
\end{equation}	
 We show a comparison of this deformation to the mixed correlator numerical bounds in figure \ref{fig:isingislanddeform}, which is simply a zoom-in of fig.~\ref{fig:anteater}(d). 

The allowed region appears to be spanned by the combined deformation, being particularly elongated towards the direction where the $\TTb$ coupling takes the allowed sign $g_{\TTb}<0$. As the derivative order increases, the top part of the bound should converge to the Ising point, once again consistently with the exclusion of the wrong sign of $\TTb$. While it is interesting that we can understand the space of solutions to crossing by considering combined deformations in AdS$_2$, this is also a limitation if we want to focus on physical RG flows.

\subsubsection{Deformations of tricritical Ising Model with $(2,1)_4$ boundary condition}

Next, we study the vicinity of the tricritical Ising model with $(2,1)_4$ boundary condition using the mixed-correlator system.

\begin{figure}
	\centering
	\includegraphics[width=0.55\textwidth]{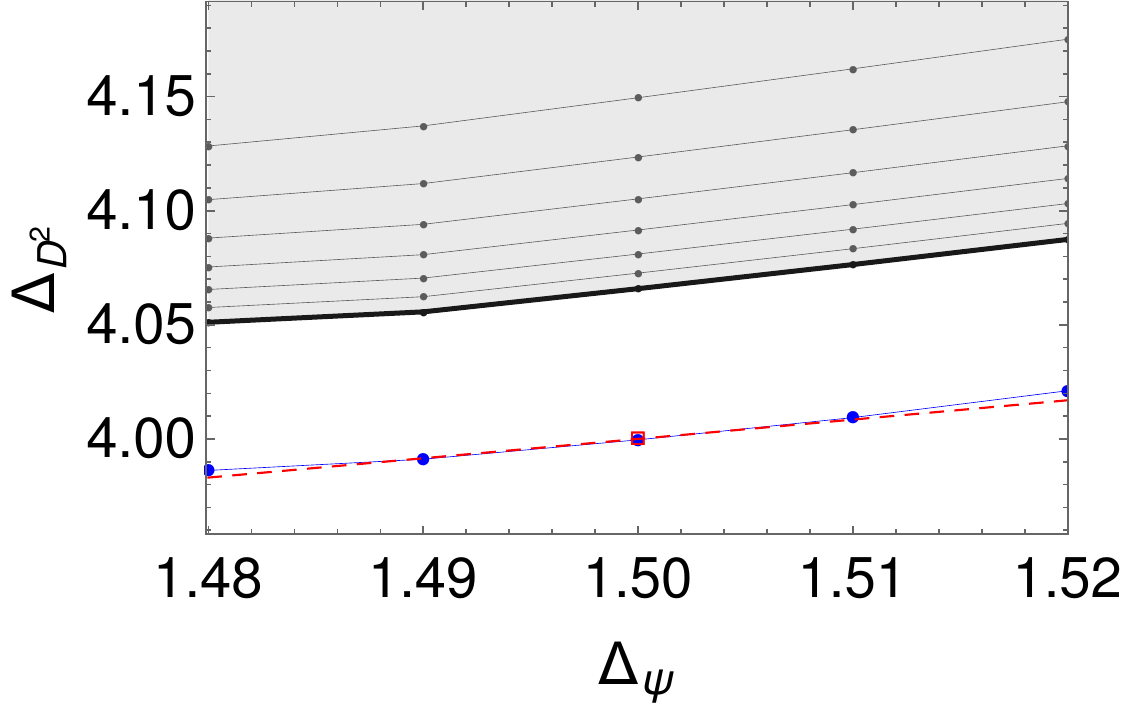} 
	\caption{ Bounds on $\D_{\Disp^2}$ as a function of $\D_\psi$, where $\D_\psi$ and $\D_\Disp$ follow the perturbative prediction of eq.~(\ref{eq:tricritphi13}). The grey dots denote bounds at increasing derivative order, which are then extrapolated to the blue dots. The red line corresponds to the analytic results of eq.~(\ref{eq:tricritphi13}). The bound is saturated for both signs of the coupling.}
	\label{fig:phi13alongpertmulti}
\end{figure}

\paragraph{(I) The relevant deformations\\}
There are two $\mathbb{Z}_2$-even relevant bulk deformations of tricritical Ising model: $\phi_{(1,3)}$ and $\phi_{(3,3)}$. For the former let us first consider the mixed correlator system, but this time varying $\D_\psi$ and $\D_\Disp$ along the one-loop prediction (see Table~\ref{tbl:an21})
\begin{equation}
	\label{eq:tricritphi13}
	\D_\psi =\frac{3}{2}+\tilde{g}_{(1,3)}, \quad \D_\Disp = 2+ \frac{4}{5}\tilde{g}_{(1,3)}\,,
\end{equation}
and comparing with
\begin{align}\label{RGpred13tricD2}
	\D_{\Disp^2} = 4+ \frac{72}{85}\tilde{g}_{(1,3)}\,.
\end{align}
As shown in figure \ref{fig:phi13alongpertmulti}, which was obtained with the gaps of Table~\ref{tbl:gapstric}, the (extrapolated) upper bound is saturated by the RG prediction of eq.~\eqref{RGpred13tricD2} for both signs of the coupling. This is not a surprise: taking $\tilde{g}_{(1,3)}<0$
 should lead to the Ising model in the bulk, while taking it positive is expected to gap out the bulk, hence making all the boundary scaling dimensions become large. In flat space these massive $\phi_{(1,3)}$ deformations of minimal models correspond to a family of integrable massive theories known as RSOS models, which are related to certain restricted versions of the sine-Gordon theory, as discussed in~\cite{Smirnov:1990vm,Eguchi:1989hs}.

\begin{figure}
	\centering
	\includegraphics[width=0.55\textwidth]{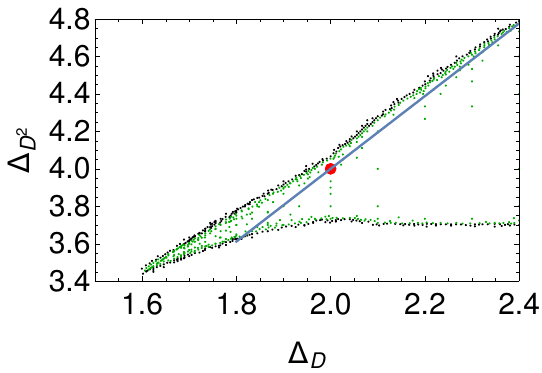} 
	\caption{The combined $\phi_{(3,3)}+\phi_{(1,3)}$ deformation of the $(2,1)_4$ b.c. for $\D_\psi=3/2$. The solid line is the perturbative result of eq.~\eqref{epsepsppdef}, which appears to saturate the upper bound to the right of the $(2,1)_4$ point, represented by the red circle.}
	\label{fig:3depsepspp}
\end{figure}

Having two relevant singlet couplings, we can consider the combined RG flow in AdS, in analogy to what we have done for the Ising model boundary condition. We study the family of 1d CFTs obtained by turning on small couplings to $\phi_{(1,3)}$ and $\phi_{(3,3)}$, keeping the ratio $g_{(1,3)}/g_{(3,3)}$ fixed.\footnote{The individual $\phi_{(3,3)}$ deformation in the complex plane is known to lead to an integrable massive system: Zamolodchikov's $E_7$ theory \cite{Zamolodchikov:1989hfa}.}
For the  $(2,1)_4$ boundary condition, using the result of Table~\ref{tbl:an21}, we can trade $(g_{(3,3)},g_{(1,3)})$ for $(\D_\psi,  \D_\Disp)$ and write
\begin{align}\label{epsepsppdef}
	\D_{\Disp^2}=\frac{1}{17} (20+33 \D_{\Disp}-12 \D_{\psi} )\,.
\end{align}
In the vicinity of the $(2,1)_4$ boundary condition, we can tune $g_{(1,3)}/g_{(3,3)}$  to keep $\D_\psi$ fixed at $1.5$ and then compare this one-loop prediction with the bootstrap bounds of section~\ref{ss:mixedcorr}, in particular with those presented in fig.~\ref{fig:anteater1p5}. As shown in fig.~\ref{fig:3depsepspp}, this combined flow appears to saturate the boundary of the anteater, with a slightly better fit to the right of the $(2,1)_4$ boundary condition. 

\begin{figure}
	\centering
	\subfloat[]{
		\begin{minipage}{0.5\textwidth}
			\hspace{-0.7cm}\includegraphics[width=1\textwidth]{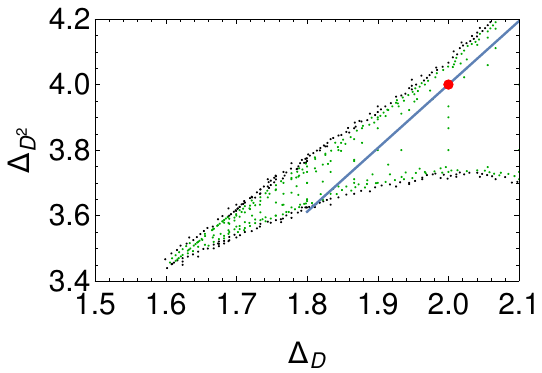} 
	\end{minipage}}
	\hspace{0.5cm}
	\subfloat[]{
		\begin{minipage}{0.5\textwidth}
				\hspace{-0.7cm}\includegraphics[width=1\textwidth] {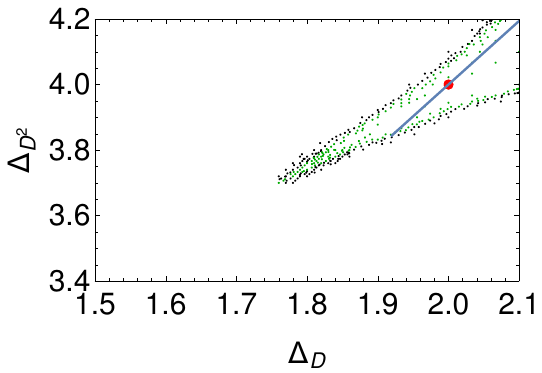}
	\end{minipage}}
	\subfloat[]{
		\begin{minipage}{0.5\textwidth}
				\hspace{-0.7cm}\includegraphics[width=1\textwidth] {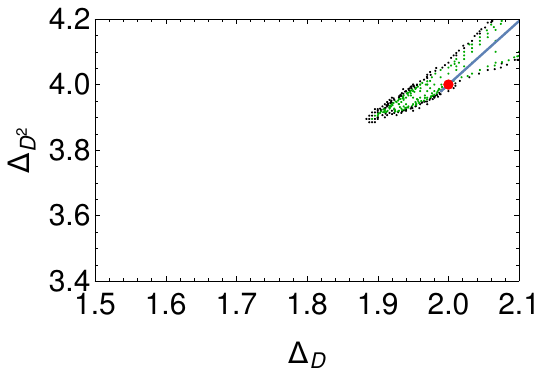}
	\end{minipage}}~
	\caption{The combined $\phi_{(3,3)}+\phi_{(1,3)}$ deformation of the $(2,1)_4$ b.c. for $\D_\psi=3/2$ at $\Lambda=45$. In (a) ${\tt gap}_{(+,+)}=5.5$, in (b) ${\tt gap}_{(+,+)}=5.8$ and in (c) ${\tt gap}_{(+,+)}=6.0$. The solid line is the perturbative result of eq.~\eqref{epsepsppdef}.}
	\label{fig:anteater_tric_gaps}
\end{figure}

It is also interesting to investigate how the bounds change as we vary ${\tt gap}_{(+,+)}$. Figure~\ref{fig:3depsepspp} was obtained by choosing ${\tt gap}_{(+,+)}=5.5$, and allows for deformations with both signs of the (combined) coupling. As shown in figure \ref{fig:anteater_tric_gaps}, upon increasing the gap to 5.8 the allowed region shrinks while still allowing for both signs of the coupling.  Taking the gap all the way to 6 leads to near saturation of the tip by the $(2,1)_4$ boundary condition, and a positive sign of the combined deformation is in near contradiction with the gap assumption. In other words, as we lower the gap from 6, the `nose' of the anteater grows, and the location of the tip gives a heuristic definition for how big `the combined coupling' can be for a fixed value of the gap. We also note that the top part of the bound is insensitive to the gap assumption. This means it is meaningful to identify the tricritical Ising as a theory saturating the upper bound on $\D_{\Disp^2}$.

\paragraph{(II) The leading irrelevant deformation\\}

\begin{figure}
	\centering
	\includegraphics[width=0.55\textwidth]{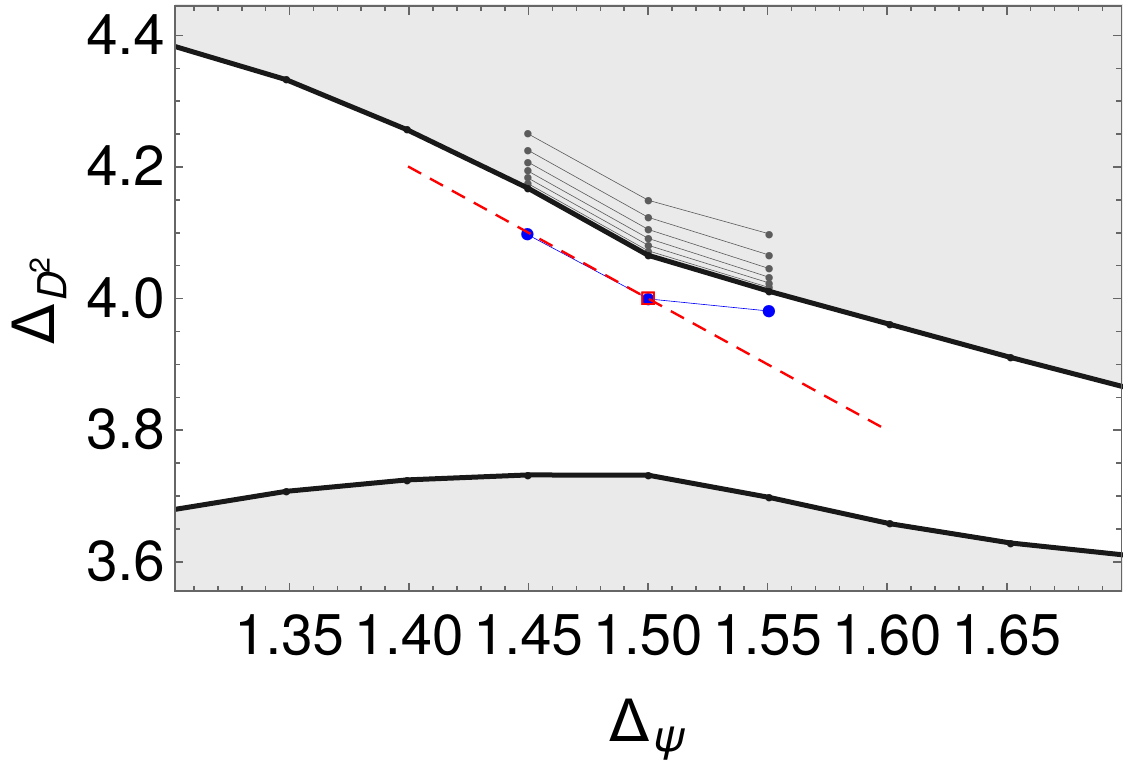} 
	\caption{Bounds on $\D_{\Disp^2}$ as a function of $\D_\psi$, along eq.~\eqref{eq:tricritphi31}. The grey dots denote bounds at increasing derivative order, which are then extrapolated to the blue dots. The red line corresponds to the analytic results of eq.~\eqref{RGpred31tric}. The bound is saturated for $\tilde{g}_{(3,1)}<0$.}
	\label{fig:epspp_tricr_L30_1}
\end{figure}

We consider the $\phi_{(3,1)}$ deformation of $(2,1)_4$, which is $\mathbb{Z}_2$-preserving. We run again the mixed-correlator system, varying $\D_\psi$ and $\D_\Disp$ along the one-loop prediction (see Table~\ref{tbl:an21})
\begin{equation}
	\label{eq:tricritphi31}
	\D_\psi =\frac{3}{2}+\tilde{g}_{(3,1)}, \quad \D_{\Disp} = 2-\frac{2}{3}\tilde{g}_{(3,1)}\,,
\end{equation}
and comparing with
\begin{align}\label{RGpred31tric}
	\D_{\Disp^2} = 4 -2 \tilde{g}_{(3,1)}\,.
\end{align}
The results are shown in fig.~\ref{fig:epspp_tricr_L30_1} (the chosen gaps are those of Table~\ref{tbl:gapstric}). While for $\tilde{g}_{(3,1)}<0$ the bound appears to be saturated, the other sign points towards the interior of the allowed region. This is possible due to a quick change in the slope of the bound around the $(2,1)_4$ point.
The $m$'th minimal model flows under the relevant $\phi_{(1,3)}$ deformation to the $m-1$'th minimal model in the IR. From the IR point of view, it is the \emph{irrelevant} $\phi_{(3,1)}$ operator that begins the flow back up to the UV. Interestingly, it is precisely when $\tilde{g}_{(3,1)}<0$ that the tricritical $\phi_{(1,3)}$ operator is becoming less irrelevant, which means that it is exactly the flow up to the tetracritical Ising model that saturates the bound in the perturbative region.

\begin{figure}
	\centering
	\includegraphics[width=0.5\textwidth]{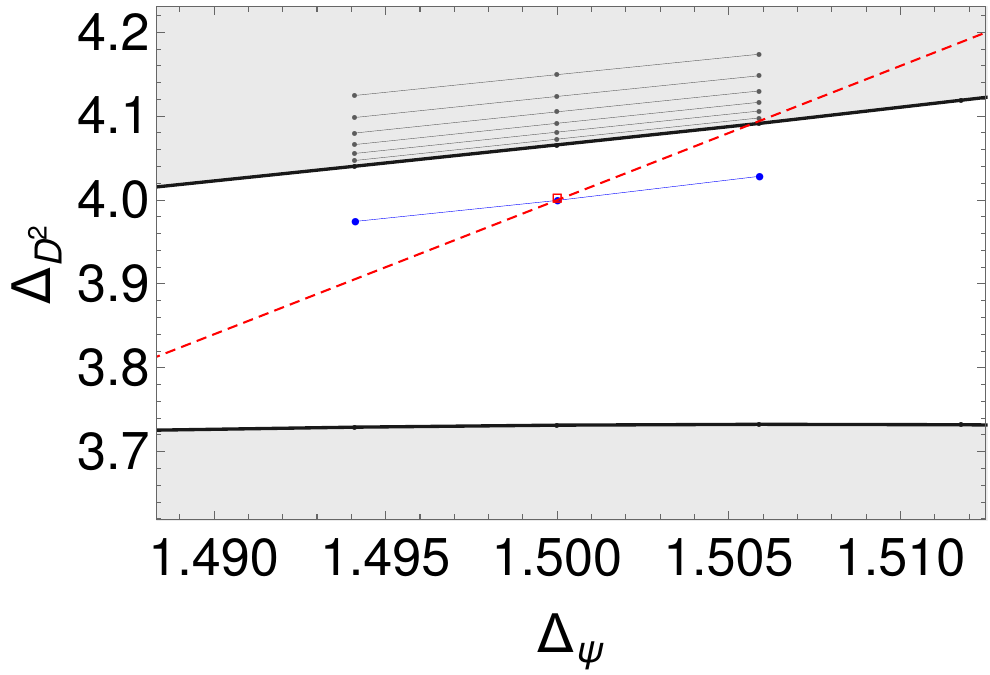} 
	\caption{Bounds on $\D_{\Disp^2}$ as a function of $\D_\psi$, along eq.~\eqref{eq:tricritphiTTb}. The grey dots denote bounds at increasing derivative order, which are then extrapolated to the blue dots. The red line corresponds to the analytic results of eq.~\eqref{RGpredTTbtric}. The bound is violated for $\tilde{g}_{\TTb}>0$.}
	\label{fig:TTb_tricr_L30_1}
\end{figure}

\paragraph{(III) The $\TTb$ deformation\\}
Finally we consider the $\TTb$ deformation. This time we vary $\D_\psi$ and $\D_\Disp$ along the one-loop prediction (see Table~\ref{tbl:an21})
\begin{equation}
	\label{eq:tricritphiTTb}
	\D_\psi =\frac{3}{2}+ \tilde{g}_{\TTb}, \quad \D_\Disp = 2+ \frac{8}{3}\tilde{g}_{\TTb}\,,
\end{equation}
and comparing with
\begin{align}\label{RGpredTTbtric}
	\D_{\Disp^2} = 4+16\tilde{g}_{\TTb}\,.
\end{align}
As shown in fig.~\ref{fig:TTb_tricr_L30_1}, there is a clear sign constraint, so it must be that $g_{\TTb}\leq 0$. This is the same sign determined in section \ref{sec:Displacement}, consistent with causality. It is once again reassuring to find out that different boundary correlators  lead to the same inconsistency.

\subsubsection{Explorations around tricritical Ising Model with $(2,2)_4$ boundary condition}
\label{sec:22}

We conclude this section with an exploration of the `neighborhood' of tricritical Ising with $(2,2)_4$ boundary condition. 
To this end, we employ again the agnostic approach of section~\ref{ss:trictocrit}, but this time we consider a $\mathbb{Z}_2$-invariant system of correlators with two global boundary primaries, one $\mathbb{Z}_2$-odd ($\psi$) and one $\mathbb{Z}_2$-even ($\chi$). Hence we consider:
\begin{equation}
	\langle\psi\psi\psi\psi\rangle\,, \quad 
	\langle\chi\chi\chi\chi\rangle\,, \quad
	\langle\psi\psi\chi\chi\rangle\,, \quad
	\langle\psi\chi\psi\chi\rangle\,.
\end{equation}
In order to bootstrap this system of correlators it is useful to impose gaps.
\begin{table}
	\begin{center}
		\begin{tabular}{c||c|c}
			${\tt gap}_{(\mathbb{Z}_2,P)}$& $P \, +$ & $P \, -$ \\
			\hline
			\hline
			$\mathbb{Z}_2 \, +$ & $2.4$ & -\\
			\hline
			$\mathbb{Z}_2 \, -$ & $1.3$ & $2.8$ \\
		\end{tabular}
	\end{center}
	\caption{Our assumed values for ${\tt gap}_{({\mathbb Z}_2, P)}$.}
	\label{tbl:gaps22}
\end{table}
At the $(2,2)_4$ conformal boundary condition, we can identity $\psi$ with $\psi_{(3,3)}$ and $\chi$ with $\psi_{(1,3)}$. With this in mind, and recalling the OPEs of eq.~\eqref{leadingOPEstric22}, we assume
\begin{align}
	\psi \times \psi &\sim \hid + \chi+ \Disp +{\tt gap}_{(+,+)}+ \dots \, , \nonumber \\
	\chi \times \chi &\sim \hid + \chi+ \Disp + {\tt gap}_{(+,+)}+ \dots \, , \nonumber \\
	\psi \times \chi &\sim \psi  + {\tt gap}_{(-,+)} + {\tt gap}_{(-,-)} + \dots \, .
\end{align}
A possible choice for the gaps (leaving some leeway to deformations) is displayed in Table~\ref{tbl:gaps22}.

\begin{figure}
	\centering
	\includegraphics[width=0.55\textwidth]{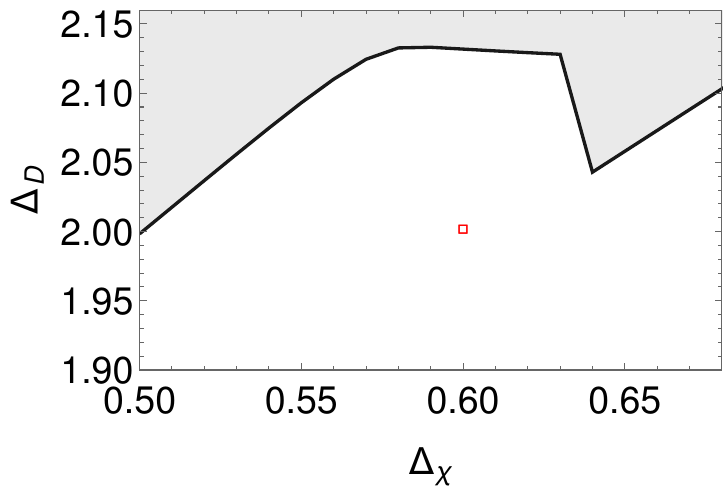} 
	\caption{Allowed region in $(\D_{\chi},\D_{\Disp})$ space for $\D_{\psi}=1/10$ with mild gap assumptions. The red dots is the location of $(2,2)_4$. The plot was obtained at derivative order $\Lambda=33$.}
	\label{fig:tricr_22_kink}
\end{figure}

\begin{figure}
	\centering
	\includegraphics[width=0.55\textwidth]{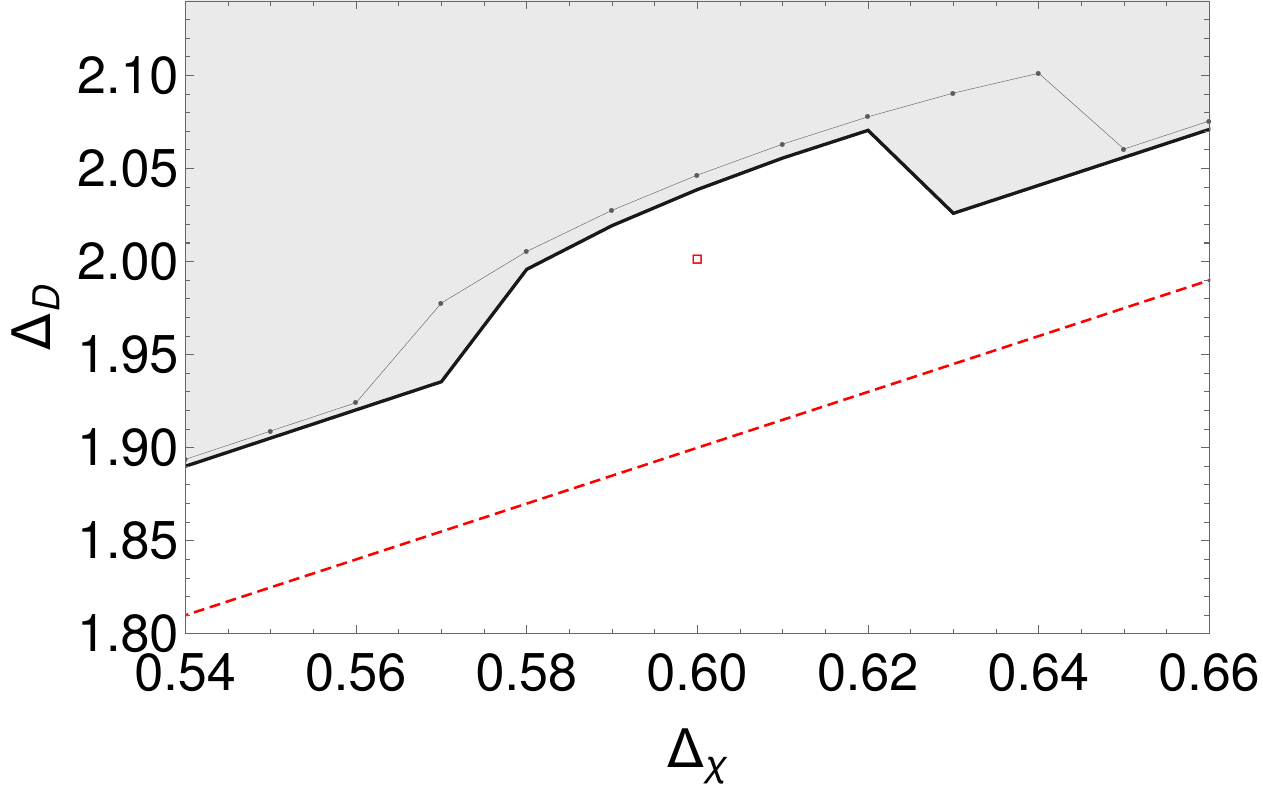} 
	\caption{Allowed region in $(\D_{\chi},\D_{\Disp})$ space for $\D_{\psi}=1/10$, for gap assumptions that saturate the spectrum of $(2,2)_4$ (red dot), at $\Lambda=33,41$. The red dashed line is the family of solution described by \eqref{eq:funnycorrelator}.
	} 
	\label{fig:weirdbump}
\end{figure}

In order to explore the vicinity of $(2,2)_4$ we choose $\D_{\psi}=1/10$ and explore the allowed values of $\D_\chi$ and $\D_{\Disp}$ around their $(2,2)_4$ values which are $0.6$ and $2$, respectively. This results in the kinky bound of fig.~\ref{fig:tricr_22_kink}, which shows that the $(2,2)_4$ boundary condition is deep inside the allowed region. In order to gain some insights on the two kinks in this figure, we can modify slightly our gap assumptions. With all the gaps set at their exact values corresponding to the $(2,2)_4$ boundary condition, we find the upper bound in figure~\ref{fig:weirdbump}. This time the upper bound is almost linear, except for a small bump precisely around the $(2,2)_4$ boundary condition. 
As we increase the number of derivatives, the linear part of the bound appears to converge towards the red dashed line in the figure. This line corresponds to a spurious solution to crossing given by
\begin{equation}
	\label{eq:funnycorrelator}
	\mathcal{G}(\eta)= \frac{\eta^{3\D_\chi/2}}{(1-\eta)^{\D_\chi}}+\frac{\eta^{\D_\chi}}{(1-\eta)^{\D_\chi/2}}-\frac{\eta^{3\D_\chi/2}}{(1-\eta)^{\D_\chi/2}}\,.
\end{equation}
This correlator can be obtained by taking linear combinations of fully connected Wick contractions of $\langle\phi^4\phi^4\phi^4\phi^4\rangle$, where $\phi$ is a GFF with scaling dimension $\D_\phi = \D_\chi/4$.
Just like the solution of equation \eqref{solution_without_identity}, it has the properties that no identity is exchanged. The leading (subleading) exchanged operator has dimension $\D_\chi$ ($\D_{\Disp}=3\D_{\chi}/2 +1$). If we identify $\chi$ with $\psi_{(1,3)}$ of  the $(2,2)_4$ boundary condition, the gap in \eqref{eq:funnycorrelator} is at $3\D_{(1,3)}/2 +1=  1.9$, just below the displacement at 2.


\subsection{Correlator maximization and the conformal staircase}\label{ss:corrmax}
In this section we will find a different quantity to extremize such that the $(2,2)_4$ b.c saturates the bound. It turns out that a natural object is the four-point correlator $\mathcal{G}(\eta)=x_{12}^{2\D_\psi}x_{34}^{2\D_\psi}\langle\psi\psi\psi\psi\rangle$ of the $\mathbb{Z}_2$-odd operator $\psi$ and its second derivative evaluated at the crossing symmetric point:
\begin{equation}
	\label{staircasetarget}
	\{\mathcal{G}(z=1/2),\mathcal{G}''(z=1/2)\}.
\end{equation}
Our motivation for studying the allowed region in this plane comes from the observation that the RG flows between minimal models can be embedded in the so-called `staircase' RG flows, which are connected to the sinh-Gordon/staircase model and its flat space S-matrix \cite{Zamolodchikov:1992ulx}. These S-matrices of a single massive particle without bound-states were recently shown to saturate bounds in the space \cite{Chen:2021pgx}: $$\{S(2m^2),S''(2m^2)\}.$$ Using the connection between S-matrices and correlators in the flat-space limit \cite{Paulos:2016fap} we arrive at equation \eqref{staircasetarget} as the natural uplift of these bounds to the QFT in AdS setup. A review of the staircase model and a more detailed explanation of its connection to the bounds below is given in appendix \ref{app:Staircase}. 

To bound $\mathcal{G}(1/2) \equiv g$ and $\mathcal{G}''(1/2) \equiv g''$ we fix them to a specific value and then determine whether this value is allowed\footnote{See also \cite{Paulos:2020zxx,Paulos:2021jxx} to an alternative approach to correlator extremization.}. For $g$ we can just use the recipe described in \cite{Antunes:2021abs}, where one works with a shifted identity conformal block:
\begin{equation}
	(1-\eta)^{2\D_\psi} G_\D(\eta)\to(1-\eta)^{2\D_\psi} G_\D(\eta)- \delta_{\D,0}2^{-2\D_\psi}g\equiv F^*_\D(\eta)\,,
\end{equation}
and then adds the zero derivative functional to the search space. To generalize this to the case where we fix both $g$ and $g''$ we perform two steps. First, we work with a different shifted identity block:
\begin{equation}
	(1-\eta)^{2\D_\psi} G_\D(\eta)\to F^*_\D(\eta)-\delta_{\D,0}\bigg(\eta-\frac{1}{2}\bigg)^2 2^{-1-2\D_\psi}\big(g''-8\D_\psi(2\D_\psi+1)g\big)\equiv F''_\D(\eta)\,.
\end{equation}
Then, alongside the usual odd derivative functionals, we include the zero- and two-derivative terms to the basis. We can subsequently perform a two-dimensional feasibility search in this space, for several external dimensions $\D_\psi$. We will always assume a gap of $2\D_\psi$ in the spectrum, in analogy with a theory without bound states in AdS. This is a rather conservative assumption with respect to the boundary conditions we wish to study. With this choice, the generalized free boson (GFB) and generalized free fermion (GFF) theories are always in the allowed space. By convexity that means that the line connecting these theories is also allowed, which allows one to find a line strictly in the interior of the allowed region. An efficient numerical exploration can then be performed by doing a radial/angular search around an interior point.

\begin{figure}
	\centering
	\includegraphics[width=0.5\linewidth]{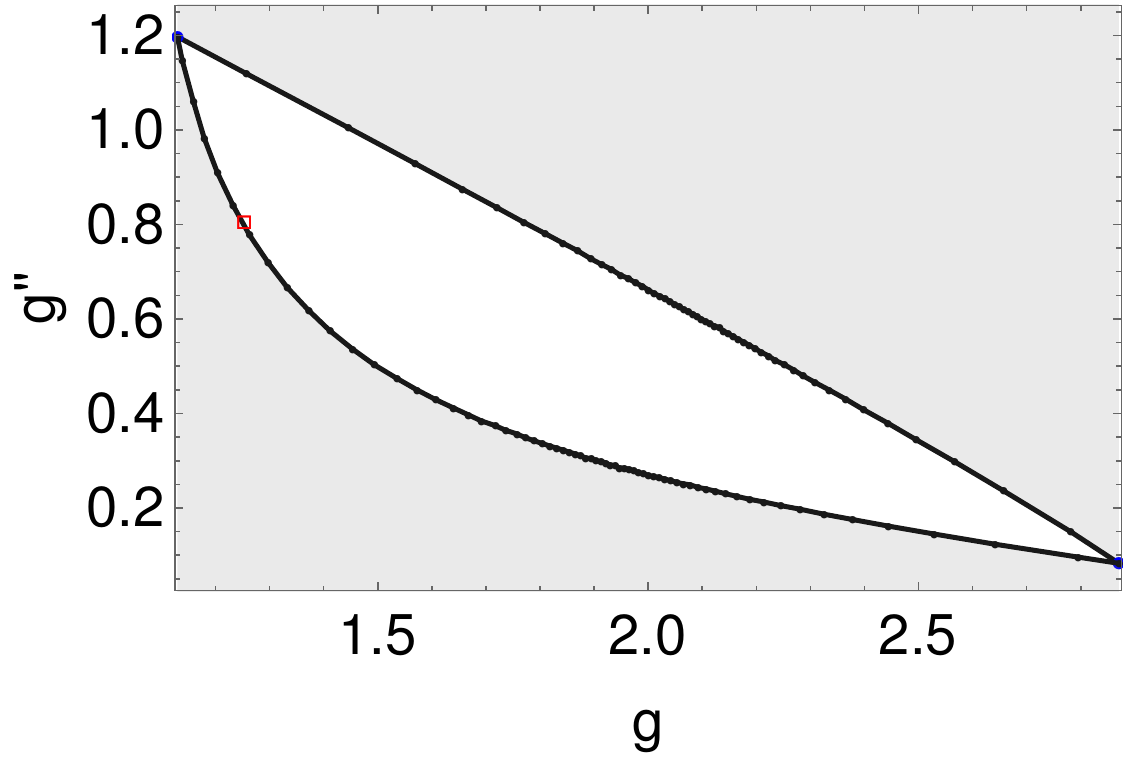}
	\caption{Bounds on the space of values of the correlator and its second derivative at the crossing symmetric point for $\D_\psi=1/10$. GFF/GFB sit on the kinks and the $(2,2)_4$ tricritical Ising point, in red, also saturates the bound. The points shown were obtained upon extrapolation to infinite $\Lambda$.}
	\label{fig:corrmaxdeltaphi01}
\end{figure}

For $\D_{\psi} =1/10$, which corresponds to the $\psi_{(1,2)}$ operator in the $(2,2)_4$ b.c., we find the bound in figure \ref{fig:corrmaxdeltaphi01} after extrapolation in the derivative order $\Lambda$.\footnote{We obtained bounds at finite derivative order $\Lambda=9,13,\dots,33$, which converged rather quickly. In particular the $(2,2)_4$ boundary condition is already very close to saturation even at finite $\Lambda$.
On the other hand, convergence close to the corners is rather slow.}
We find an island with two sharp corners corresponding to the GFB and GFF solutions, as expected. Furthermore, we can compute the four point function of the $\psi_{(1,2)}$ operator using the techniques of appendix \ref{app:examplesmm}, and plot the result. This is the red square which neatly saturates the bound.

\begin{figure}
	\centering
	\includegraphics[width=0.6\linewidth]{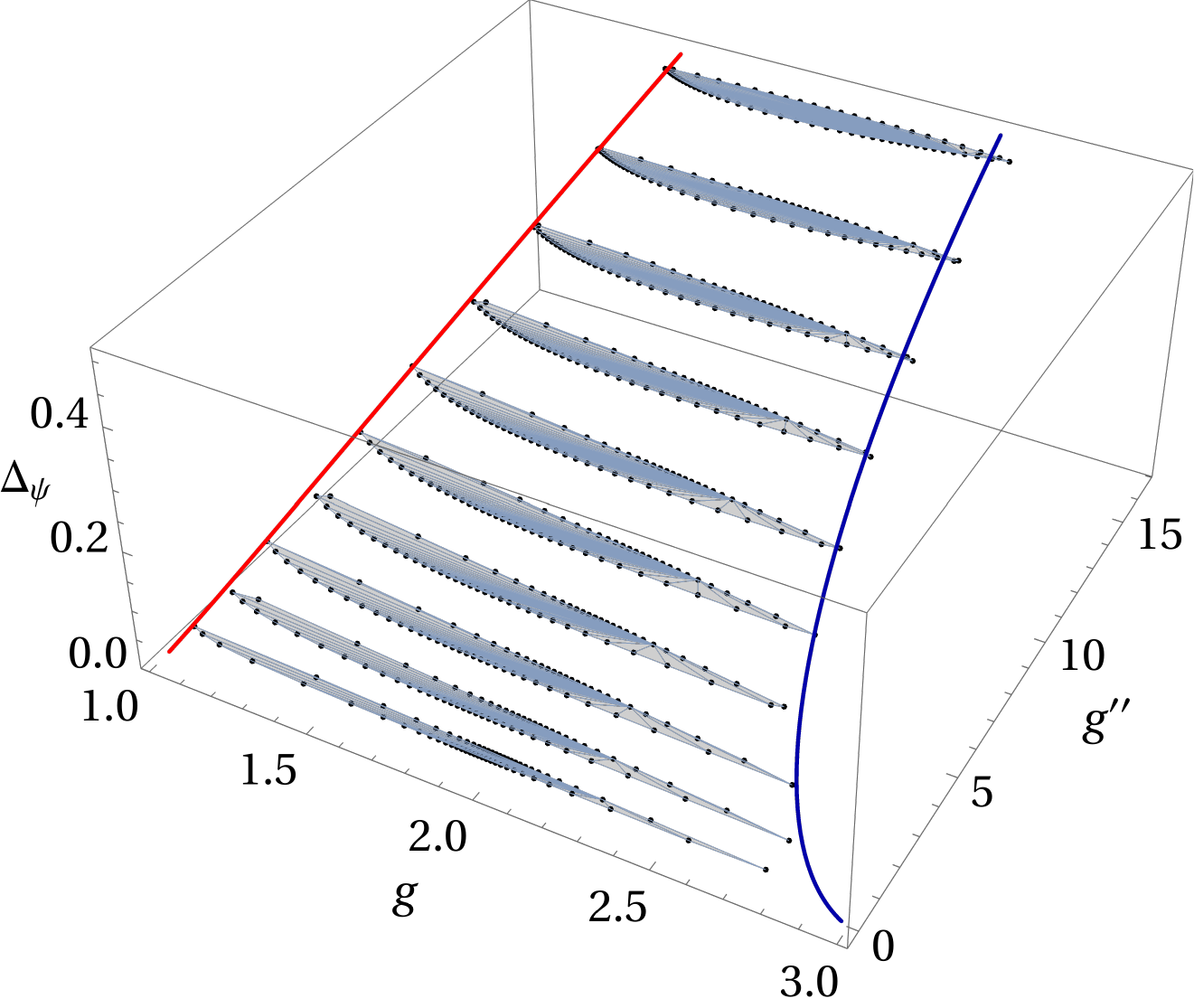}
	\caption{Allowed space of $\mathbb{Z}_2$ symmetric correlators parametrized by $g$ and $g''$ for varying $\D_\psi$. We see a tower of island shaped regions drifting in the $g''$ direction. The red and blue lines respectively correspond to the GFF and GFB families of correlators which not only saturate the bounds, but actually sit in kinks at the endpoints of these islands.}
	\label{fig:CFTboundsvarydeltaphi}
\end{figure}

\begin{figure}[H]
	\centering
	\subfloat{
		\begin{minipage}{0.51\textwidth}
			\includegraphics[width=1\textwidth]{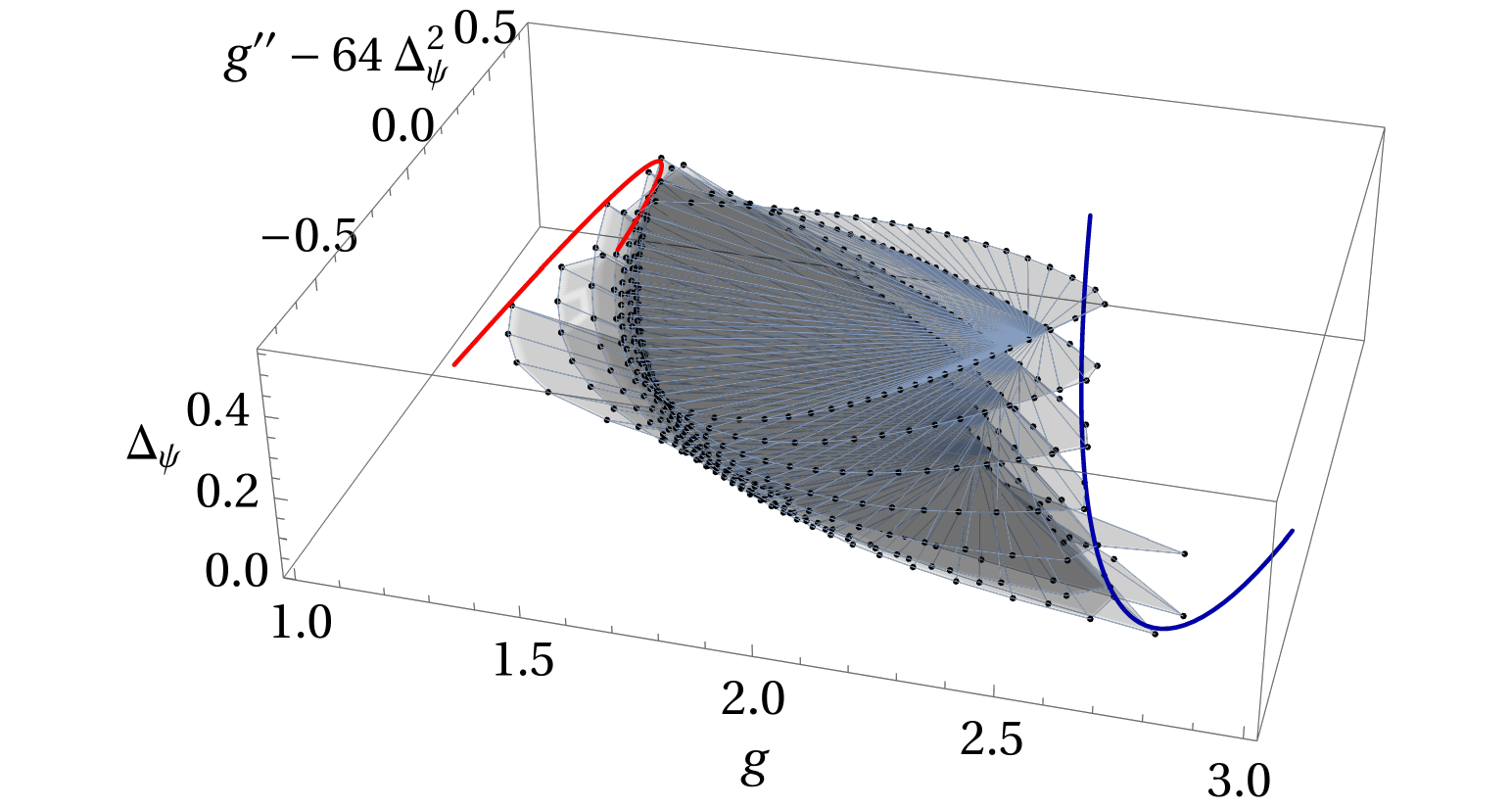} 
	\end{minipage}}
	\hspace{-1cm}
	\subfloat{
		\begin{minipage}{0.51\textwidth}
			\includegraphics[width=1\textwidth]{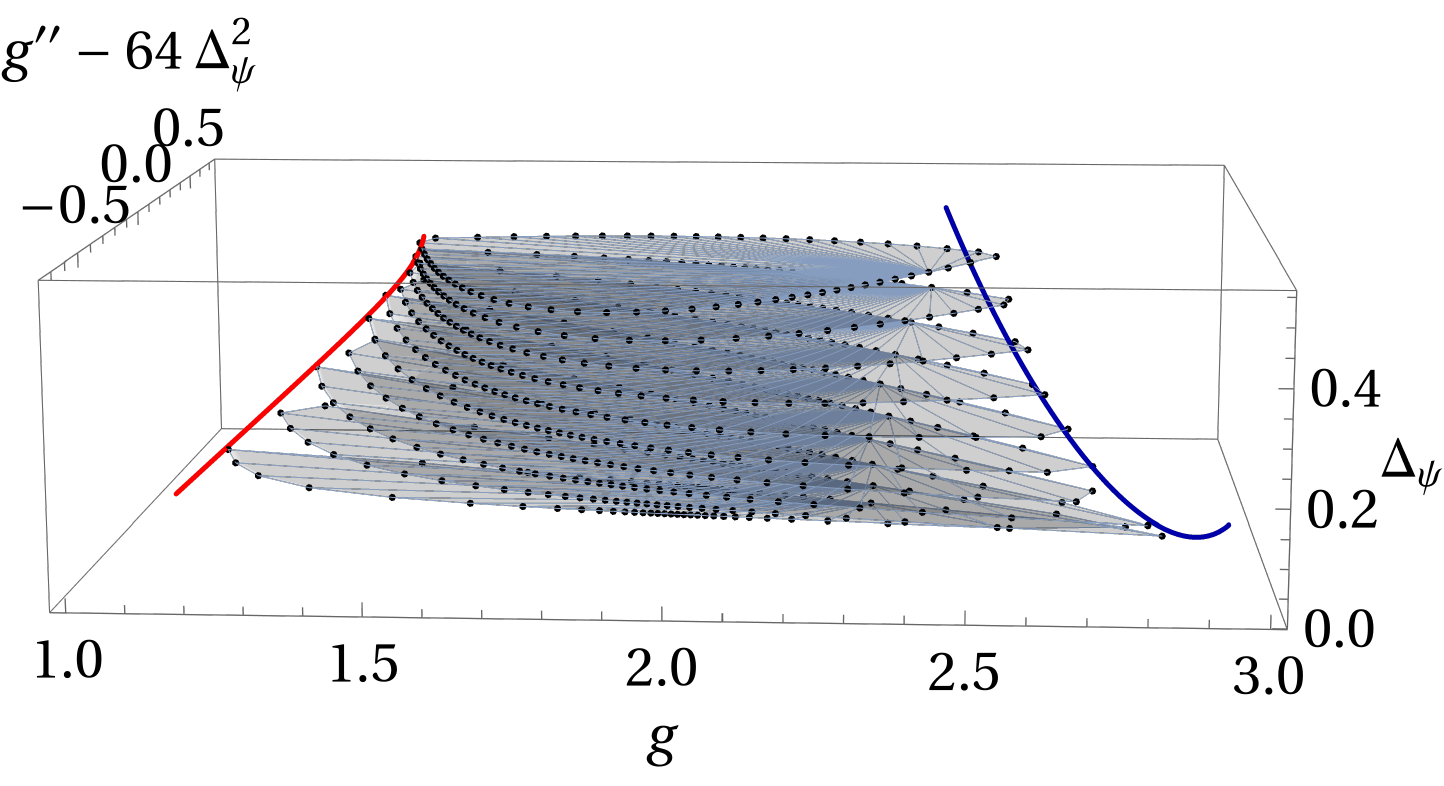} 
	\end{minipage}}~
	\caption{Subtracting the center of mass drift of the previous figure. We now see interesting rotation and stretching patterns of the islands.}
	\label{fig:CFTboundsvarydeltaphisub}
\end{figure}

Computing deformations of these bounds perturbatively is challenging because it involves integrating five-point functions in AdS. In theory it should however be possible to follow the RG flow between the $(2,2)_4$ boundary condition and the $(1,2)_3$ boundary condition by studying the same bounds for different values of $\D_{\psi}.$\footnote{It would also be interesting to understand how these bounds change for higher $m$ minimal models. This presumably sheds some light on the UV of the staircase model.} This leads to a family of islands similar to the one above which we present in figure \ref{fig:CFTboundsvarydeltaphi}.
These islands always have two sharp kinks corresponding to the GFB and GFF solutions and drift in the direction of increasing $g''$.
This drift is related to the "center of mass" of the GFF and GFB solutions. For explicitness we write the values of $\{g,g''\}$ for these solutions:
\begin{align}
	GFF(\D_\psi)&=\left\{2-4^{-\D _{\psi }},2^{3-2 \D _{\psi }}
	\D _{\psi } \left(2 \left(4^{\D _{\psi
		}+1}-1\right) \D _{\psi }+1\right)\right\}\,,\nonumber\\
	GFB(\D_\psi)&=\left\{2+4^{-\D _{\psi }},8 \D _{\psi } \left(8
	\D _{\psi }+4^{-\D _{\psi }} \left(2 \D
	_{\psi }-1\right)\right)\right\}\,,	
\end{align}
whose center of mass is $\{2,64\D_\psi^2\}$. After subtracting the quadratically growing second component we find figure \ref{fig:CFTboundsvarydeltaphisub}.

\begin{figure}[H]
	\centering
	\includegraphics[width=0.5\linewidth]{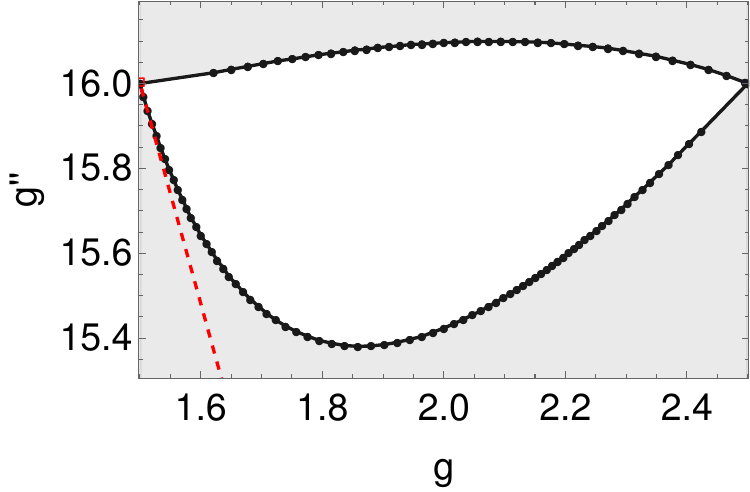}
	\caption{Bounds for $\D_\psi=1/2$. GFF/Ising sits on the kink and the $T\bar{T}$ deformation is tangent. As always, only one sign of the $\TTb$ coupling is consistent with the bounds. The points shown are obtained after extrapolation in $\Lambda$.}
	\label{fig:deltaphi05}
\end{figure}

Finally, we focus on $\D_\psi=1/2$ which contains the $(1,2)_3$ boundary condition. We show the bounds in figure \ref{fig:deltaphi05}, where we see that the Ising theory sits in the kink (red square), since it coincides with GFF. We also plot the $T\bar{T}$ deformation which is tangent to bound, with only one sign being allowed, as we have seen many times by now. In this case, the perturbative calculation is feasible since we can use a fermionic contact Witten diagram to compute the correction to the correlator, as discussed in more detail in appendix \ref{app:Staircase}.

These results suggest that we can track two RG flows that end on the same $(1,2)_3$ BCFT by studying different bootstrap problems. Starting with the $(2,1)_4$ BCFT, we follow gap maximization. Instead, starting with $(2,2)_4$, we can study correlator maximization.

\section{Outlook}
\label{sec:outlook}

We conclude with a discussion on future directions.

A natural extension of our work is to consider $\mathbb{Z}_2$-breaking deformations of minimal model boundary conditions in AdS. The simplest example would be the magnetic deformation of the Ising model, which was studied in AdS with Hamiltonian truncation in~\cite{Hogervorst:2021spa}. More generally, we could combine the thermal and magnetic deformations hence studying Ising field theory \cite{Fonseca:2001dc}. The regime of weak magnetic field with only one stable particle in the infrared is particularly amenable to the bootstrap, since we can parametrize the $\mathbb{Z}_2$ breaking through a self OPE coefficient $\lambda_{\psi\psi\psi}$ of the $\mathbb{Z}_2$-odd field. Another option is to include multiple boundary conditions, separated by boundary condition-changing operators.\footnote{Such `changing operators' received some recent interest in \cite{Zhou:2023fqu}, where the authors studied one-dimensional defects in a higher $d$-dimensional bulk. It would be interesting to apply numerical bootstrap techniques to the corresponding $\AdStwo \times S^{d-2}$ setup.}

One limitation of our setup is that we cannot impose locality of the bulk theory along the RG flow. Our strategy in this work is to look for bootstrap bounds which are saturated at the fixed points of the RG flow when the bulk theory is Weyl-equivalent to a local BCFT, but we are not guaranteed saturation throughout the flow. In order to make progress on this, a promising strategy recently discussed in~\cite{Levine:2023ywq,meineri2023renormalization} is to include bulk observables into the bootstrap. These authors considered sum rules on the three-point functions between one bulk operator and two boundary operators (called AdS 2-particle form factors) which capture bulk locality. In \cite{meineri2023renormalization} sum rules stemming from the two-point function of the bulk stress-energy tensor were also found, allowing the formulation of a positive semi-definite system of correlators involving bulk and boundary data. This extended the formalism of \cite{Karateev:2019ymz,Correia:2022dyp} to AdS, where bounds as a function of the UV central charge were obtained. It would be very interesting to see to what extent these additional constraints would improve the bounds obtained in this paper.

Our strategy works well for computing certain non-perturbative numerical bounds on irrelevant couplings of generic CFTs. There has been a lot of progress in the context of EFT corrections for free bosons and fermions in AdS~\cite{Caron-Huot:2021enk,Knop:2022viy}, and our approach could be used in order to search for such constraints in interacting CFTs. As an application, we have found a constraint on the sign of the $\TTb$ deformation. For other types of deformations, for instance those of section~\ref{sec:reviewDD2}, we found no sign constraints along our one-loop perturbation theory, and higher-order corrections might be useful to detect further bulk inconsistencies.

Finally, it would be interesting to understand if integrability, which plays a key role in the solution of minimal model RG flows both in flat space and on the upper half-plane, survives in AdS. To understand this it would be necessary to follow the behavior of bulk higher-spin currents which exist at the fixed point, due to Virasoro symmetry. Another notion of solvability for conformal correlators is extremality, i.e. saturation of the bootstrap bounds, which is known to lead to sparse spectra. While the relation between these concepts is well understood in the flat-space limit, it is still an open problem to understand this connection at finite AdS radius.

\section*{Acknowledgements }

We would like to thank C. Behan, C. Bercini, M. Bill\`o, A. Cavagli\`a, M. Costa, L. Di Pietro, A. Gimenez-Grau, M. Hogervorst, A. Kaviraj, P. Liendo, A. Manenti, M. Meineri, M. Milam, M. Paulos, J. Penedones, V. Schomerus, A. Tilloy, E. Trevisani, and P. van Vliet for discussions. We also thank V. Fofana and Y. Fitamant for assistance in the use of the Cholesky cluster at the Ecole Polytechnique. AA thanks the University of Turin, where part of this work was presented, for hospitality. AA received funding from the German Research Foundation DFG under Germany’s Excellence Strategy – EXC 2121 Quantum Universe – 390833306.  Centro  de  F\'isica  do  Porto  is  partially  funded  by  Funda\c{c}\~ao  para  a  Ci\^encia  e  Tecnologia (FCT) under the grant UID04650-FCUP. BvR is supported by the Simons Foundation grant $\#$488659 (Simons Collaboration on the non-perturbative bootstrap). This
project is funded by the European Union: for EL by ERC “QFT.zip” with Project ID 101040260
(held by A. Tilloy) and for BvR by ERC “QFTinAdS” with Project ID 101087025. Views and opinions expressed are however those of the author(s) only and do not necessarily reflect those of the European Union or the European Research Council Executive Agency. Neither the European Union nor the granting authority can be held responsible for them.

\appendix

\section{Conventions}
\label{app:conventions}

\begin{table}[h]
	\begin{center}
		$\begin{array}{|c|c|c|l|}
			\hline
			\hline
			m &(a_1,a_2)_m&  \mathbb{Z}_2 \text{ - preserving}& \text{Boundary spectrum} \\
			\hline
			\hline
			3 &(1,2)_3 &\checkmark& \hid, \psi_{(1,3)} \\
			\hline
			4 &(1,2)_4 && \hid, \psi_{(1,3)}   \\
			&(1,3)_4 && \hid, \psi_{(1,3)}   \\
			&(2,1)_4 &\checkmark& \hid, \psi_{(3,1)}   \\
			&(2,2)_4 &\checkmark& \hid, \psi_{(1,3)}, \psi_{(3,1)}, \psi_{(3,3)}   \\
			\hline
			5 &(1,2)_5 && \hid, \psi_{(1,3)}   \\
			&(1,3)_5 &\checkmark& \hid, \psi_{(1,3)}, \psi_{(1,5)}  \\
			&(1,4)_5 && \hid, \psi_{(1,3)}   \\
			&(2,1)_5 && \hid, \psi_{(3,1)}   \\
			&(2,2)_5 && \hid, \psi_{(1,3)}, \psi_{(3,1)}, \psi_{(3,3)}   \\
			&(2,3)_5 &\checkmark& \hid, \psi_{(1,3)}, \psi_{(1,5)}, \psi_{(3,1)}, \psi_{(3,3)}, \psi_{(3,5)}  \\
			&(2,4)_5 && \hid, \psi_{(1,3)}, \psi_{(3,1)}, \psi_{(3,3)}   \\
			&(2,5)_5 && \hid, \psi_{(3,1)}   \\
			\hline
			\hline
		\end{array}$
	\end{center}
	\caption{Elementary conformal boundary conditions for diagonal and unitary minimal models with $m \leq 5$. We dropped the conformal b.c. labeled by the identity (and its $\mathbb{Z}_2$-conjugate $(1, m)_m$): they are always possible, and allow only for $\hid$ at the boundary.}
	\label{tbl:smallconfbc}
\end{table}

\subsection{OPEs and basic correlation functions}\label{app:OPEcorr}
Consider a generic BCFT on the upper half-plane. Here $\phi_i(z,\bar{z})$ denotes a scalar bulk global primary with dimension $\D_i$, and $\psi_i(x)$ denotes a scalar global boundary primary with dimension $\hD_i$. Unless otherwise specified, primary bulk and boundary operators are taken to be unit-normalized. The bulk and boundary identity operators satisfy: $\langle \id \rangle =\langle \hid \rangle =1$.

The bulk-bulk, bulk-boundary, and boundary-boundary OPEs are
\begin{align}\label{OPEs}
	\phi_i(z_1,\bar{z}_1)\phi_j(z_2,\bar{z}_2)
	&=\sum_{k}C_{ij}{}^k\phi_k(z_2,\bar{z}_2)|z_1-z_2|%
	^{\D_k-\D_i-\D_j}+\cdots\,,\nonumber\\
	\phi_i(x+iy,x-iy)&=\sum_k B_i{}^k\psi_k(x)(2y)^{\hD_k-\D_i}
	+\cdots\,,\nonumber\\
	\psi_i(x_1)\psi_j(x_2)&=\sum_k\bdc_{ij}{}^k\psi_k(x_2)
	(x_1-x_2)^{\hD_k-\hD_i-\hD_j} + \cdots \qquad(x_1>x_2)\,.
\end{align}
The ellipsis in the first (second and third) line above denotes $SL(2,\mathbb{C})$ ($SL(2,\mathbb{R})$) descendants. 
Indices are raised and lowered using the Zamolodchikov metric. 

The simplest correlation functions on the upper half-plane are:
\begin{align}\label{correlatorsuhpCFTconventions}
	\langle \psi(x_1)\psi(x_2)\rangle_{\uhp}&=\frac{1}{(x_{12}^2)^{\hD}}\,,\qquad\qquad\qquad\qquad\quad(x_1>x_2)~\nonumber\\
	\langle\phi(x+iy,x-iy)\rangle_{\uhp}&=\frac{B_\phi}{(2y)^{\D}}\,,\qquad\qquad\qquad\qquad\quad\quad(y>0)~\nonumber\\ 
	\langle\phi(x_1+iy_1,x_1-iy_1)\psi(x_2)\rangle_{\uhp} &=
	\frac{B_{\phi \psi}}{
		(2y_1)^{\D-\hD}(x_{12}^2+y_1^2)^{\hD}}\,,~\qquad\quad(y_1>0)~\nonumber\\ 
	\langle\psi_i(x_1)\psi_j(x_2)\psi_k(x_3)\rangle_{\uhp} &=\frac{\bdc_{ijk}}{
		(x_{12})^{\hD_{ijk}} (x_{23})^{\hD_{jki}}
		(x_{13})^{\hD_{ikj}}}\,,\quad(x_{i}>x_{i+1})\,.
\end{align}
with $x_{ij}\equiv x_i-x_j$ and $\hD_{ijk}\equiv \hD_i+\hD_j-\hD_k$.
In a generic BCFT, the coefficients $B$, $\bdc$ and $C$ are determined via the `sewing' constraints of Lewellen~\cite{Lewellen:1991tb}. For unitary and diagonal minimal models with elementary conformal boundary conditions, the solution to these constraints appeared in \cite{Runkel:1998he} (see \cite{Runkel:1999dz} for the extension to the D-series of minimal models), where $\bdc$ and $B$ are written in terms of F-matrices (the one-point function of $\phi_{(r,s)}$ is determined by the Cardy state, as written in \eqref{bulkboundaryidentity}).
The $C$ coefficients are the same of the homogeneous minimal model, and they can be computed via the `Coulomb gas formalism' of refs.~\cite{Dotsenko:1984nm,Dotsenko:1984ad,Dotsenko:1985hi} (see also the Mathematica notebook attached to the submission of~\cite{Esterlis:2016psv}, where many Coulomb gas formulae are implemented.)

\subsection{Global conformal blocks on the upper half-plane}\label{ss:blockexpansions}
Next, we discuss the four-point correlation function between four boundary conformal primaries $\psi_i$ (not necessarily Virasoro primaries) with scaling dimensions $\hD_i$ in a generic 2d BCFT. By $SL(2,\mathbb{R})$ symmetry we have
\begin{align}
	\langle \psi_1 (x_1)\psi_2 (x_2)\psi_3 (x_3)\psi_4 (x_4)\rangle_{\uhp} =\left(\frac{x_{14}}{x_{24}}\right)^{\hD_{21}} \left(\frac{x_{14}}{x_{13}}\right)^{\hD_{34}} \frac{\mathcal{G}^{1234}(\eta)}{(x_{12})^{\hD_{1}+\hD_{2}} (x_{34})^{\hD_{3}+\hD_{4}}}\,,\quad x_{i}>x_{i+1}\,,
\end{align}
with $\hD_{ij}\equiv \hD_i-\hD_j$ and the cross-ratio $\eta$ is defined as in eq.~\eqref{eta_def_line} and repeated here for convenience
\begin{align}
	\eta= \frac{x_{12}x_{34}}{x_{13}x_{24}}\,,\quad 0<\eta<1\,.
\end{align}
We have the following s-channel expansion
\begin{align}\label{schanelblocksdec}
	\mathcal{G}^{1234}(\eta) = \sum_{k} \bdc_{12}{}^k \bdc_{34k}G(\hD_{21},\hD_{34},\hD_k,\eta)\,,
\end{align}
with global blocks given by~\cite{Dolan:2011dv}
\begin{align}\label{schanelblocksgen}
	G(a,b,\D,\eta)= \eta^{\D } \, _2F_1(a+\D ,b+\D ;2 \D ;\eta)\,.
\end{align}
We will sometimes use the following notation
\begin{align}
	G_\D(\eta)\equiv G(0,0,\D,\eta)\,.
\end{align}

\section{Correlation functions for generalized free theories}
\label{app:GFFGFB}

\subsection{Generalized free fermion}
Consider a 1d generalized free fermion $\psi$ of scaling dimension $\D$. We take $\psi$ to be unit normalized. By Wick's theorem the four-point correlation function is
\begin{align}
	\mathcal{G}^{\psi\psi\psi\psi}(\eta)=1-\eta^{2\D}+\left(\frac{\eta}{1-\eta}\right)^{2\D}\,.
\end{align}
The cross-ration $\eta$ is defined as in eq.~\eqref{eta_def_line}.
The global conformal blocks expansion reads~\cite{Gaiotto:2013nva}
\begin{align}
	\mathcal{G}^{\psi\psi\psi\psi}(\eta) =1+\sum_{n=0}^{\infty}\frac{2(2\D)^2_{2n+1}}{(2n+1)!(4\D+2n)_{2n+1}}G_{2\D+2n+1}(\eta)\,,
\end{align}
with global blocks given in~\eqref{schanelblocksgen}.
We want to compute mixed four-point correlation functions between $\psi$ and the leading primary in the $\psi\times\psi$ OPE, i.e.
\begin{align}
	\psi(x)\psi(0) = \frac{\hid}{x^{2\D}}+x\, (\psi \partial \psi)(0)+\dots\,.
\end{align}
From the four-point function above we can easily obtain
\begin{align}
	\langle \psi\partial\psi(x)\psi\partial\psi(0)\rangle = \frac{2\D}{x^{4\D+2}}\,.
\end{align}
Using Wick's theorem, from the eight-point correlation function of $\psi$ we can obtain the four-point correlation function of $\psi\partial\psi$
\begin{align}\label{psidpsi2pt}
	\langle \psi\partial\psi(x_1)\psi\partial\psi(x_2)\psi\partial\psi (x_3)\psi\partial\psi(x_4)\rangle\,.
\end{align}
When $\D=1/2$ we get
\begin{align}
	\mathcal{G}^{[\psi\partial\psi][\psi\partial\psi][\psi\partial\psi][\psi\partial\psi]}(\eta)=1+\frac{\eta ^2 \left(\eta ^6-4 \eta ^5+22 \eta ^4-52 \eta ^3+66 \eta ^2-48 \eta +16\right)}{(\eta -1)^4}\,.
\end{align}
The r.h.s. above can be expanded into s-channel conformal blocks of eq.~\eqref{schanelblocksgen}  as follows
\begin{align}
	\mathcal{G}^{[\psi\partial\psi][\psi\partial\psi][\psi\partial\psi][\psi\partial\psi]}(\eta)& = 1 +16G_{2}(\eta) +\frac{98}{5}G_{4}(\eta)+\frac{512}{63}G_{6}(\eta)+\frac{1270}{429}G_{8}(\eta)+\dots\,.
\end{align}
Hence the first parity-even primary operator after the identity is the displacement operator $\Disp$, and after it there is $\Disp^2$.

Consider now the following mixed four-point function
\begin{align}
	\langle \psi(x_1)\psi(x_2)\psi\partial\psi (x_3)\psi\partial\psi(x_4)\rangle \,.
\end{align}
When $\D = 1/2$ we find
\begin{align}
	\mathcal{G}^{\psi\psi[\psi\partial\psi][\psi\partial\psi]}(\eta)= 1+\frac{(\eta -2)^2 \eta ^2}{(\eta -1)^2}\,.
\end{align}
In the s-channel we find the following decomposition
\begin{align}
	\mathcal{G}^{\psi\psi[\psi\partial\psi][\psi\partial\psi]}(\eta) &= 1+4 G_{2}(\eta)+\frac{7}{5} G_{4}(\eta)+\frac{16}{63} G_{6}(\eta)+\frac{29}{858} G_{8}(\eta)+\frac{46G_{10}(\eta)}{12155}+\dots\,.
\end{align}
The first parity-even primary operator after the identity is $\Disp$, and after it there is $\Disp^2$. 
For generic $\D$ we find
\begin{align}
	\mathcal{G}^{\psi\psi[\psi\partial\psi][\psi\partial\psi]}(\eta) = 2 \D \left(1+(2 \D (\eta -1)-1) \eta ^{2 \D}+(1-\eta)^{-1}{(2 \D-\eta +1) \left(\frac{\eta }{1-\eta }\right)^{2 \D}}\right)\,,
\end{align}
with the following  s-channel decomposition
\begin{align}
	\mathcal{G}&^{\psi\psi[\psi\partial\psi][\psi\partial\psi]}(\eta)= 2 \D +4 (2 \D +3) \D ^2 G_{2\D+1}(\eta)\nonumber\\
	&+\frac{4 (\D +1) (\D +3) (2 \D +1)^2 \D ^2 }{12 \D +9}G_{2\D+3}(\eta)+\nonumber\\
	&+\frac{(\D +1) (\D +2) (2 \D +1)^2 (2 \D +3) (\D  (2 \D +11)+10) \D ^2 }{15 (4 \D +5) (4 \D +7)}G_{2\D+5}(\eta)\nonumber\\
	&+\frac{(\D +1)^2 (\D +2) (\D +3) (2 \D +1)^2 (2 \D +3) (2 \D +5) (\D  (2 \D +15)+21) \D ^2 }{315 (4 \D +7) (4 \D +9) (4 \D +11)}G_{2\D+7}(\eta)\nonumber\\
	&+\dots\,.
\end{align}
In order to investigate on parity-odd operators we consider the s-channel decomposition of
\begin{align}
	\langle \psi(x_1)\psi\partial\psi(x_2)\psi(x_3)\psi\partial\psi(x_4)\rangle \,.
\end{align}
When $\D =1/2$ we find
\begin{align}
	\mathcal{G}^{\psi[\psi\partial\psi]\psi[\psi\partial\psi]}(\eta) = 	\frac{\sqrt{\eta } \left(\eta ^4-2 \eta ^3+5 \eta ^2-4 \eta +1\right)}{(\eta -1)^2}\,,
\end{align}
whose s-channel decomposition gives
\begin{align}\label{GFFparityodd}
	\mathcal{G}^{\psi[\psi\partial\psi]\psi[\psi\partial\psi]}(\eta) &= G\left(\frac{3}{2},-\frac{3}{2},\frac{1}{2},\eta \right)+G\left(\frac{3}{2},-\frac{3}{2},\frac{9}{2},\eta \right)\nonumber\\
	&+\frac{1}{5} G\left(\frac{3}{2},-\frac{3}{2},\frac{13}{2},\eta \right)-\frac{4}{429} G\left(\frac{3}{2},-\frac{3}{2},\frac{15}{2},\eta \right)\nonumber\\
	&+\dots\,.
\end{align}
Squared OPE coefficients associated with parity-odd, global primary boundary exchanges appear with a minus sign, see e.g. appendix K of~\cite{Homrich:2019cbt}. 
Hence, after $\psi$ itself we have a parity-even operator of dimension $3\D + 3$ and a parity-odd operator of dimension $3\D + 6$. 
For generic $\D$ we find
\begin{align}
	\mathcal{G}^{\psi[\psi\partial\psi]\psi[\psi\partial\psi]}(\eta) = -2 \D  \left((2 \D  (\eta -1)+\eta ) \eta ^{\D }-\eta ^{3 \D +1}- \left(\frac{1}{\eta -1}\right)^{2 \D +1} (2 \D  \eta +\eta -1) \right)\,,
\end{align}
with the following  s-channel decomposition ($e^{2i \pi \D} = -1$)
\begin{align}
	\mathcal{G}^{\psi[\psi\partial\psi]\psi[\psi\partial\psi]}(\eta)&= 4 \D^2 G\left(\D _{21},\D _{34},\D,\eta \right)+2 (2 \D+1) \D^2 G\left(\D _{21},\D _{34},3 \D+3,\eta \right)\nonumber\\
	&+\frac{\D^2 (\D+1) (2 \D+1)^2 (2 \D+3) }{3 (6 \D+7)}G\left(\D _{21},\D _{34},3 \D+5,\eta \right)\nonumber\\
	&-\frac{2 \D^3 (\D+1) (\D+2) (2 \D+1)^2 (2 \D+3) }{45 (3 \D+4) (3 \D+5)}G\left(\D _{21},\D _{34},3 \D+6,\eta \right)\nonumber\\
	&+\dots\,.
\end{align}

\subsection{Generalized free boson}
Consider a 1d generalized free boson $\phi$ of scaling dimension $\D$. We take $\phi$ to be unit normalized. By Wick's theorem the four-point correlation function is
\begin{align}
	\mathcal{G}^{\phi\phi\phi\phi}(\eta)=1+\eta^{2\D}+\left(\frac{\eta}{1-\eta}\right)^{2\D}\,.
\end{align}
The cross-ration $\eta$ is defined as in eq.~\eqref{eta_def_line}. The global conformal blocks expansion reads~\cite{Gaiotto:2013nva}
\begin{align}
	\mathcal{G}^{\phi\phi\phi\phi}(\eta) =1+\sum_{n=0}^{\infty}\frac{2(2\D)^2_{2n}}{(2n)!(4\D+2n-1)_{2n}}G_{2\D+2n}(\eta)\,.
\end{align}
Next, we compute
\begin{align}
	\langle \phi^2(x_1)\phi^2(x_2)\phi^2(x_3)\phi^2(x_4)\rangle\,,
\end{align}
to find
\begin{align}
	\mathcal{G}^{\phi^2\phi^2\phi^2\phi^2}(\eta)=
	4 \left(1+4 \eta ^{2 \D }+\left(4 (1-\eta )^{-2 \D }+1\right) \eta ^{4 \D }+4 \left(\frac{\eta }{1-\eta}\right)^{2 \D }+\left(\frac{\eta }{1-\eta}\right)^{4 \D }\right)\,.
\end{align}
The r.h.s. above has the following s-channel decomposition
\begin{align}
	\mathcal{G}^{\phi^2\phi^2\phi^2\phi^2}(\eta)&= 4  +32 G_{2\D}(\eta)+24 G_{4\D}(\eta)+\frac{32 (2 \D +1) \D ^2}{4 \D +1}G_{2\D+2}(\eta)\nonumber\\
	&+\frac{64 \D ^2 (2 \D +1)}{8 \D +1} G_{4\D+2}(\eta)+\frac{8 \D ^2 (\D +1) (2 \D +1)^2 (2 \D +3)}{3 (4 \D +3) (4 \D +5)}G_{2\D+4}(\eta)\nonumber\\
	&+\frac{16 \D ^2 (2 \D +1) (\D  (8 \D  (4 \D +5)+17)+3)}{3 (8 \D +3) (8 \D +5)}G_{4\D+4}(\eta)\nonumber\\
	&+\dots\,.
\end{align}
When $\D=1$ the first parity-even primary operator after the identity is the displacement operator. Note that there are two dimension-four operators:  $\phi^4$ and $\phi \partial^2 \phi$. Consider now
\begin{align}
	\langle \phi(x_1)\phi(x_2)\phi^2(x_3)\phi^2(x_4)\rangle \,.
\end{align}
Using Wick's theorem we find
\begin{align}
	\mathcal{G}^{\phi\phi\phi^2\phi^2}(\eta)=2+4 \eta ^{2 \D }+4 \left(\frac{\eta }{1-\eta}\right)^{2 \D }\,.
\end{align}
The s-channel decomposition reads
\begin{align}
	\mathcal{G}^{\phi\phi\phi^2\phi^2}(\eta)&= 2 +8 G_{2\D}(\eta)+\frac{8 (2 \D +1) \D ^2}{4 \D +1}G_{2\D+2}(\eta)\nonumber\\
	&+\frac{2 \D ^2 (\D +1) (2 \D +1)^2 (2 \D +3)}{3 (4 \D +3) (4 \D +5)}G_{2\D+4}(\eta)+\dots\,.
\end{align}
When $\D=1$ the first parity-even primary operator after the identity is the displacement operator, and after it there is $\phi \partial^2 \phi$. In order to uncover the spectrum of parity-odd operators we consider
\begin{align}
	\langle \phi(x_1)\phi^2(x_2)\phi(x_3)\phi^2(x_4)\rangle \,.
\end{align}
Using Wick's theorem we find
\begin{align}
	\mathcal{G}^{\phi\phi^2\phi\phi^2}(\eta)=2 \eta ^{\D } \left(2+2 \left(1-\frac{1}{\eta }\right)^{-2 \D }+\eta ^{2 \D }\right)\,.
\end{align}
The s-channel decomposition reads ($e^{-2\pi i \D}=1$)
\begin{align}
	\mathcal{G}^{\phi\phi^2\phi\phi^2}(\eta) &=4 G(\D ,-\D ,\D ,\eta )+6 G(\D ,-\D ,3 \D ,\eta )+\frac{8 \D ^2 (2 \D +1)}{6 \D +1}G(\D ,-\D ,3 \D +2,\eta )\nonumber\\
	&-\frac{8 \D ^3 (\D +1) (2 \D +1)}{27 \D  (\D +1)+6}G(\D ,-\D ,3 \D +3,\eta )+\dots\,.
\end{align}
When $\D=1$ the first parity-even primary operator after $\phi$ itself is $\phi^3$, while the first parity-odd primary is constructed from three derivatives acting on $\phi^3$.

\section{Parity-odd channel in correlators with the displacement}\label{app:displcorrelatorsDecn}

In this appendix we discuss the spectrum of parity-odd primaries in the $\Disp \times \psi$ OPE, being $\psi$ a Virasoro primary of scaling dimension $\hD$. To this end we consider the conformal block expansion of the correlator
\begin{align}\label{DphiDphiuhp}
	\langle \Disp(x_1) \psi(x_2) \Disp(x_3) {\psi}&(x_4)\rangle_{\uhp} =\left(\frac{x_{14}}{x_{24}}\right)^{\hD-2} \left(\frac{x_{14}}{x_{13}}\right)^{2-\hD} \frac{\mathcal{G}^{\Disp\psi\Disp\psi}(\eta)}{(x_{12})^{2+\hD} (x_{34})^{2+\hD}} \,,\quad x_i > x_{i+1}\nonumber\\
	&\mathcal{G}^{\Disp\psi\Disp\psi}(\eta)=\eta ^{\hD+2} \left(\frac{c}{2}+ \frac{\hD (2 (\eta -1) \eta +\hD)}{\eta^2(\eta-1)^2}\right)\,,
\end{align}
which was computed in appendix A.3 of \cite{Lauria:2023uca}. For unit-normalized operators and generic $\hD$, the s-channel block decomposition reads:
\begin{align}\label{schanelblocksdecDhphiDhphi}
	\frac{\mathcal{G}^{\Disp\psi\Disp\psi}(\eta)}{c/2} = \sum_{n=0,2,3,4,\dots} \bdc_{\Disp\psi}{}^n {\bdc}_{\Disp\psi n}G(\hD-2,2-\hD,\hD+n,\eta)\,,
\end{align}
where the first non-zero coefficients read
\begin{align}
	\bdc_{\Disp\psi}{}^0 \bdc_{\Disp\psi 0} &= \frac{2\hD^2}{c}\,,\nonumber\\
	\bdc_{\Disp\psi}{}^2 \bdc_{\Disp\psi 2} &= 1+\frac{2\hD}{c} \left(4-\frac{9}{2 \hD+1}\right)\,,\nonumber\\
	\bdc_{\Disp\psi}{}^3 \bdc_{\Disp\psi 3} &=-\frac{4\hD}{c(\hD+1) (\hD+2)}\left((c-7) \hD+c+3 \hD^2+2\right) \,,\nonumber\\
	\bdc_{\Disp\psi}{}^4 \bdc_{\Disp\psi 4} &= \frac{2\hD (5 c (4 \hD (\hD+2)+3)+4 \hD (\hD (8 \hD-19)+26)-15)}{c(\hD+3) (2 \hD+3) (2 \hD+5)}\,.
\end{align}
Hence, for generic $\hD$, the first parity-even global primary (after $\psi$ itself) appears at level 2, while the first parity-odd appears at level 3. For $c=1/2$ when $\hD =1/2$ (i.e. $\psi_{(1,3)}$ in the Ising model with $(1,2)_3$ boundary condition) the first few coefficients vanish and the first parity-even global primary after $\psi_{(1,3)}$ appears at level four, while the first parity-odd appears at level seven, correspondingly
\begin{align}
	\bdc_{\Disp\psi}{}^4 \bdc_{\Disp\psi4}=1\,,\quad \bdc_{\Disp\psi}{}^7 \bdc_{\Disp\psi 7}=-\frac{4}{429}\,,\quad \psi=\psi_{(1,3)}\,.
\end{align}
For $c=7/10$, leading parity-even (odd) quasi-primaries have
\begin{align}
	\bdc_{\Disp\psi}{}^2 \bdc_{\Disp\psi 2}&=\frac{17}{2}\,,		& \bdc_{\Disp\psi}{}^5 \bdc_{\Disp\psi5}&=-\frac{1360}{1617}\,,		&\psi&=\psi_{(3,1)}\,,\nonumber\\
	\ \bdc_{\Disp\psi}{}^4 \bdc_{\Disp\psi 4}&=\frac{39}{868}\,,		& \bdc_{\Disp\psi}{}^3 \bdc_{\Disp\psi 3}&=-\frac{40}{77}\,,	&\psi&=\psi_{(3,3)}\,,\nonumber\\
	\bdc_{\Disp\psi}{}^2 \bdc_{\Disp\psi 2}&=\frac{65}{77}\,,	& \bdc_{\Disp\psi}{}^5 \bdc_{\Disp\psi5}&=-\frac{80}{897}\,, &\psi&=\psi_{(1,3)}\,.
\end{align}

\section{Correlators in minimal model boundary conditions}
\label{app:examplesmm}

In this section we discuss some correlation functions in minimal model boundary conditions ${\bf a} = (a_1,a_2)_m$. Computing correlation functions with bulk Virasoro primaries is a standard application of the method of images~\cite{Cardy:1984bb} (see also Chapter 11.2 of~\cite{DiFrancesco:1997nk}). Boundary Virasoro primaries behave as holomorphic Virasoro primaries as far as the Ward identities are concerned, hence for their correlation functions we will not need to employ the method of images.

\subsection{Bulk two-point function of \texorpdfstring{$\phi_{(1,2)}$}{phi12}}\label{app:phi122pt}
We start with the bulk two-point functions of $\phi_{(1,2)}$ on the upper half-plane:
\begin{align}
	\langle \phi_{(1,2)}(x_1+i y_1,x_1-i y_1) \phi_{(1,2)}(x_2+i y_2,x_2-i y_2)\rangle_{\uhp}\,.
\end{align}
In order to compute this correlator we employ the method of images and consider the differential equation satisfied by the four-point function of $\phi_{(1,2)}(z)$ in the homogeneous theory, i.e:
\begin{align}\label{12diffeq}
	\left({\cal L}_{-2}^{(z_4)}-\frac{3}{2(2h_{1,2}+1)}{\cal L}_{-1}^2\right)	\langle \phi_{(1,2)}(z_1)\phi_{(1,2)}(z_2)\phi_{(1,2)}(z_3)\phi_{(1,2)}(z_4)\rangle=0\,,
\end{align}
where $h_{1,2}$ is given in eq.~\eqref{VirasoroPrimariesmain}, ${\cal L}_{-1}=\partial_{z_4}$, and ${\cal L}_{-2}^{(\cdot)}$ is the following differential operator:
\begin{align}\label{calLdiffopLn}
	\mathcal{L}_{-k}^{(w)}\equiv\sum_{i=1}^{3}\left(\frac{ (k-1)h_i}{(z_i-w)^k}-\frac{1}{(z_i-w)^{k-1}}\partial_i\right)\,.
\end{align}
By $SL(2,\mathbb{R})$ symmetry, the holomorphic correlator takes the following form
\begin{align}
	\langle \phi_{(1,2)}(z_1)\phi_{(1,2)}(z_2)\phi_{(1,2)}(z_3)\phi_{(1,2)}(z_4)\rangle=\frac{	\tcG(\tieta)}{(z_{12} z_{34})^{2h_{1,2}}}\,.
\end{align}
The cross-ratio $\tieta$ is
\begin{align}\label{tietadef1234}
	\tieta= \frac{\eta^2}{1-\eta}\,,\quad \eta=\frac{z_{12}z_{34}}{z_{13}z_{24}}\,.
\end{align}
It is not difficult to solve the differential equation~\eqref{12diffeq} for generic $m$ and $\bf a$. The particular solution is obtained by imposing bulk-boundary crossing symmetry, upon setting $z_2=z_1^*$, $z_4=z_3^*$ (being $z_1 = x_1+ i y_1$ and $z_3 = x_2+ i y_2$). For the case at hand, the two Virasoro blocks that correspond to the exchange of $\id$ and $\phi_{(1,3)}$ in the bulk are
\begin{align}\label{phi12twoptblocksbulk}
	V^{\text{bulk}}_{(1,1)}(\tieta)&=\tieta^{2h_{1,2}}\, _2F_1\left(\frac{2-m}{2 m+2},\frac{1}{2 m+2};\frac{2}{m+1};-\frac{4}{\tieta}\right) \,,\nonumber\\
	V^{\text{bulk}}_{(1,3)}(\tieta)&=\tieta^{2h_{1,2}-h_{1,3}}\, _2F_1\left(\frac{m}{2 m+2},\frac{2 m-1}{2 m+2};\frac{2 m}{m+1};-\frac{4}{\tieta}\right)\,,
\end{align}
where
\begin{align}\label{tietadef}
	\tieta\equiv\frac{16 y_1^2 y_2^2}{\left((y_1-y_2)^2+(x_1-x_2)^2\right) \left((y_1+y_2)^2+(x_1-x_2)^2\right)}\,.
\end{align}
The final result is
\begin{align}
	\langle \phi_{(1,2)}(x_1 + i y_1,x_1 - i y_1)&\phi_{(1,2)}(x_2+ i y_2,x_2 - i y_2)\rangle_{\uhp}=\nonumber\\
	&\frac{1}{(4y_1 y_2)^{2h_{1,2}}}\left(V^{\text{bulk}}_{(1,1)}(\tilde\eta)+	B_{(1,3)}^{{\bf a}} C_{(1,2)(1,2)(1,3)}V^{\text{bulk}}_{(1,3)}(\tilde\eta)\right)\,,
\end{align}
where
\begin{align}
B_{(1,3)}^{{\bf a}} &=\left(1+2 \cos \left(\frac{2 \pi  a_2 m}{m+1}\right)\right)\sqrt{\frac{{\sin \left(\frac{\pi }{m}\right) \sin \left(\frac{\pi  m}{m+1}\right)} }{{\sin \left(\frac{\pi }{m}\right) \sin \left(\frac{3 \pi  m}{m+1}\right)}}}\,,\nonumber\\
C_{(1,2)(1,2)(1,3)}&=\sqrt{\frac{\Gamma \left(\frac{2}{m+1}\right) \Gamma \left(\frac{m}{m+1}\right) \Gamma \left(2-\frac{3}{m+1}\right) \Gamma \left(\frac{2}{m+1}-1\right)}{\Gamma \left(\frac{1}{m+1}\right) \Gamma \left(\frac{m-1}{m+1}\right) \Gamma \left(\frac{2 m}{m+1}\right) \Gamma \left(\frac{3}{m+1}-1\right)}}\,.
\end{align}
In the boundary channel, corresponding to the exchange of $\hid$ and $\psi_{(1,3)}$ we have
\begin{align}\label{phi12twoptblocks}
	V_{(1,1)}(\tieta)&=\, _2F_1\left(\frac{2-m}{2 m+2},\frac{m}{2 m+2};\frac{m+3}{2 m+2};-\frac{\tieta}{4}\right)\,,\nonumber\\
	V_{(1,3)}(\tieta)&=\tieta^{{h_{1,3}}/{2}}\, _2F_1\left(\frac{2 m-1}{2 m+2},\frac{1}{2 m+2};\frac{3 m+1}{2 m+2};-\frac{\tieta}{4}\right)\,.
\end{align}
The final result is
\begin{align}\label{phi12twopt}
	\langle \phi_{(1,2)}(x_1 + i y_1,x_1 - i y_1)&\phi_{(1,2)}(x_2+ i y_2,x_2 - i y_2)\rangle_{\uhp}=\nonumber\\
	&\frac{1}{(4y_1 y_2)^{2h_{1,2}}}\left((B_{(1,2)}^{{\bf a}})^2V_{(1,1)}(\tilde\eta)+	(B_{(1,2)}^{{\bf a}\,(1,3)})^2V_{(1,3)}(\tilde\eta)\right)\,,
\end{align}
with
\begin{align}
	B_{(1,2)}^{{\bf a}}&=	2 (-1)^{a_1}  \cos \left(\frac{\pi a_2 m}{m+1}\right)\sqrt{-\frac{\sin \left(\frac{\pi  m}{m+1}\right)}{\sin \left(\frac{2 \pi  m}{m+1}\right)}}\,,
\end{align}
and, as it follows from bulk-boundary crossing symmetry
\begin{align}\label{B1211ope}
	(B_{(1,2)}^{{\bf a}\,(1,3)})^2&=\frac{\Gamma \left(\frac{2 m-1}{2 m+2}\right) \Gamma \left(\frac{3 m}{2 m+2}\right)}{2^{\frac{1}{m+1}} \Gamma \left(\frac{m-1}{m+1}\right) \Gamma \left(\frac{3 m+1}{2 m+2}\right)}-\frac{(B_{(1,2)}^{{\bf a}})^2\Gamma \left(\frac{3 m}{2 m+2}\right) \Gamma \left(1-\frac{3}{2 m+2}\right) \Gamma \left(\frac{1}{2}+\frac{1}{m+1}\right)}{2^{1-\frac{2}{m+1}} \Gamma \left(\frac{m}{2 m+2}\right) \Gamma \left(\frac{3}{2}-\frac{1}{m+1}\right) \Gamma \left(1-\frac{1}{2 m+2}\right)}\,.
\end{align}
This result is consistent with the F-matrices computation of ref.~\cite{Runkel:1998he}. As a particular case, for the Ising model with $\mathbb{Z}_2$-preserving conformal boundary conditions we find
\begin{align}
	(B_{\sigma}^{{\bf a}\,(1,3)} )^2=\frac{1}{\sqrt{2}}\,,
\end{align}
as predicted by~\cite{Lewellen:1991tb,Liendo:2012hy}. Hence $\psi_{(1,3)}$ is $\mathbb{Z}_2$-odd in this boundary condition.

\subsection{Bulk two-point function of \texorpdfstring{$\phi_{(2,1)}$}{phi21}}\label{app:phi212pt}
Next, we consider the bulk two-point functions of $\phi_{(2,1)}$ on the upper half-plane:
\begin{align}
	\langle \phi_{(2,1)}(x_1+i y_1, x_1- i y_1)\phi_{(2,1)}(x_2+i y_2, x_2- i y_2)\rangle_{\uhp}\,.
\end{align}
By the method of images, this correlator satisfies the same second order differential equation as the four-point function of $\phi_{(2,1)}(z)$ in the homogeneous theory
\begin{align}
	\left({\cal L}_{-2}^{(z_4)}-\frac{3}{2(2h_{2,1}+1)}{\cal L}_{-1}^2\right)	\langle \phi_{(2,1)}(z_1)\phi_{(2,1)}(z_2)\phi_{(2,1)}(z_3)\phi_{(2,1)}(z_4)\rangle=0\,.
\end{align}
By $SL(2,\mathbb{R})$ symmetry, this holomorphic correlator takes the following form
\begin{align}\label{21diffeq}
	\langle\phi_{(2,1)}(z_1)\phi_{(2,1)}(z_2)\phi_{(2,1)}(z_3)\phi_{(2,1)}(z_4)\rangle=\frac{\tcG(\tieta)}{(z_{12} z_{34})^{2h_{2,1}}}\,.
\end{align}
The cross-ratio $\tieta$, defined as in eq.~\eqref{tietadef1234}, becomes \eqref{tietadef} on the upper half-plane. The two Virasoro blocks that correspond to the exchange of $\hid$ and $\psi_{(3,1)}$ on the boundary read
\begin{align}\label{phi21twoptblocks}
	V_{(1,1)}(\tieta)&=\, _2F_1\left(\frac{m+1}{2 m},-\frac{m+3}{2 m};\frac{m-2}{2 m};-\frac{\tieta}{4}\right)\,,\nonumber\\
	V_{(3,1)}(\tieta)&=\tieta^{{h_{3,1}}/{2}} \, _2F_1\left(\frac{2 m+3}{2 m},-\frac{1}{2 m};\frac{3 m+2}{2 m};-\frac{\tieta}{4}\right)\,.
\end{align}
In the bulk channel, corresponding to the exchange of $\id$ and $\phi_{(3,1)}$ we have
\begin{align}\label{phi21twoptblocksbulk}
	V^{\text{bulk}}_{(1,1)}(\tieta)&=\tieta^{2h_{1,2}}\, _2F_1\left(-\frac{m+3}{2 m},-\frac{1}{2 m};-\frac{2}{m};-\frac{4}{\tieta}\right)\,,\nonumber\\
	V^{\text{bulk}}_{(3,1)}(\tieta)&=\tieta^{2h_{2,1}-h_{3,1}}\, _2F_1\left(\frac{m+1}{2 m},\frac{2 m+3}{2 m};\frac{2 m+2}{m};-\frac{4}{\tieta}\right)\,.
\end{align}
The final solution is
\begin{align}
	\langle \phi_{(2,1)}(x_1 + i y_1,x_1 - i y_1)&\phi_{(2,1)}(x_2+ i y_2,x_2 - i y_2)\rangle_{\uhp}=\nonumber\\
	&\frac{1}{(4y_1 y_2)^{2h_{2,1}}}\left(V^{\text{bulk}}_{(1,1)}(\tilde\eta)+	B_{(3,1)}^{{\bf a}} C_{(2,1)(2,1)(3,1)}V^{\text{bulk}}_{(3,1)}(\tilde\eta)\right)\,,
\end{align}
where
\begin{align}
	B_{(3,1)}^{{\bf a}} &=\left(1+2 \cos \left(\frac{2 \pi  a_1 (m+1)}{m}\right)\right)\sqrt{\frac{\sin \left(\frac{\pi }{m}\right) \sin \left(\frac{\pi  m}{m+1}\right)}{\sin \left(\frac{3 \pi }{m}\right) \sin \left(\frac{\pi  m}{m+1}\right)}},\nonumber\\
	C_{(2,1)(2,1)(3,1)}&=\sqrt{\frac{\Gamma \left(1+\frac{1}{m}\right) \Gamma \left(2+\frac{3}{m}\right) \Gamma \left(-\frac{2}{m}\right) \Gamma \left(-\frac{m+2}{m}\right)}{\Gamma \left(2+\frac{2}{m}\right) \Gamma \left(-\frac{1}{m}\right) \Gamma \left(\frac{m+2}{m}\right) \Gamma \left(-\frac{m+3}{m}\right)}}\,.
\end{align}
In the boundary channel we have
\begin{align}\label{phi21twopt}
	\langle \phi_{(2,1)}(x_1 + i y_1,x_1 - i y_1)&\phi_{(2,1)}(x_2+ i y_2,x_2 - i y_2)\rangle_{\uhp}=\nonumber\\
	&\frac{1}{(4y_1 y_2)^{2h_{2,1}}}\left((B_{(2,1)}^{{\bf a}})^2V_{(1,1)}(\tilde\eta)+	(B_{(2,1)}^{{\bf a}\,(3,1)})^2V_{(3,1)}(\tilde\eta)\right)\,,
\end{align}
with
\begin{align}
	B_{(2,1)}^{{\bf a}}&=2 (-1)^{a_2} \sqrt{\frac{\sin \left(\frac{\pi }{m}\right) \sin \left(\frac{\pi  m}{m+1}\right)}{\sin \left(\frac{2 \pi }{m}\right) \sin \left(\frac{\pi  m}{m+1}\right)}} \cos \left(\frac{\pi  a_1 (m+1)}{m}\right)\,,
\end{align}
and, as determined by bulk-boundary crossing symmetry
\begin{align}
	(B_{(2,1)}^{{\bf a}\,(3,1)})^2&=\frac{\sqrt{\pi } \Gamma \left(2+\frac{3}{m}\right)}{2^{\frac{m+2}{m}} \Gamma \left(\frac{3}{2}+\frac{1}{m}\right) \Gamma \left(\frac{m+2}{m}\right)}-\frac{(B_{(2,1)}^{{\bf a}})^2 \Gamma \left(\frac{1}{2}-\frac{1}{m}\right) \Gamma \left(2+\frac{3}{m}\right)}{\sqrt{\pi } 2^{\frac{m+2}{m}} \Gamma \left(2+\frac{2}{m}\right)}\,.
\end{align}
This result is consistent with the F-matrices computation of ref.~\cite{Runkel:1998he}. 
For the tricritical Ising model with $\mathbb{Z}_2$-preserving conformal boundary conditions one finds
\begin{align}
	(B_{\sigma'}^{{\bf a}\,(3,1)})^2=\frac{7}{4 \sqrt{2}}\,.
\end{align}
Hence $\psi_{(3,1)}$ is $\mathbb{Z}_2$-odd in this boundary condition.

\subsection{Tricritical Ising model with \texorpdfstring{$\mathbb{Z}_2$}{Z2}-preserving conformal b.c.}\label{app:tric}

\subsubsection{Bulk two-point functions of $\epsilon'$}
Consider the bulk two-point correlation function of $\epsilon'=\phi_{(1,3)}$ in the tricritical Ising model with $\mathbb{Z}_2$-preserving conformal boundary conditions
\begin{align}
	\langle \epsilon'(x_1+i y_1, x_1- i y_1)\epsilon'(x_2+i y_2, x_2- i y_2)\rangle_{\uhp}\,.
\end{align}
This correlator satisfies a third order differential equation, whose solution is known in closed form (for a generic minimal model~\cite{Lauria:2023uca}). In the boundary channel, the Virasoro blocks that are relevant to the $m=4$ case are\footnote{Another linearly independent solution corresponds to the exchange of $\psi_{(1,5)}$, which however does not exists for $m \leq 4$.}
\begin{align}\label{phi13twoptblocks}
	V_{(1,1)}(\tilde\eta)&=\, _3F_2\left(\frac{1-m}{m+1},\frac{2m}{m+1},-\frac{2m-2}{m+1};\frac{3+m}{2m+2},\frac{2-m}{m+1};-\frac{\tieta}{4}\right)\,,\nonumber\\
	V_{(1,3)}(\tilde\eta)&=\tieta^{{h_{1,3}}/{2}} \, _3F_2\left(\frac{5 m-1}{2m+2},\frac{1-m}{2m+2},-\frac{3m-3}{2m+2};\frac{3-m}{2m+2},\frac{3 m+1}{2m+2};-\frac{\tieta}{4}\right)\,,
\end{align}
with $\tieta$ defined as in eq.~\eqref{tietadef1234}.
The final correlator reads
\begin{align}\label{phi13twopt}
	\langle \phi_{(1,3)}(x_1 + i y_1,x_1 - i y_1)&\phi_{(1,3)}(x_2+ i y_2,x_2 - i y_2)\rangle_{\uhp}=\nonumber\\
	&\frac{1}{(4y_1 y_2)^{2h_{1,3}}}\left((B_{(1,3)}^{{\bf a}})^2V_{(1,1)}(\tilde\eta)+	(B_{(1,3)}^{{\bf a}\,(1,3)})^2V_{(1,3)}(\tilde\eta)\right)\,,
\end{align}
where the coefficient $B_{(1,3)}^{{\bf a}}$ is given in Tables~\ref{tbl:tric214}-\ref{tbl:tric224} and crossing symmetry implies for the $(2,2)_4$ boundary condition that \cite{Lauria:2023uca}
\begin{align}
	(B_{\epsilon'}^{{\bf a}\,(1,3)})^2&\simeq 0.663053\,.
\end{align}
This results implies that in the $(2,2)_4$ conformal boundary condition $\psi_{(1,3)}$ is $\mathbb{Z}_2$-even. Note that $\psi_{(1,3)}$ does not exist in the $(2,1)_4$ b.c., and correspondingly $(B_{\epsilon'}^{{\bf a}\,(1,3)})^2=0$.

\subsubsection{The boundary four-point function of $\psi_{(3,1)}$}\label{app:bd4pt31}
Consider the boundary four-point correlation function of $\psi_{(3,1)}$ in the tricritical Ising model with $\mathbb{Z}_2$-preserving conformal boundary conditions
\begin{align}\label{epspp4pt}
	\langle \psi_{(3,1)}(x_1) \psi_{(3,1)}(x_2)\psi_{(3,1)}(x_3) \psi_{(3,1)}(x_4)\rangle_{\uhp}\,,\quad x_i > x_{i+1}\,.
\end{align}
This correlator satisfies the same third-order differential equation as the four-point correlation function $\psi_{(3,1)}(z)$ in the homogeneous tricritical Ising model, and its solution is known in closed form for any $m$ (see for instance \cite{Lauria:2023uca}). The most generic solution for $m = 4$ is
\begin{align}\label{psi31fourpt}
	\langle \psi_{(3,1)}(x_1)&\psi_{(3,1)}(x_2)\psi_{(3,1)}(x_3)\psi_{(3,1)}(x_4)\rangle_{\uhp}=\frac{1}{(x_{12}x_{34})^{2h_{3,1}}}\left(V_{(1,1)}(\tilde\eta)+({\bdc}_{(3,1)(3,1)(3,1)}^{{\bf a}})^2V_{(3,1)}(\tilde\eta)\right)\,,
\end{align}
(we have set to zero the coefficient of the third linearly independent solution which would correspond to the exchange of $\psi_{(5,1)}$ on the boundary), with
\begin{align}\label{phi31twoptblocks}
	V_{(1,1)}(\tieta)&=\, _3F_2\left(-\frac{4}{m}-2,-\frac{2}{m}-1,\frac{2}{m}+2;-\frac{3}{m}-1,\frac{1}{2}-\frac{1}{m};-\frac{\tieta}{4}\right)\,,\nonumber\\
	V_{(3,1)}(\tieta)&=\tieta^{{h_{3,1}}/{2}}\, _3F_2\left(-\frac{3}{m}-\frac{3}{2},-\frac{1}{m}-\frac{1}{2},\frac{3}{m}+\frac{5}{2};-\frac{2}{m}-\frac{1}{2},\frac{1}{m}+\frac{3}{2};-\frac{\tieta}{4}\right)\,,
\end{align}
and
\begin{align}
	\tieta=\frac{x_{12}^2 x_{34}^2}{x_{12}x_{14}  x_{23}x_{24}}\,,
\end{align}
which is positive if $x_i > x_{i+1}$. The remaining coefficient in eq.~\eqref{psi31fourpt} is fixed by crossing symmetry to be \cite{Lauria:2023uca}
\begin{align}\label{bdC313131}
	({\bdc}_{(3,1)(3,1)(3,1)}^{{\bf a}})^2=\frac{ 2^{\frac{m-2}{m}}\pi \cos \left(\frac{2 \pi }{m}\right) \Gamma \left(3+\frac{4}{m}\right) \Gamma \left(-\frac{m+2}{2 m}\right)}{\left(2 \cos \left(\frac{2 \pi }{m}\right)+1\right) \Gamma \left(\frac{1}{2}+\frac{1}{m}\right) \Gamma \left(\frac{3}{2}+\frac{1}{m}\right) \Gamma \left(2+\frac{3}{m}\right) \Gamma \left(-\frac{m+4}{2 m}\right)}\,,
\end{align}
which vanishes identically for $m=4$, but is positive otherwise. Hence, the self-OPE of $\psi_{(3,1)}$ in the tricritical Ising model contains only the identity.

\section{One-loop computations for the \texorpdfstring{$\TTb$}{TTb} deformation}
\label{app:pertTTb}

In this section we compute the following correlation functions at one loop in the $\TTb$ deformation of eq.~\eqref{TTbads}
\begin{align}
	&\langle \Disp(x_1) \Disp(x_2)\rangle\,,~ \langle \Disp^2(x_1) \Disp^2(x_2)\rangle\,,\nonumber\\
	\langle \Disp(x_1)& \Disp(x_2)\Disp(x_3)\rangle\,,~\langle \Disp(x_1) \Disp(x_2)\Disp^2(x_3)\rangle\,.
\end{align}
Here and in the following of this section, the ordering along the boundary of $\AdStwo$ is taken such that $x_i >x_{i+1}$. As we are dealing with covariant bulk RG flows, these correlators take the form of 1d conformal correlation functions, i.e.\footnote{Notice the slight change of conventions: here (as well as in section \ref{app:pertVir}) boundary primaries are generically \emph{not} unit-normalized.}
\begin{align}\label{1dcorrRG}
	\langle \psi_i(x_1) \psi_j(x_2)\rangle = \delta_{ij}\frac{\hC_{i}(g_{\TTb})}{(x^2)^{\hD_{i}(g_{\TTb})}}\,,~ 		\langle\psi_i(x_1)\psi_j(x_2)\psi_k(x_3)\rangle=\frac{\hC_{ijk}(g_{\TTb})}{
		(x_{12})^{\hD_{ijk}(g_{\TTb})} (x_{23})^{\hD_{jki}(g_{\TTb})}
		(x_{13})^{\hD_{ikj}(g_{\TTb})}}\,.
\end{align}
In particular, they are specified by the scaling dimensions of $\Disp$ and $\Disp^2$ which we parametrize along the RG as follows
\begin{align}\label{scalingdimparam}
	\D_{\Disp}(g_{\TTb})&= 2 + g_{\TTb}\,\dhD_{\Disp}+O(g_{\TTb}^2)\,,\nonumber\\
	\D_{\Disp^2}(g_{\TTb})&= 4 + g_{\TTb}\,\dhD_{\Disp^2}+O(g_{\TTb}^2)\,,
\end{align}
as well by  as their normalizations (tree-level values are computed in appendix A.3 of \cite{Lauria:2023uca})
\begin{align}\label{OPEcoeffsparam}
	\hC_{\Disp}(g_{\TTb})\equiv \hC_{\Disp\Disp}(g_{\TTb})&= c/2 + g_{\TTb}\,\delta \hC_{\Disp}+O(g_{\TTb}^2)\,,\nonumber\\
	\hC_{\Disp^2}(g_{\TTb})\equiv \hC_{\Disp^2\Disp^2}(g_{\TTb})&=\frac{c}{10} (22+5 c) + g_{\TTb}\,\delta \hC_{\Disp^2}+O(g_{\TTb}^2)\,,\nonumber\\
	\hC_{\Disp\Disp\Disp}(g_{\TTb})&=c+ g_{\TTb}\,\delta \hC_{\Disp\Disp\Disp}+O(g_{\TTb}^2)\,,\nonumber\\
	\hC_{\Disp\Disp\Disp^2}(g_{\TTb})&=\frac{c}{10} (22+5 c)+ g_{\TTb}\,\delta \hC_{\Disp\Disp\Disp^2}+O(g_{\TTb}^2)\,.
\end{align}
For conformal b.c. that support a boundary Virasoro primary $\psi$ of scaling dimension $\hD_\psi$ we will also compute, along the same $\TTb$ deformation in $\AdStwo$
\begin{align}
	\langle \psi(x_1)\psi(x_2)\rangle\,,~ \langle \Disp(x_1) \psi(x_2)\psi(x_3)\rangle\,,~\langle \Disp^2(x_1) \psi(x_2)\psi(x_3)\rangle\,.
\end{align}
We will assume $\psi$ to be unit-normalized at tree level, and so we will let
\begin{align}
	\D_{\psi}(g_{\TTb})&= \hD_\psi+ g_{\TTb}\,\delta \hD_{\psi}+O(g_{\TTb}^2)\,,\nonumber\\
	\hC_{\psi}(g_{\TTb})\equiv \hC_{\psi\psi}(g_{\TTb})&= 1+ g_{\TTb}\,\delta \hC_{\psi}+O(g_{\TTb}^2)\,,\nonumber\\
	\hC_{\Disp \psi \psi}(g_{\TTb})&= \hD_\psi+ g_{\TTb}\,\delta \hC_{\Disp \psi \psi}+O(g_{\TTb}^2)\,,\nonumber\\
	\hC_{\Disp^2 \psi \psi}(g_{\TTb})&= \frac{\hD_\psi}{5} (5 \hD_\psi+1)+ g_{\TTb}\,\delta \hC_{\Disp^2 \psi \psi}+O(g_{\TTb}^2)\,,
\end{align}
where for the tree-level OPE coefficients we used the results of appendix A.3 of \cite{Lauria:2023uca}.

\subsection{Two-point functions}
Starting with the two-point correlation functions, Poincaré coordinates of AdS with radius $R$ the one-loop corrections are computed by
\begin{align}\label{DD1loop}
	-g_{\TTb}R^4&\int_{y>a}^\infty \frac{\mathrm{d} y} {y^2}\int_{-\infty}^{\infty}\mathrm{d} x~\langle \Disp(x_1)\Disp(x_2)\TTb(x+i y,x-iy)\rangle^{\text{c}}_{0, \AdStwo}\,,\nonumber\\
	-g_{\TTb}R^4&\int_{y>a}^\infty \frac{\mathrm{d} y} {y^2}\int_{-\infty}^{\infty}\mathrm{d} x~\langle \Disp(x_1)\Disp^2(x_2)\TTb(x+i y,x-iy)\rangle^{\text{c}}_{0, \AdStwo}\,,\nonumber\\
	-g_{\TTb}R^4&\int_{y>a}^\infty \frac{\mathrm{d} y} {y^2}\int_{-\infty}^{\infty}\mathrm{d} x~\langle \Disp^2(x_1)\Disp^2(x_2)\TTb(x+i y,x-iy)\rangle^{\text{c}}_{0, \AdStwo}\,,\nonumber\\
	-g_{\TTb}R^4&\int_{y>a}^\infty \frac{\mathrm{d} y} {y^2}\int_{-\infty}^{\infty}\mathrm{d} x~\langle \psi(x_1)\psi(x_2)\TTb(x+i y,x-iy)\rangle^{\text{c}}_{0, \AdStwo}\,,
\end{align}
where $\langle \dots \rangle^{c}$ means `connected'. We have introduced a IR cut-off at $y=a$ and omitted the counterterms. The correlation functions in the integrands above are obtained from those on the upper half-plane $\uhp$, upon Weyl rescaling to $\AdStwo$. These are (limits) of correlation functions with many insertions of $T$ on $\uhp$: four, five, six, and two, respectively.
Explicit expressions for such correlation functions on the $\uhp$ can be found, for example, in the appendices of~\cite{Lauria:2023uca}. 

Computing the integral is not difficult, and up to $O(a,g_{\TTb}^2)$ corrections we find for the one-loop contribution
\begin{align}
	(x_{12})^4\delta\langle \Disp(x_1) \Disp(x_2)\rangle  &= - \pi\,\hC_{\Disp\Disp\Disp}(0)\left(1+\frac{c}{24}+\frac{1}{2} \log \left(\frac{x_{12}^2}{4a^2}\right)\right) g_{\TTb}\,,\nonumber\\
	(x_{12})^6\delta\langle \Disp(x_1) \Disp^2(x_2)\rangle  &= -\pi \hC_{\Disp^2}(0) \left(1- \frac{5x_{12}^2}{8 a^2}\right) g_{\TTb}\,,\nonumber\\
	(x_{12})^8\delta\langle \Disp^2(x_1) \Disp^2(x_2)\rangle  &=-6\pi \hC_{\Disp^2}(0) \left(\frac{131}{90}+\frac{c}{36}+\log \left(\frac{x_{12}^2}{4 a ^2}\right)\right) g_{\TTb}\,,\nonumber\\
	(x_{12}^2)^{\hD_\psi}\delta\langle \psi(x_1) \psi(x_2)\rangle &=- \frac{\pi \hD_\psi}{2} \left[\hD_\psi +(\hD_\psi  -1) \log \left(\frac{x_{12}^2}{4a^2}\right)\right] g_{\TTb}\,.
\end{align}
Note that an off-diagonal two-point function is generated at one-loop.

\subsubsection{Renormalization and mixings}

We define the renormalized operators
\begin{align}\label{mixings}
	\Disp_R = Z_\Disp \Disp +Z_{\Disp \hid}\hid+\dots\,,\quad 	{\Disp^2}_R= Z_{\Disp^2} \Disp^2+Z_{\Disp^2 \Disp} \Disp+Z_{\Disp^2 \Disp''} \Disp''+Z_{\Disp^2 \hid}\hid+\dots
\end{align}
($\Disp''$ denotes the second derivative of $\Disp$). The wave-functions $ Z_{AA'}$ are fixed by requiring correlation functions with insertions of $\Disp_R$ and $\Disp_R^2$ to be finite and to match the expected form of 1d conformal correlators. By letting
\begin{align}
	Z_\Disp &= 1 + g_{\TTb}  z_{\Disp} \log(a/R)\,,&&  Z_{\Disp^2} = 1 + g_{\TTb}z_{\Disp^2}  \log(a/R)\,,\nonumber\\
	Z_{\Disp^2\Disp} &= g_{\TTb} \frac{z_{\Disp^2\Disp}}{a^2}\,,&& Z_{\Disp^2\Disp''} = g_{\TTb} z_{\Disp^2\Disp''}\,,
\end{align}
being $R$ the $\AdStwo$ radius we see for example that
\begin{align}
	\langle \Disp_R (x_1) \Disp_R(x_2)\rangle = Z_\Disp^2 \langle \Disp (x_1) \Disp(x_2)\rangle\,,
\end{align}
is completely finite if $z_{\Disp} = -\pi$, and in particular (taking the $a\rightarrow 0$ limit) we get the following one-loop correction
\begin{align}
	(x_{12})^4\delta\langle \Disp_R (x_1) \Disp_R(x_2)\rangle  &  = - {c \pi }{}\left(1+\frac{c}{24}-\log 2+\frac{1}{2} \log \left({x_{12}^2/R^2}\right)\right)g_{\TTb}\,.
\end{align}
Hence, by comparing to the $O(g_{\TTb})$ expansion of eq.~\eqref{1dcorrRG} we find
\begin{align}
	\dhD_{\Disp}= \pi\,,\quad 
	\delta \hC_{\Disp}= -\pi c\,\left(1+\frac{ c}{24}-\log 2\right)\,.
\end{align}

Analogously,
\begin{align}
	\langle {\Disp^2}_R (x_1) {\Disp^2}_R(x_2)\rangle = Z_{\Disp^2}^2 \langle \Disp^2 (x_1) \Disp^2(x_2)\rangle~
\end{align}
is completely finite in the $a\rightarrow 0$ limit if we set $z_{\Disp} = -6\pi$. By comparing to the $O(g_{\TTb})$ expansion of eq.~\eqref{1dcorrRG} we find 
\begin{align}
	\dhD_{\Disp^2}= 6 \pi\,,\quad 
	\delta\hC_{\Disp^2}= -\frac{c (5 c+22)\pi}{10}  \left(\frac{c}{6}+\frac{131}{15}-12 \log 2\right)\,.
\end{align}

For the mixed two-point function, requiring that
\begin{align}
	0&=	\langle {\Disp}_R (x_1) {\Disp^2}_R(x_2)\rangle \nonumber\\
	&=Z_{\Disp^2}Z_{\Disp} \langle \Disp (x_1) \Disp^2(x_2)\rangle+Z_{\Disp^2}Z_{\Disp^2\Disp} \langle \Disp (x_1) \Disp(x_2)\rangle+Z_{\Disp^2}Z_{\Disp^2\Disp''} \langle \Disp (x_1) \Disp''(x_2)\rangle\,,
\end{align}
fixes
\begin{align}
	z_{\Disp^2\Disp} =-\frac{\pi  (5 c+22)}{8}\,,\quad z_{\Disp^2\Disp''} =\frac{\pi  (5 c+22)}{100}\,.
\end{align}

Finally, in order to renormalize the two-point function of $\psi$  we introduce $\psi_R = Z_\psi \psi$ with
\begin{align}
	Z_\psi = 1 +  g_{\TTb}  z_{\psi} \log(a/R)\,, \quad z_{\psi} = -\frac{\pi}{2}  (\hD_\psi -1) \hD_\psi\,.
\end{align}
By plugging into the renormalized correlator
\begin{align}
	\langle {\psi}_R (x_1) {\psi}_R(x_2)\rangle = Z_{\psi}^2 \langle \psi(x_1) \psi(x_2)\rangle\,,
\end{align}
and comparing to the $O(g_{\TTb})$ expansion of eq.~\eqref{1dcorrRG} we find
\begin{align}
	\delta\hD_\psi= - z_{\psi}\,,\quad 
	\delta\hC_{\psi}= -\frac{\pi \hD_\psi}{2}   (2\log 2-\hD_\psi  (2\log 2-1))\,.
\end{align}

\subsection{Three-point functions}

For the one-loop corrections to three-point functions we shall compute
\begin{align}\label{3pt1LoopTTb}
	-g_{\TTb}R^4&\int_{y>a}^\infty \frac{\mathrm{d} y} {y^2}\int_{-\infty}^{\infty}\mathrm{d} x~\langle \Disp(x_1)\Disp(x_2)\Disp(x_3)\TTb(x+i y,x-iy)\rangle^{\text{c}}_{0, \AdStwo}\,,\nonumber\\
	-g_{\TTb}R^4&\int_{y>a}^\infty \frac{\mathrm{d} y} {y^2}\int_{-\infty}^{\infty}\mathrm{d} x~\langle \Disp(x_1)\Disp(x_2)\Disp^2(x_3)\TTb(x+i y,x-iy)\rangle^{\text{c}}_{0, \AdStwo}\,,\nonumber\\
	-g_{\TTb}R^4&\int_{y>a}^\infty \frac{\mathrm{d} y} {y^2}\int_{-\infty}^{\infty}\mathrm{d} x~\langle \Disp(x_1)\psi(x_2)\psi(x_3)\TTb(x+i y,x-iy)\rangle^{\text{c}}_{0, \AdStwo}\,,\nonumber\\
	-g_{\TTb}R^4&\int_{y>a}^\infty \frac{\mathrm{d} y} {y^2}\int_{-\infty}^{\infty}\mathrm{d} x~\langle \Disp^2(x_1)\psi(x_2)\psi(x_3)\TTb(x+i y,x-iy)\rangle^{\text{c}}_{0, \AdStwo}\,.
\end{align}
The integrands are again obtained from appropriate limits of correlation functions with many insertions of $T$ on $\uhp$: five, six, three, four, respectively. Computing the integral is easy if we set for example $x_3 =0$ and $x_1=1$, so up to $O(a,g_{\TTb}^2)$ corrections we find
\begin{align}
	&(x-1)^2 x^2\delta\langle \Disp(1) \Disp(x)\Disp(0)\rangle=   \nonumber\\
	&-\frac{\pi  c^2 g_{\TTb}}{4} \left(1-\frac{c \left(x^8-4 x^7+6 x^6-4 x^5+3 x^4-4 x^3+6 x^2-4 x+1\right)}{8 c (x-1)^2 x^2 a^2}+\frac{2}{c}\log \left(\frac{(1-x)^2 x^2}{64 a^6}\right)\right)\,,\nonumber\\
	\nonumber\\
	&x^4\delta\langle \Disp(1) \Disp(x)\Disp^2(0)\rangle=   \nonumber\\
	& -\frac{\pi c(5 c+22)}{10}g_{\TTb}   \left(\frac{5 c x^4}{128 (x-1)^4 a ^4}-\frac{5 x^2}{4 (x-1)^2 a ^2}\right)~\nonumber\\
	&-\frac{\pi  c (5 c+22)}{10}g_{\TTb}\left(\frac{(c+20) x^2-2 (c+14) x+c+20}{6 (x-1)^2}-2 \log ((1-x)^2)+3 \log x^2-8 \log (2 a )\right)\,,\nonumber\\
	\nonumber\\
	&(x-1)^2(x^2)^{\hD_\psi-1}\delta\langle \Disp(1) \psi(x)\psi(0)\rangle= \nonumber\\
	&+ \frac{\pi  c g_{\TTb} }{16}\frac{(x-1)^2}{x^2a^2}-{\pi g_{\TTb}\hD_\psi}{}  \left(\frac{c}{12}+\frac{1}{2} (\hD_\psi-2)^2-((\hD_\psi-1) \hD_\psi+1) \log 2\right)\nonumber\\
	&+\pi  g_{\TTb} \hD_\psi   \left(\frac{1}{2}\left(1-\hD_\psi^2+\hD_\psi\right) \log x^2-\frac{1}{2}\log ((1-x)^2)+(1+(\hD_\psi-1) \hD_\psi) \log a\right)\,,\nonumber
\end{align}
\begin{align}\label{1loopresTTb}
	&(x-1)^4(x^2)^{\hD_\psi-2}\delta\langle \Disp^2(1) \psi(x)\psi(0)\rangle = \\
	&+\frac{\pi  c (5 c+22)}{10} \left(\frac{5 \hD_\psi  (x-1)^2}{4 c a ^2}-\frac{5 (x-1)^4}{64 x^2 a ^4}\right)g_{\TTb}\nonumber\\
	&-\frac{\pi  \hD_\psi}{30}  (5 c (x ((\hD_\psi +2)x-6)+6)+x ((\hD_\psi  (3 \hD_\psi  (5 \hD_\psi -39)+238)+92)x-132)+132)g_{\TTb}\nonumber\\
	&+\frac{\pi  \hD_\psi  (5 \hD_\psi +1) x^2}{5}  \left(((\hD_\psi -1) \hD_\psi +6) \log (2 a )+\frac{1}{2}\left(6+\hD_\psi-\hD_\psi ^2 +6\right) \log x^2-3 \log ((1-x)^2)\right)g_{\TTb} \nonumber\,.
\end{align}

\subsubsection{Renormalization and mixings}

To resolve the infrared divergences we include the mixings of eq.~\eqref{mixings} i.e. write
\begin{align}
	\langle {\Disp}_R (x_1) \Disp_R(x_2) \Disp_R(x_3)\rangle &=Z^3_{\Disp} \langle \Disp (x_1) \Disp(x_2)\Disp(x_3)\rangle\nonumber\\
	&+Z^2_{\Disp} Z_{\Disp\hid} \left(\langle \Disp (x_1) \Disp(x_2)\hid\rangle+\langle \Disp (x_1) \Disp(x_3)\hid\rangle+\langle \Disp (x_2) \Disp(x_3)\hid\rangle\right)\nonumber\\
	&+Z_{\Disp\hid}^3 \langle \hid \hid \hid \rangle\,,
\end{align}
with $Z_{\Disp}$ as in the previous section and
\begin{align}
	Z_{\Disp\hid} =	g_{\TTb} \frac{\pi  c}{16 a ^2}\,.
\end{align}
With these chosen counterterms we find (up to $O(g_{\TTb}^2)$ corrections)
\begin{align}
	(x-1)^2 x^2\langle \Disp_R (1) \Disp_R(x)\Disp_R(0)\rangle &  = c- \frac{c \pi g_{\TTb}}{4}\left({c+2 \log (x^2(1-x^2)/R^4)-12 \log 2}{}\right)\,,
\end{align}
and so, comparing to the $O(g_{\TTb})$ expansion of eq.~\eqref{1dcorrRG} we find
\begin{align}
	\delta\hC_{\Disp\Disp\Disp} = \pi  c \left( 3\log 2-\frac{c}{4}\right)\,.
\end{align}
The choices above are of course enough in order to renormalize
\begin{align}
	\langle \Disp_R(1) \psi_R(x)\psi_R(0)\rangle  =Z_{\Disp} Z_\psi^2\langle \Disp (x_1) \psi(x_2)\psi(x_3)\rangle+Z^2_{\psi} Z_{\Disp\hid} \langle \psi (x_1) \psi(x_2)\hid\rangle\,,
\end{align}
from which we can extract
\begin{align}
	\delta\hC_{\psi\psi\Disp} = -\frac{\pi \hD_\psi}{2} \left(\frac{c}{6}-(1-\hD_\psi+\hD_\psi ^2 )\log 4+(\hD_\psi -2)^2\right)\,.
\end{align}

More mixings are needed in order to remove divergences as well as spurious finite terms in the second of eq.~\eqref{1loopresTTb}, but the essence is the same. We write
\begin{align}
	\langle {\Disp}_R (x_1) \Disp_R(x_2) \Disp^2_R(x_3)\rangle &=Z^2_{\Disp}Z_{\Disp^2} \langle \Disp (x_1) \Disp(x_2)\Disp^2(x_3)\rangle\nonumber\\
	&+Z^2_{\Disp} Z_{\Disp^2\Disp} \langle \Disp (x_1) \Disp(x_2)\Disp(x_3)\rangle+Z^2_{\Disp} Z_{\Disp^2\Disp''} \langle \Disp (x_1) \Disp(x_2)\Disp''(x_3)\rangle\nonumber\\
	&+Z^2_{\Disp} Z_{\Disp^2\hid} \langle \Disp (x_1) \Disp(x_2)\hid\rangle + \text{subleading}\,.
\end{align}
Almost all divergent terms, and all the spurious finite terms, are subtracted with the previous choices for the wave functions. We are left with one divergent contribution, which we subtract by including a mixing term with the identity, i.e.
\begin{align}
	Z_{\Disp^2\hid} =	g_{\TTb} \frac{\pi  c (5 c+22)}{128 a ^4}\,.
\end{align}
The counterterms above will also renormalize
\begin{align}
	\langle {\Disp^2}_R (x_1) \psi_R(x_2) \psi_R(x_3)\rangle &=Z^2_{\psi}Z_{\Disp^2} \langle \Disp^2 (x_1) \psi(x_2)\psi(x_3)\rangle\nonumber\\
	&+Z^2_{\psi} Z_{\Disp^2\Disp} \langle \Disp (x_1) \psi(x_2)\psi(x_3)\rangle+Z^2_{\psi} Z_{\Disp^2\Disp''} \langle \Disp'' (x_1) \psi(x_2)\psi(x_3)\rangle\nonumber\\
	&+Z^2_{\psi} Z_{\Disp^2\hid} \langle \hid \psi (x_2) \psi(x_3)\rangle \,.
\end{align}
The above renormalized correlators have the right conformal structure, and we from them we extract 
\begin{align}
	\delta\hC_{\Disp\Disp\Disp^2} &=-\frac{\pi  c (5 c+22)}{10}\left(\frac{c}{6}+\frac{41}{15}-8 \log 2\right)\,,\nonumber\\
	\delta\hC_{\psi\psi\Disp^2} &=\frac{\pi \hD_\psi (5 \hD_\psi+1)}{5\times 30} \left(180 \log 2-5 c+15 \hD_\psi (8-\hD_\psi+(\hD_\psi-1) \log 4)-262\right)\,.
\end{align}

\subsubsection{Final results for the OPE coefficients}
Taking into account the renormalization of the external operators computed earlier, the OPE coefficients for unit-normalized operators are
\begin{align}
	\bdc_{\Disp\Disp\Disp}(g_{\TTb})\equiv \frac{\hC_{\Disp\Disp\Disp}(g_{\TTb})}{\hC_{\Disp\Disp}(g_{\TTb})^{3/2}}&=\frac{2 \sqrt{2}}{c} \left(1-\frac{\pi (c-24)}{8}g_{\TTb}+O(g_{\TTb}^2)\right)\,,\nonumber\\
	\bdc_{\psi\psi\Disp}(g_{\TTb})\equiv\frac{\hC_{\psi\psi\Disp}(g_{\TTb})}{\hC_{\psi}(g_{\TTb})\hC_{\Disp}(g_{\TTb})^{1/2}}&=\frac{\sqrt{2} \hD_\psi}{\sqrt{c}}\left(1-\pi  \left(1+\frac{c}{24}-2 \hD_\psi\right)g_{\TTb}+O(g_{\TTb}^2)\right)\,,\nonumber\\
	\bdc_{\psi\psi\Disp^2}(g_{\TTb})\equiv\frac{\hC_{\psi\psi\Disp^2}(g_{\TTb})}{\hC_{\psi}(g_{\TTb})\hC_{\Disp^2}(g_{\TTb})^{1/2}}&=\frac{\sqrt{\frac{2}{5}} \hD_\psi (5 \hD_\psi+1)}{\sqrt{c (5 c+22)}}\nonumber\\
	&\times\left(1-\frac{\pi}{60}  (5 c-240 \hD_\psi+262)g_{\TTb}+O(g_{\TTb}^2)\right)\,,\nonumber\\
	\bdc_{\Disp\Disp\Disp^2}(g_{\TTb})\equiv\frac{\hC_{\Disp\Disp\Disp^2}(g_{\TTb})}{\hC_{\Disp\Disp}(g_{\TTb})\hC_{\Disp^2\Disp^2}(g_{\TTb})^{1/2}}&=\frac{1}{c}{\sqrt{\frac{2}{5}} \sqrt{c (5 c+22)}}{}\left(1+\frac{109 \pi }{30}g_{\TTb}+O(g_{\TTb}^2)\right)\,.
\end{align}

\section{One-loop computations for Virasoro deformations}
\label{app:pertVir}

In this section we compute the following correlation functions at one loop in the Virasoro deformation of eq.~\eqref{defgenconformalbc}
\begin{align}
	&\langle \Disp(x_1) \Disp(x_2)\rangle\,,~ \langle \Disp^2(x_1) \Disp^2(x_2)\rangle\,,\nonumber\\
	\langle \Disp(x_1)& \Disp(x_2)\Disp(x_3)\rangle\,,~\langle \Disp(x_1) \Disp(x_2)\Disp^2(x_3)\rangle\,.
\end{align}
For covariant RG flows in $\AdStwo$, these take the form of 1d correlation functions. The ordering along the boundary of AdS is taken such that $x_i >x_{i+1}$. Following the conventions of appendix~\ref{app:pertTTb}, we parametrize the CFT data along the RG as
\begin{align}
	\D_{\Disp}(g_{\phi})&= 2 + g_{\phi}\,\dhD_{\Disp}+O(g_{\phi}^2)\,,\nonumber\\
	\D_{\Disp^2}(g_{\phi})&= 4 + g_{\phi}\,\dhD_{\Disp^2}+O(g_{\phi}^2)\,,
\end{align}
\begin{align}
	\hC_{\Disp}(g_{\phi})\equiv \hC_{\Disp\Disp}(g_{\phi})&= c/2 + g_{\phi}\,\delta \hC_{\Disp}+O(g_{\phi}^2)\,,\nonumber\\
	\hC_{\Disp^2}(g_{\phi})\equiv \hC_{\Disp^2\Disp^2}(g_{\phi})&=\frac{c}{10} (22+5 c) + g_{\phi}\,\delta \hC_{\Disp^2}+O(g_{\phi}^2)\,,\nonumber\\
	\hC_{\Disp\Disp\Disp}(g_{\phi})&=c+ g_{\phi}\,\delta \hC_{\Disp\Disp\Disp}+O(g_{\phi}^2)\,,\nonumber\\
	\hC_{\Disp\Disp\Disp^2}(g_{\phi})&=\frac{c}{10} (22+5 c)+ g_{\phi}\,\delta \hC_{\Disp\Disp\Disp^2}+O(g_{\phi}^2)\,.
\end{align}

\subsection{Two-point functions}
The one-loop corrections are computed by
\begin{align}\label{DD1loopVir}
	-g_{\phi}R^{\D_\phi}&\int_{y>a}^\infty \frac{\mathrm{d} y} {y^2}\int_{-\infty}^{\infty}\mathrm{d} x~\langle \Disp(x_1)\Disp(x_2)\phi(x+i y,x-iy)\rangle^{\text{c}}_{0, \AdStwo}\,,\nonumber\\
	-g_{\phi}R^{\D_\phi}&\int_{y>a}^\infty \frac{\mathrm{d} y} {y^2}\int_{-\infty}^{\infty}\mathrm{d} x~\langle \Disp(x_1)\Disp^2(x_2)\phi(x+i y,x-iy)\rangle^{\text{c}}_{0, \AdStwo}\,,\nonumber\\
	-g_{\phi}R^{\D_\phi}&\int_{y>a}^\infty \frac{\mathrm{d} y} {y^2}\int_{-\infty}^{\infty}\mathrm{d} x~\langle \Disp^2(x_1)\Disp^2(x_2)\phi(x+i y,x-iy)\rangle^{\text{c}}_{0, \AdStwo}\,,
\end{align}
where $\langle \dots \rangle^{c}$ means `connected', $y=a$ is an IR cut-off, and we omitted counterterms. The correlation functions in the integrands above are obtained from correlation functions on the upper half-plane $\uhp$ via Weyl rescaling. These are in turn computed from appropriate limits of correlation functions with one insertion of the Virasoro primary $\phi$ and (respectively) two, three and four insertions of $T$ on $\uhp$. Explicit expressions of such can be found for example in~\cite{Lauria:2023uca}.

Computing the integral we find (up to $O(a,g_{\phi}^2)$ corrections)
\begin{align}\label{twoptVir}
	(x_{12})^4 \delta\langle \Disp(x_1) \Disp(x_2)\rangle&= - \frac{\pi \D_{\phi} B_\phi}{2^{\D_\phi-1}}\left(\D_\phi+(\D_\phi-2) \log \left(\frac{x_{12}^2}{4a^2}\right)\right) g_{\phi}\,,\nonumber\\
	(x_{12})^6 \delta\langle \Disp(x_1) \Disp^2(x_2)\rangle&= - \frac{\pi \D_{\phi} B_\phi}{5\times 2^{\D_\phi+2}}\left(8 (\D_{\phi}-2) (5 \D_{\phi}+2)-{5  (2 c+5 (\D_{\phi}-2) \D_{\phi}+4)}{}\frac{x_{12}^2}{a^2}\right) g_{\phi}\,,\nonumber\\
	(x_{12})^8\delta\langle \Disp^2(x_1) \Disp^2(x_2)\rangle&= - \frac{\pi \D_{\phi} B_\phi}{75\times 2^{\D_\phi+1}}\left[1344+\D_{\phi} (600 c+5 \D_{\phi} (185 \D_{\phi}-692)+4468)\right.\nonumber\\
	&\quad\left.+30 (\D_{\phi}-2) (20 c+25 (\D_{\phi}-2) \D_{\phi}+64) \log \left(\frac{x_{12}^2}{4 a ^2}\right)\right] g_{\phi}\,.
\end{align}
Note that an off-diagonal two-point function is generated at one-loop.

\subsubsection{Renormalization and mixings}
In order to renormalize the two-point functions above we shall essentially repeat the analysis of appendix~\ref{app:pertTTb}. We define the renormalized operators
\begin{align}
	\Disp_R = Z_\Disp \Disp +Z_{\Disp \hid}\hid+\dots\,,\quad 	{\Disp^2}_R= Z_{\Disp^2} \Disp^2+Z_{\Disp^2 \Disp} \Disp+Z_{\Disp^2 \Disp''} \Disp''+Z_{\Disp^2 \hid}\hid+\dots
\end{align}
and fix the wave-functions $ Z_{AA'}$ by requiring correlation functions with insertions of $\Disp_R$ and $\Disp_R^2$ to be finite and to match the expected form of 1d conformal correlators. By letting
\begin{align}\label{counterVIr}
	Z_\Disp &= 1 + g_{\phi}  z_{\Disp} \log(a/R)\,, \quad Z_{\Disp^2} = 1 + g_{\phi} ~z_{\Disp^2}  \log(a/R)\,,\nonumber\\
	&Z_{\Disp^2\Disp} = g_{\phi}~ \frac{z_{\Disp^2\Disp}}{a^2}\,,\quad Z_{\Disp^2\Disp''} = g_{\phi}~ z_{\Disp^2\Disp''}\,,
\end{align}
being $R$ the AdS radius, the wave-functions are found to be
\begin{align}
	z_{\Disp}&= -\frac{B_\phi}{2^{\D_{\phi}}}\frac{4\pi}{c}   (\D_\phi -2) \D_\phi\,,\quad z_{\Disp^2} = -\frac{B_\phi}{2^{\D_\phi}}  \frac{2\pi \D_\phi (\D_\phi -2) (20 c+25 (\D_\phi -2) \D_\phi +64)}{c (5 c+22)}\,,\nonumber\\
	z_{\Disp^2\Disp} &=\frac{\pi  B_\phi  \D_{\phi}}{2^{\D_\phi+1}c}(2 c+5 (\D_{\phi}-2) \D_{\phi}+4)\,,\quad z_{\Disp^2\Disp''} =\frac{\pi  B_\phi}{2^\D_{\phi} \times 25 c}(\D_{\phi}-2) \D_{\phi} (5 \D_{\phi}+2)\,.
\end{align}
From the renormalized two-point correlation functions, comparing to the $O(g_{\TTb})$ expansion of eq.~\eqref{1dcorrRG} we find
\begin{align}
	\dhD_{\Disp}&= - z_{\Disp}\,,\qquad 
	\dhD_{\Disp^2}= - z_{\Disp^2}\,,\nonumber\\
	\delta\hC_{\Disp}&= - \frac{\pi  B_\phi  \D_{\phi}}{2^{\D_{\phi}-1}}{ (\D_{\phi}-2 (\D_{\phi}-2) \log 2)}\,,\nonumber\\
	\delta\hC_{\Disp^2}&= -\frac{\pi  B_\phi\D_{\phi}}{75\times 2^{\D_{\phi}+1}}\left(1344+\D_{\phi} (600 c+5 \D_{\phi} (185 \D_{\phi}-692)+4468)\right.\nonumber\\
	&\quad\left. -60 (\D_{\phi}-2) (20 c+25 (\D_{\phi}-2) \D_{\phi}+64)\log 2\right)\,.
\end{align}

\subsection{Three-point functions}
For the one-loop corrections to three-point functions we shall compute
\begin{align}\label{3pt1LoopVir}
	-g_{\phi}R^{\D_\phi}&\int_{y>a}^\infty \frac{\mathrm{d} y} {y^2}\int_{-\infty}^{\infty}\mathrm{d} x~\langle \Disp(x_1)\Disp(x_2)\Disp(x_3)\phi(x+i y,x-iy)\rangle^{\text{c}}_{0, \AdStwo}\,,\nonumber\\
	-g_{\phi}R^{\D_\phi}&\int_{y>a}^\infty \frac{\mathrm{d} y} {y^2}\int_{-\infty}^{\infty}\mathrm{d} x~\langle \Disp(x_1)\Disp(x_2)\Disp^2(x_3)\phi(x+i y,x-iy)\rangle^{\text{c}}_{0, \AdStwo}\,.
\end{align}
The integrand are again obtain from appropriate limits of correlation functions with many insertions of $T$ and one insertion of $\phi$ on $\uhp$.
From the integrals we find
\begin{align}\label{threeptVir}
	&(x-1)^2 x^2\delta\langle \Disp(1) \Disp(x)\Disp(0)\rangle  = \nonumber\\
	&\qquad\qquad\qquad+\frac{12 \pi  B_\phi (\D_{\phi}-4) \D_{\phi}^2}{2^{\D_{\phi}+2}}\left[1-\frac{2 (\D_{\phi}-2)}{3 (\D_{\phi}-4) \D_{\phi}} \log \left(\frac{(x-1)^2 x^2}{64 a ^6}\right)\right.~\nonumber\\
	&\qquad\qquad\qquad+ \left. \frac{c ((x-1) x+1) \left((x-1) x \left(x^4-2 x^3+x+3\right)+1\right)}{12 (\D_{\phi}-4) \D_{\phi} (x-1)^2 x^2 a ^2}\right]g_{\phi}\,,\nonumber\\
	&\nonumber\\
	&x^4\delta\langle \Disp(1) \Disp(x)\Disp^2(0)\rangle   = \nonumber\\
	&\qquad\qquad\frac{\pi  B_\phi\D_\phi }{15\times 2^{\D_\phi +5}}\left(\frac{240 (5 (\D_\phi -2) \D_\phi +4) x^2 + 480 c x^2}{(x-1)^2 a ^2}-\frac{15 c (5 \D_\phi +2) x^4}{(x-1)^4 a ^4}\right.\nonumber\\
	&\qquad\qquad+\left. \frac{\left(c_0+c_1 x+c_2 x^2\right)}{(x-1)^2} + d_0+d_1 \log a+d_2 \log((x-1)^2) +d_3 \log (x^2)\right)g_{\phi}\,,
\end{align}
with
\begin{align}
	c_0&= -32 (\D_\phi  (\D_\phi  (40 \D_\phi -229)+466)+120)\,,\nonumber\\
	c_1&=64 (\D_\phi  (\D_\phi  (40 \D_\phi -259)+514)+144)\,,\nonumber\\
	c_2  &= -32 (\D_\phi  (\D_\phi  (40 \D_\phi -229)+466)+120)\,,\nonumber\\
	d_0 &= 96 (20 c (\D_\phi  (2\log 2-1)-\log 16)+(\D_\phi -2) (25 (\D_\phi -2) \D_\phi +152) \log 2)\,,\nonumber\\
	d_1 &= 96 (\D_\phi -2) (40 c+25 (\D_\phi -2) \D_\phi +152)\,,\nonumber\\
	d_2 &= 48 (\D_\phi -2) (25 (\D_\phi -2) \D_\phi -24)\,,\nonumber\\
	d_3&= 48 (\D_\phi -2) (-20 c-25 (\D_\phi -2) \D_\phi -64)\,.
\end{align}

\subsubsection{Renormalization and mixings}

One can verify that the wave-functions in the previous section, together with the following mixing terms with the identity
\begin{align}
	Z_{\Disp \hid} = g_\phi \frac{2\pi B_\phi \D_\phi }{2^{2+\D_\phi } a^2}\,,\quad 	Z_{\Disp^2 \hid} = g_\phi \frac{\pi B_\phi \D_\phi  (5 \D_\phi +2)}{2^{4+\D_\phi } a^4},
\end{align}
are enough to remove all divergences in eq.~\eqref{threeptVir}. The resulting functions are conformally covariant, and by comparing them to the $O(g_{\TTb})$ expansion of eq.~\eqref{1dcorrRG} we extract
\begin{align}
	\delta \hC_{\Disp \Disp \Disp} &= \frac{3 \pi B_\phi\D_\phi }{2^{\D_\phi }} \left((\D_\phi -4) \D_\phi +4 (\D_\phi -2) \log 2 \right)\,,\nonumber\\
	\delta \hC_{\Disp \Disp \Disp^2} &=	\frac{\pi B_\phi\D_\phi}{75 \times 2^{\D_\phi }}  \left(-\D_\phi  (300 c+5 \D_\phi  (40 \D_\phi -247)+2474)\right.\nonumber\\
	&+\left.15 (\D_\phi -2)  (40 c+25 (\D_\phi -2) \D_\phi +152)\log 2-672\right)\,.
\end{align}

\subsubsection{Final results for the OPE coefficients}
Taking into account the renormalization of the external operators computed earlier, the OPE coefficients for unit-normalized operators are

\begin{align}
\bdc_{\Disp\Disp\Disp}(g_{\phi})&\equiv	\frac{\hC_{\Disp\Disp\Disp}(g_{\phi})}{\hC_{\Disp\Disp}(g_{\phi})^{3/2}}=\frac{2 \sqrt{2}}{c} \left(1+\frac{B_\phi}{2^{\D_\phi } }\frac{3\pi}{c} (\D_\phi -2) \D_\phi^2\,g_{\phi}+O(g_{\phi}^2)\right)\,,\nonumber\\
\bdc_{\Disp\Disp\Disp^2}(g_{\phi})&\equiv	\frac{\hC_{\Disp\Disp\Disp^2}(g_{\phi})}{\hC_{\Disp\Disp}(g_{\phi})\hC_{\Disp^2\Disp^2}(g_{\phi})^{1/2}}\nonumber\\
	&=\frac{1}{c}{\sqrt{\frac{2}{5}} \sqrt{c (5 c+22)}}{}\left(1+\frac{B_\phi}{2^{\D_\phi }}\frac{\pi(\D_\phi -2) \D_\phi  (5 \D_\phi +2) (25 \D_\phi +336)}{30 c (5 c+22)}g_{\phi}+O(g_{\phi}^2)\right)\,.
\end{align}

\section{\texorpdfstring{$\phi_{(1,2)}$}{phi12} deformations of minimal models}\label{app:phi12def}

In this section we study the $\phi\equiv\phi_{(1,2)}$ deformation of a diagonal minimal model with elementary conformal boundary condition ${\bf a}=(a_1,a_2)_m$, on $\AdStwo$
\begin{align}\label{phi12def}
	\delta S = gR^{\D-2}& \int d^2 x \sqrt{g}\,\phi(x+ i y, x - iy)+\text{counterterms}\,.
\end{align}
The scaling dimension of $\phi$ is, from eq.~\eqref{VirasoroPrimariesmain}
\begin{align}
	\D \equiv \D_{1,2}= 2h_{1,2} = \frac{m-2}{2 (m+1)}\,.
\end{align}
We will work with $m$ finite.
The main result of this section pertains the one-loop anomalous dimension of the boundary Virasoro primary $\psi_{(r,s)}$ (assuming it exists) in a generic conformal boundary condition $\bf a$ and at finite $m$ i.e. 
\begin{align}
	\D_{r,s}=h_{r,s}+g\, \delta\hD_{r,s}+O(g^2)\,.
\end{align}
The tree-level scaling dimension $h_{(r,s)}$ is given in eq.~\eqref{VirasoroPrimariesmain}. In Poincaré coordinates of $\AdStwo$ we shall then study
\begin{align}\label{phi12loop}
	G_1(x_{12})&=R^{\D}\int_{y>a}^\infty \frac{\mathrm{d} y} {y^2}\int_{-\infty}^{\infty}\mathrm{d} x~\langle \psi_{(r,s)}(x_1)\psi_{(r,s)}(x_2)\phi_{(1,2)}(x+i y,x-iy)\rangle^{\text{c}}_{0, \AdStwo}+\text{counterterms}\,.
\end{align}
As usual, here $\langle \dots \rangle^{c}$ means `connected' and $y=a$ is an IR cut-off. The correlation functions in the integrand above is obtained from correlation functions on the upper half-plane via Weyl rescaling.

\subsection{Correlator between two \texorpdfstring{$\psi_{(r,s)}$}{psirs} and one \texorpdfstring{$\phi_{(1,2)}$}{phi12} }\label{app:bdbdbulk3pt}
Our first task is to compute 
\begin{align}\label{bdbdphi12gen}
	\langle \psi_{(r,s)}(x_1)\psi_{(r,s)}(x_2)\phi_{(1,2)}(x+i y,x-iy)\rangle_{\uhp}\,,\quad x_1>x_2\,.
\end{align}
By the method of images, this correlator satisfies the following second order differential equation
\begin{align}\label{thirdgengen}
	\left({\cal L}_{-2}^{(z_4)}-\frac{3}{2(2h_{r,s}+1)}{\cal L}_{-1}^2\right)\langle \psi_{(r,s)}(z_1)\psi_{(r,s)}(z_2)\phi_{(1,2)}(z_3)\phi_{(1,2)}(z_4)\rangle=0\,,
\end{align}
where ${\cal L}_{-n}^{(\cdot)}$ is the differential operator defined in eq.~\eqref{calLdiffopLn} and ${\cal L}_{-1}=\partial_{z_4}$. By $SL(2,\mathbb{R})$ symmetry we have
\begin{align}
	\langle \psi_{(r,s)}(z_1)\psi_{(r,s)}(z_2)\phi_{(1,2)}(z_3)\phi_{(1,2)}(z_4)\rangle=\frac{\cG(\eta)}{(z_{12})^{2h_{r,s}} (z_{34})^{2h_{1,2}}}\,.
\end{align}
The cross-ratio $\eta$ is
\begin{align}\label{eta_def2}
	\eta&= \frac{z_{12}z_{34}}{z_{13}z_{24}}=\frac{2 i y x_{12}}{(x_1-z) (x_2-z^*)}\,.
\end{align}
From \eqref{thirdgengen} we get
\begin{align}
0&=4 \eta  (\eta -1)^2 (m+1)^2 \cG''(\eta )+4 (\eta -1) (m+1) (\eta  (m+2)-2) \cG'(\eta )\nonumber\\
&-\eta  \cG(\eta ) (m r-m s+r-1) (m r-m s+r+1)\,.
\end{align}
In order to solve this equation, it is convenient to define another function
\begin{align}
	\cG(\eta) =\tcG(\tieta)\,,
\end{align}
where
\begin{align}
	\tieta=\frac{\eta^2}{\eta-1}=\frac{4 y^2 (x_{12})^2}{\left((x_1-x)^2+y^2\right) \left((x_2-x)^2+y^2\right)}\,,\quad 0\leq \tieta\leq 4\,. 
\end{align}
The Virasoro blocks corresponding to the exchange of $\hid$ and $\psi_{(1,3)}$ read
\begin{align}\label{allblocks}
	V_{(1,1)}(\tilde\eta)&=\, _2F_1\left(\frac{m (r-s)+r+1}{2 m+2},\frac{m (s-r)-r+1}{2 m+2};\frac{m+3}{2 m+2};\frac{\tieta}{4}\right)\,,\nonumber\\
	V_{(1,3)}(\tilde\eta)&=\tieta^{h_{1,3}/2}  \, _2F_1\left(\frac{m(r-s)+m+r}{2 m+2},\frac{m(s-r)+m-r}{2 m+2};\frac{3 m+1}{2 m+2};\frac{\tieta}{4}\right)\,.
\end{align}
The final solution is then
\begin{align}\label{finalcorrectgen}
	\langle \psi_{(r,s)}(x_1)&\psi_{(r,s)}(x_2)\phi_{(1,2)}(x+iy,x-iy)\rangle_{\uhp}=\frac{\tcG(\tilde\eta)}{(x_{12})^{2h_{r,s}} (2y)^{2h_{1,2}}}\,,\nonumber\\
	&\tcG(\tilde\eta)=B_{(1,2)}^{{\bf a}}V_{(1,1)}(\tilde\eta)+	\bdc_{(r,s)(r,s)(1,3)}^{{\bf a}}B_{(1,2)}^{{\bf a}\,(1,3)}V_{(1,3)}(\tilde\eta)\,,
\end{align}
where $B_{(1,2)}^{{\bf a}}$ was given in eq.~\eqref{B1211ope}.
The remaining coefficient in the equation above is determined by the following requirement. The function $\tcG(\tieta)$ has a branch cut along $\tieta\in [4,\infty]$. None of these singularities correspond to an OPE channel: they are unphysical and so they should disappear~\cite{Lewellen:1991tb}.\footnote{This condition has been exploited in higher dimensions as well: to prove `triviality' of certain free theory conformal defects~\cite{Lauria:2020emq,Herzog:2022jqv}, to constrain the space of conformal boundary conditions for a theory of a free massless scalar field~\cite{Behan:2020nsf,Behan:2021tcn}, to compute perturbative data in $O(N)$ models with boundaries of defects~\cite{Nishioka:2022odm,Nishioka:2022qmj} and in the context of QFTs in AdS \cite{Levine:2023ywq}.} Requiring that $\text{Disc}~\tcG=0$ across the cut one finds
\begin{align}\label{alpha13}
\bdc_{(r,s)(r,s)(1,3)}^{{\bf a}}B_{(1,2)}^{{\bf a}\,(1,3)}&=-B_{(1,2)}^{{\bf a}}\frac{2^{\frac{2}{m+1}-1} \Gamma \left(\frac{1}{2}+\frac{1}{m+1}\right) \Gamma \left(\frac{r m-s m+m+r}{2 m+2}\right) \Gamma \left(\frac{-r m+s m+m-r}{2 m+2}\right)}{\Gamma \left(\frac{3}{2}-\frac{1}{m+1}\right) \Gamma \left(\frac{m r+r-m s+1}{2 m+2}\right) \Gamma \left(\frac{-((m+1) r)+m s+1}{2 (m+1)}\right)}\,.
\end{align}
This formula is consistent with the results from F-matrices~\cite{Runkel:1998he}.

\subsection{Anomalous dimensions}
We have all the ingredients to compute the anomalous dimension of boundary Virasoro primaries along the deformation of eq.~\eqref{phi12def}. Following the same steps as those of section 4.2 in \cite{Lauria:2023uca} we arrive at the following result for the anomalous dimension of $\psi_{(r,s)}$
\begin{align}
	\delta \hD_{r,s} =  \delta\hD_{1,1}+\delta\hD_{1,3}\,,
\end{align}
where
\begin{align}
\delta\hD_{1,1} &= B_{(1,2)}^{{\bf a}}\frac{(m r-m s+r-1) (m r-m s+r+1)}{2^{\frac{3m}{2 m+2}}(m+1) (m+3) }\nonumber\\
&\times\sum_{n=0}^\infty \frac{ \pi\left(1/2\right)_n}{n! (2)_n \left(\frac{3}{2}+\frac{1}{m+1}\right)_n} \left(\frac{3+r+m (r-s+2)}{2 m+2}\right)_n \left(\frac{3-r+m (-r+s+2)}{2 m+2}\right)_n\,,\nonumber\\
\delta\hD_{1,3} &=\bdc_{(r,s)(r,s)(1,3)}^{{\bf a}}B_{(1,2)}^{{\bf a}\,(1,3)}\frac{2^{\frac{m}{2 m+2}}  \Gamma \left(-\frac{1}{m+1}\right)}{\Gamma \left(\frac{1}{2}-\frac{1}{m+1}\right) }\nonumber\\
&\times\sum_{n=0}^\infty\frac{\sqrt{\pi} \left(-\frac{1}{m+1}\right)_n \left(\frac{r m-s m+m+r}{2 m+2}\right)_n \left(\frac{-r m+s m+m-r}{2 m+2}\right)_n}{n! \left(\frac{1}{2}-\frac{1}{m+1}\right)_n \left(\frac{3}{2}-\frac{1}{m+1}\right)_n}\,.
\end{align}

\section{Review of the Staircase model}
\label{app:Staircase}

In the main text we discussed how to study the $(2,2)_4$ b.c. of the tricritical Ising by analyzing the space of  values of the four-point function and its derivatives at the crossing symmetric point. We motivated this by recalling that RG flows between minimal models can be embedded in the so-called staircase RG flows which are associated to the S-matrix of the staircase model.
In this appendix we review the definition and properties of this model and flesh out the connection to our original problem of minimal model RG flows in AdS.

\subsection{Defining properties}
The staircase model is an integrable 2-dimensional quantum field theory, whose S-matrix is obtained by analytic continuation in the coupling of sinh-Gordon (shG) theory as first done by Alyosha Zamolodchikov in an unpublished paper \cite{Zamolodchikov:1992ulx}. It describes the scattering of a single massive scalar without bound-states. The shG S-matrix is a pure CDD-zero:
\begin{equation}
	\label{eq:S-MatrixshG}
	S_{\textrm{shG}}(\theta) = \frac{\sinh\theta-i \sin \gamma}{\sinh\theta+i \sin \gamma}\,,
\end{equation}
where $\gamma$ is related to the sinh-Gordon coupling and $\theta$ is the usual rapidity defined in terms of Mandelstam's $s=(p_1+p_2)^2$ through $ s=2m^2(1+\cosh \theta)$. This S-matrix is invariant under the duality $\gamma \to \pi-\gamma$, which is a weak-strong duality in the original coupling. One then goes to the self-dual point and gives the coupling an imaginary part: $\gamma \to \frac{\pi}{2}+i \theta_0$, leading to the S-matrix:
\begin{equation}
	\label{eq:S-Matrixstaircase}
	S_{\textrm{stc}}(\theta) = \frac{\sinh\theta-i \cosh \theta_0}{\sinh\theta+i \cosh \theta_0}\,.
\end{equation}
This is a perfectly healthy S-matrix with all the right reality and crossing properties. However the Lagrangian nature of the UV theory is completely obscured by this procedure, as it would correspond to a sine-Gordon theory with a complex potential.
It is important to recall that since this is a purely elastic theory, one has access to some off-shell quantities through the Thermodynamic Bethe Ansatz (TBA). In particular, one can obtain the ground state energy on a circle of radius $R$ which is related to the effective central-charge of the theory \cite{Zamolodchikov:1989cf}:
\begin{equation}
	E(R)= -\frac{\pi c_{\textrm{eff}}(x)}{6R}\,,
\end{equation}
where $x= \log(m R/2)$ is a convenient dimensionless scale. This quantity, in the UV and IR matches the central charges of the UV and IR CFTs (which can of course be trivial and have $c=0$), and is an RG monotone. In fact it is the monotonic quantity defined in Sasha Zamolodchikov's $c$-theorem \cite{Zamolodchikov:1986gt}. Solving the TBA equations (numerically), one finds that the IR central charge is 0, as it should for a massive theory, and the UV one is 1, as one might expect from the relation to the shG model. We emphasize that this does not mean that the theory can be described by a UV lagrangian with a massless scalar, since such a theory would have a rather sick potential. 
\begin{figure}
	\centering
	\includegraphics[width=0.7\linewidth]{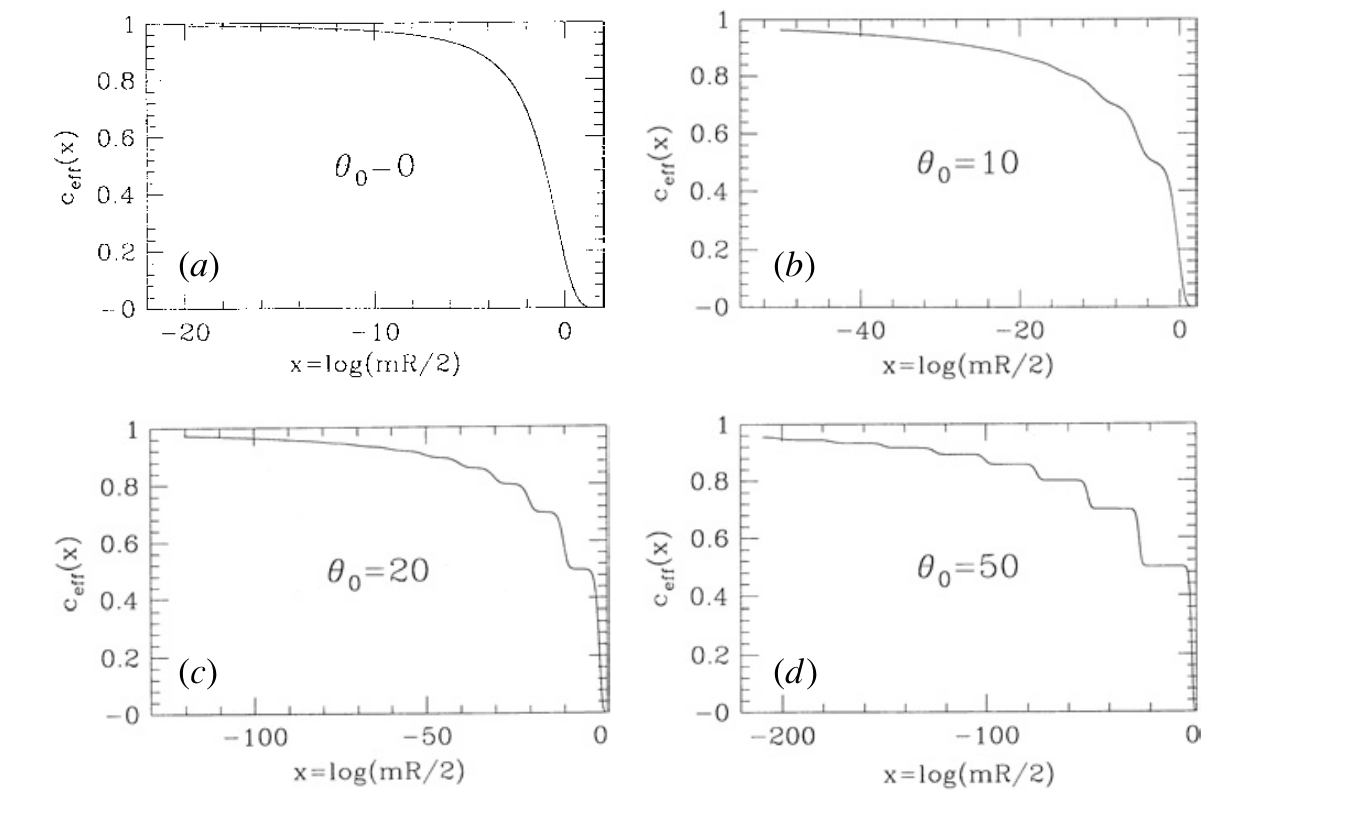}
	\caption{Effective central charge from the TBA equation for the staircase model at several values of $\theta_0$, taken from \cite{Zamolodchikov:1992ulx}. As $\theta_0$ increases, the function develops plateaus which take precisely the minimal model values.}
	\label{fig:centralcharge}
\end{figure}

Looking at the explicit solutions replicated in figure \ref{fig:centralcharge}, one sees that the central charge develops a staircase pattern, spending RG time at central charge plateaus which precisely match the central charges of the unitary minimal models $\MM_m$. Indeed, as $\theta_0\to \infty$, the RG flow of this theory approaches the integrable RG flows between $\MM_m \to \MM_{m-1}$ which are triggered by the integrable, nearly marginal deformation driven by the $\phi_{(1,3)}$ operator in the UV. This operator obviously becomes irrelevant and manifests itself as the $\phi_{(3,1)}$ operator in the IR. In the Ising model (the last plateau), this operator does not exist, and the irrelevant deformation is instead driven by the $T\bar{T}$ operator, as we have seen in the main text.

\subsection{A hint from the S-matrix Bootstrap}
\begin{figure}
	\centering
	\includegraphics[width=0.5\linewidth]{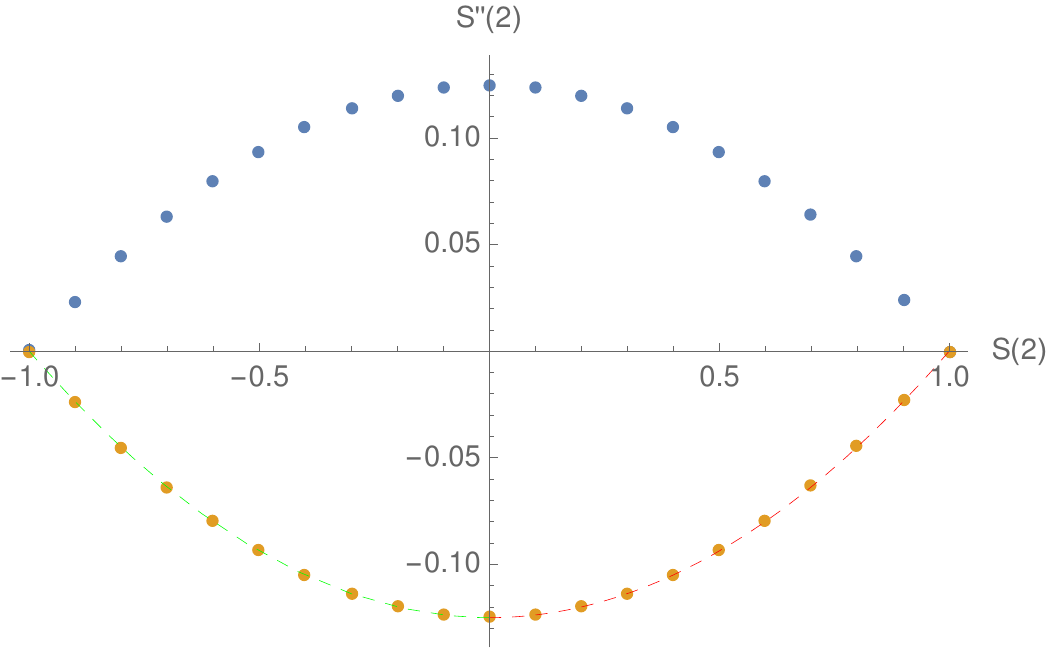}
	\caption{Allowed space of S-matrices for a single scalar particle without bound states projected to the $\{S(2),S''(2)\}$ subspace. The right and left endpoints are a real free boson and fermion, respectively. The bottom right section, in dashed red, are the shG S-matrices as $\gamma$ varies from $0$ to $\pi/2$. Going beyond this brings us back, by duality. Reaching the self-dual point and giving an imaginary part $\theta_0$ from $0$ to $\infty$ builds the green dashed line, corresponding to the staircase models. At large $\theta_0$ we recover the gapped fermion.}
	\label{fig:Smatrixbounds}
\end{figure}

As a simple CDD factor, we expect the shG and staircase S-matrices to saturate S-matrix bounds \cite{Paulos:2016but}. Since they have no poles/bound-states, the natural observable is the effective quartic coupling which we can take to be $S(s=2m^2)$. However this leads to very trivial bounds $-1 \leq S(2)\leq 1$ which are saturated by a free Majorana fermion on the left and a free boson on the right. A natural extension of this is to consider a low energy expansion, which we can take to be the Taylor series around $s=2m^2$. Crossing ensures that $S'(2)=0$, so we can focus on the two dimensional space of parameters $\{S(2),S''(2)\}$. Using the standard numerical S-matrix bootstrap we find the region in figure \ref{fig:Smatrixbounds}. Indeed, the staircase model saturates the bounds, interpolating between the self-dual point of shG and the massive Majorana. A deformed version of this plot was presented previously in the work of \cite{Chen:2021pgx}.
This feature is reminiscent of the $O(2)$ symmetric S-matrices of the sine-Gordon kinks. Indeed, this one parameter family of S-matrices saturates similar bounds, where the 2-dimensional space is instead spanned by 2 components associated to different representations of the $O(2)$ symmetry (say singlet and rank 2 tensor components). In the work of \cite{Antunes:2021abs}, the authors understood how to embed such bounds as a flat space limit of 1d CFT bounds, with the role of the AdS radius being played by the dimension of the external $O(2)$ fundamentals. There it was clear that the UV is well described by free vertex operators, deformed by the sine-Gordon interaction. One can then wonder whether repeating this strategy for the $\mathbb{Z}_2$ symmetric space of correlators labeled by $g(z=1/2)$ and $g''(z=1/2)$ might illuminate the UV origin of the staircase model. This also leads us to the Minimal Model flows in AdS which we studied in the main text.

\subsection{More on Minimal Model RG flows}
From our analysis so far, one thing that remains unclear, is why it should be the specific $(2,2)_4$ b.c. of tricritical Ising and the specific $(1,2)\equiv(3,3)$ operator saturating this bound. While this is not completely obvious,
there is some evidence we can follow:
\begin{itemize}
	\item Our S-matrix has a $\mathbb{Z}_2$ symmetry, so it seems natural to keep it along the flow. Therefore we should pick $\mathbb{Z}_2$ preserving boundary conditions.
	\item The particles are $\mathbb{Z}_2$-odd. We therefore should consider $\mathbb{Z}_2$-odd boundary operators.
	\item For a massless boson in AdS, the natural $\mathbb{Z}_2$ preserving boundary conditions are the Dirichlet ones $\phi|_{\textrm{bdry}}=0$. Although the UV of the staircase model is \textbf{not} a free massless boson, one might be tempted to impose the analogue of the Dirichlet BC along the flow. For the Minimal Models this can be understood from the Landau-Ginzburg formulation. The L-G field $\phi$ corresponds to the lightest $\mathbb{Z}_2$-odd operator which is $\phi_{(2,2)}$ in the Kac table. Indeed, $(2,2)$ boundary conditions are always $\mathbb{Z}_2$ preserving. In this case, the bulk $\mathbb{Z}_2$-even operators will appear in the BOE of $\phi_{(2,2)}$. It seems tempting to consider the lightest boundary operator $\psi_{(3,3)}$ which becomes $\mathbb{Z}_2$-odd in these boundary conditions. These operators satisfy the property that their dimensions become small in the UV limit $m\to \infty$.
\end{itemize}
We can now check this in more detail for the Ising and tricritical Ising cases.
\subsubsection{Ising Model and $T\bar{T}$ deformation}
There is only one $\mathbb{Z}_2$ preserving boundary condition $(2,2)_3=(1,2)_3$ for the Ising model. The boundary theory contains only the identity and the $\psi_{(2,1)}$ modules. The 1d operator $\psi_{(2,1)}$ has dimension $\hD_{(1,2)}=1/2$, which is unsurprisingly dual to a bulk massless free fermion, which corresponds to the boundary GFF correlator. In this language, the irrelevant deformation which takes us back up the RG flow is the $T\bar{T}$ deformation which can be written as a quartic fermion interaction leading to the action
\begin{equation}
	S_{FF}+ S_{T\bar{T}}= \int_{AdS_2} \frac{dx dy}{y^2} (y \psi \bar{\partial}\psi + y \bar{\psi}\partial\bar{\psi}) + g_{\TTb} \int_{AdS_2} \frac{dx dy}{y^2} (y\psi \partial\psi) (y\bar{\psi}\bar{\partial}\bar{\psi})\,,
\end{equation}
Note that on the boundary there can only be one fermionic degree of freedom, corresponding to the identification $\hat{\psi}=-\hat{\bar{\psi}}$. Using standard, but somewhat involved Witten diagram techniques, we can find the first order deformation of the conformal data. Taking the four fermion operator to be normal-ordered is a convenient renormalization scheme in which the external operator doesn't get a leading order anomalous dimensions. Computing the full four-point function, we get:
\begin{align}
	\delta_{T\bar{T}}\mathcal{G}(\eta)&\propto -\frac{8 \eta^3 ((5-2 \eta) \eta-5) \log (\eta)}{(\eta-1)^2 \eta}\\
	& -\frac{8 \left(\left(2 \eta^2+\eta+2\right) (\eta-1)^3 \log (1-\eta)+2 \eta
		((\eta-1) \eta+1) (\eta-1)\right)}{(\eta-1)^2 \eta}\nonumber
\end{align}
The anomalous dimensions and correction to OPE coefficients of two-fermion operators corresponding to this interaction were actually bootstrapped using analytic functionals in \cite{Mazac:2018ycv}. Expanding our answer into blocks matches all their anomalous dimensions and all their OPE coefficients except for the OPE coefficient of the first non-trivial exchanged operator. This is to be expected since they include certain subtraction terms. These results lead to the saturation of the bounds of figure \ref{fig:deltaphi05}, in the main text.
\subsubsection{Tricritical Ising}
For the tricritical Ising, there are of course two different $\mathbb{Z}_2$ preserving boundary conditions, associated to $(2,1)_4$ and $(2,2)_4$. However, only the $(2,2)_4$ BC contains the lightest ($\mathbb{Z}_2$-odd) boundary operator $\psi_{(1,2)}=\psi_{(3,3)}$ of dimension $\D_{(1,2)}=1/10$. In this case, the symmetry of the Kac table means that we can still solve a second order BPZ equation of the (1,2) type as in appendix \ref{app:examplesmm}. Imposing crossing and normalization for the unit operator once again leads to a unique solution:
\begin{align}
	x_{12}^{1/5}x_{34}^{1/5}&\langle \psi_{(1,2)}(x_1)\psi_{(1,2)}(x_3)\psi_{(1,2)}(x_3)\psi_{(1,2)}(x_4)\rangle_{(2,2)_4}\nonumber\\
	&=  \frac{\,
		_2F_1\left(-\frac{2}{5},\frac{1}{5};\frac{2}{5};1-\eta\right)}{\sqrt[5]{1-\eta}}+\frac{(1-\eta)^{
			2/5} \Gamma \left(\frac{7}{10}\right) \Gamma \left(\frac{7}{5}\right) \,
		_2F_1\left(\frac{1}{5},\frac{4}{5};\frac{8}{5};1-\eta\right)}{\sqrt[10]{2}
		\sqrt{\left(3+\sqrt{5}\right) \pi } \Gamma \left(\frac{8}{5}\right)} \,.
\end{align}
We can then plot this point in the bounds for $\D_\psi=1/10$ as we did in figure \ref{fig:corrmaxdeltaphi01} in the main text. It saturates the bound, and is an expected position, somewhat close to the GFF point, in the direction predicted by the $T\bar{T}$ deformation.

If we were to try to backtrack the flows of this boundary conditions to the UV, the prediction of \cite{Lauria:2023uca} would tell us that $(a_1,a_2)$ flows to $(a_2,a_1)$. This is consistent with the picture outlined above, since the RG flows would stick to the $(2,2)$ boundary conditions, containing the lightest $\mathbb{Z}_2$-odd operator. This in contrast to the results in the UHP \cite{Fredenhagen:2009tn}, where our scattering of the lightest $\mathbb{Z}_2$-odd particle can not be consistently embedded in the chain of bulk and boundary RG flows.

\bibliography{bib}
\bibliographystyle{JHEP}

\end{document}